\documentclass[final,3p,times]{elsarticle}

\usepackage{amssymb}
\usepackage{amsmath}
\usepackage{xcolor}
\usepackage{txfonts}

\usepackage{subcaption}
\journal{Comput. Methods Appl. Mech. Engrg }

\begin{document}

\begin{frontmatter}



\title{An efficient mass lumping scheme for isogeometric analysis based on approximate dual basis functions}


\author[a]{Susanne Held}
\ead{susanne.held@b-tu.de}
\author[b]{Sascha Eisentr\"ager}
\ead{sascha.eisentraeger@ovgu.de}
\author[a]{Wolfgang Dornisch}
\ead{wolfgang.dornisch@b-tu.de}

\address[a]{Fachgebiet Statik und Dynamik, Brandenburgische Technische Universit\"at Cottbus-Senftenberg, Konrad-Wachsmann-Allee 2, 03046 Cottbus, Germany}
\address[b]{Institut f\"ur Mechanik, Otto-von-Guericke-Universit\"at Magdeburg, Universit\"atsplatz 2, 39106 Magdeburg, Germany}

\begin{abstract}
In this contribution, we propose a new mass lumping scheme for explicit dynamics in isogeometric analysis (IGA). To this end, an element formulation based on the idea of dual functionals is developed. Non-Uniform Rational B-splines (NURBS) are applied as shape functions and their corresponding dual basis functions are applied as test functions in the variational form, where two kinds of dual basis functions are compared. The first type are approximate dual basis functions (AD) with varying degree of reproduction, resulting in banded and diagonally-dominant mass matrices. Dual basis functions derived from the inversion of the Gram matrix (IG) are the second type and already yield diagonal mass matrices. The support along the knot spans is enlarged in comparison to the corresponding NURBS for the chosen kinds of dual basis functions to ensure continuity of test functions. Thus, computational costs for assembling the system matrices are higher in comparison to common Bubnov-Galerkin formulations based on NURBS. We will show that it is possible to apply the dual scheme as a transformation of the resulting system of equations based on NURBS for both -- shape and test functions. Hence, it can be easily implemented into existing IGA routines and it is also promising to retain the accuracy known from similar formulations without mass lumping. Treating the application of dual test functions by employing a transformation operator reduces the additional computational effort, but it cannot entirely erase it and the density of the stiffness matrix remains higher than in standard Bubnov-Galerkin formulations. In return, applying additional row-sum lumping to the mass matrices is either not necessary for IG or the caused loss of accuracy is lowered to a reasonable magnitude in the case of AD. Using these dual basis functions as test functions is unattractive for static computations, but within explicit dynamic calculations it yields a very promising approach. Numerical examples show a significantly better approximation of the dynamic behavior for the dual lumping scheme compared to standard NURBS approaches making use of conventional row-sum lumping. Applying IG yields accurate numerical results without additional lumping. But as result of the global support of the IG dual basis functions, fully populated stiffness matrices occur, which are entirely unsuitable for explicit dynamic simulations. The support of AD is also raised, but only locally resulting in banded stiffness and mass matrices, with bandwidths depending on the chosen reproduction order. In a nutshell, combining AD and row-sum lumping leads to efficient dynamical computations, with respect to effort and accuracy.

\end{abstract}


\begin{highlights}
\item We propose high order convergent dual test function-based mass lumping for IGA.
\item Inverse Gramian and approximate dual test functions are implemented and analyzed.
\item It is shown that the proposed approach significantly outperforms conventional row-sum lumping.
\item The implementation into existing IGA codes is achieved by means of a transformation operator.
\item The dual test function-based mass lumping can be included in any existing code as a black-box operation.
\end{highlights}


\end{frontmatter}


\section{Introduction}
\label{sec:intro}
The still dominating finite element method (FEM) requires a costly model conversion as commonly linear Lagrange polynomials are applied as shape functions, while the geometry description in computer-aided design (CAD) software is based on non-rational functions of higher order. For approximation of the original geometry, and to get results of reasonable quality, a large number of elements is needed, which raises the computational costs for two reasons. Firstly, a higher number of elements yields higher meshing costs and secondly, the costs of solving the resulting system of equations (SOE) is also increased, where it has to be noted that the solution step is a very costly part of each finite element analysis. This time consuming process depends on the sparsity and the size of the system matrices and thus, increases the numerical effort according to the number of elements. 

Therefore, in 2005 isogeometric analysis (IGA) was introduced by \citet{Hughes.2005} as a powerful method to bridge the gap between the fields of design and analysis, and to avoid the unnecessary high effort of transferring the considered problem. Its basic idea is to make use of the basis functions from the CAD geometry model in the analysis part as well. That fundamental concept offers the potential for taking the analysis geometry directly from the design model, in order to waive further steps for model conversion. Furthermore, using one common model for both processes enables an exact geometry description on element level \cite{Cottrell.2009}. As already mentioned, the CAD software builds up the geometry based on higher-order polynomials, i.e., splines. Usually Non-Uniform Rational B-splines (NURBS) are used, but in the context of IGA a variety of geometry descriptions exists. Next to familiar B-splines \cite{deBoor.1972} to name a few other geometrical descriptions, T-splines \cite{Bazilevs.2010}, S-splines \cite{Li.2019}, or U-splines \cite{Thomas.2022} can be mentioned. Within this work, NURBS, examined in detail in \cite{Piegl.1997}, are considered as basis functions. They provide high inter-element continuity and have also further advantageous mathematical properties \cite{Hughes.2017}. Raising the order of the splines also increases the convergence rate of isogeometric element formulations. Hence, computations with results of reasonable accuracy are enabled by basis functions of high order and continuity, while keeping the amount of elements on a low level. Within the concept of IGA, the computational effort of structural analysis is shifted from solution to assembly, compared to classical FEM \cite{Cottrell.2009}. This fact was already addressed in many contributions on efficient integration rules to lower the general computational effort of IGA and to alleviate locking phenomena. The most basic and widespread quadrature schemes are full and reduced Gauss integration \cite{Hughes.2005,Hughes.2010,Zou.2022}. The focus of current research activities is on integration rules, which are computed from moment fitting equations \cite{Adam.2014,Hiemstra.2017,Johannessen.2017,Dornisch.2020,Zou.2021}. Although the number of integration points and as a consequence also the computational costs of assembly can be reduced, the bandwidth of the system matrices cannot. As an effect of the high continuity provided by NURBS basis functions, raising the order leads to denser matrices. 

Dynamic calculations are more costly than static ones as a result of the additional time discretization. The solution of the SOE for dynamic analyses requires the inversion of the effective stiffness matrix, which is assembled from mass, damping and stiffness matrices according to the chosen time integration method. Hence, the density of these system matrices has a high influence on the computational costs. Geometrically nonlinear approaches are especially costly, as every single time step demands the mentioned inversion or an appropriate solution technique. In the simplified linear case with invariant system matrices, this additional computational effort can be cut down to simply occupying memory for the pre-inverted or decomposed matrix. In this contribution, we will focus on explicit dynamics for geometrically linear applications. Explicit time integration methods are generally only conditionally stable and therefore, require a very fine time resolution depending on the highest numerical eigenfrequency. Hence, explicit methods are computationally more costly than implicit ones, which are often unconditionally stable. However, the computational costs per time step of explicit time integration are easier to reduce, as for the undamped or mass-proportionally damped case only an inversion of the mass matrix is required. Diagonal mass matrices will reduce the computational effort of matrix inversions, increasingly noticeable for systems with a large amount of degrees of freedom (DOF) or large bandwidth. Thus, mass lumping is indispensable for explicit dynamic calculations based on basis functions of high order. Well-known techniques to obtain diagonal mass matrices, such as the diagonal scaling method or row-sum lumping were initially developed for the application in dynamic analyses within the classical FEM \cite{Zienkiewicz.2000,Hughes.1987,Cook.1989}. Thus, employing $k$-refinement (order elevation followed by knot insertion) to obtain results of high accuracy at a low level of computational costs, and applying standard lumping techniques in order to keep up with this aim, seems logical in the context of IGA. Unfortunately, for common element formulations using higher order basis functions in the field of dynamics, these techniques lead to highly inaccurate results \cite{Duczek.2019a}, unless spectral finite elements are used \cite{Duczek.2019b}. Lumping schemes transform the highly accurate mass matrices of isogeometric element formulations into diagonal matrices by simple algorithms and that is why they deteriorate the attainable convergence rates. Diagonalizing the mass matrix without taking other measures causes a high loss of accuracy. Hence, in the context of time-dependent IGA, it is necessary to think of other approaches like preconditioners or special kinds of element formulations.

The objective of this study is to construct mass matrices on which lumping does not have a detrimental impact. It will be shown that this method can be interpreted as a novel element formulation, but also as an extension of already established formulations. The approach is based on the use of dual basis functions. For IGA shell elements, B\'{e}zier dual basis functions were already investigated for static applications to overcome locking \cite{Zou.2020}. The most common use of dual basis functions in the context of IGA is in the isogeometric mortar method \cite{Dornisch.2017,Zou.2018,Dornisch.2021}. Already in 2009, \citet{Cottrell.2009} came up with the idea of \textit{dual lumping} with the explicitly defined dual basis functions of \citet{Schumaker.2007}. Due to its incompleteness and difficult construction, this proposed dual basis was not considered for further investigations. \citet{Anitescu.2019} introduced a dual-basis diagonal mass formulation in 2019, yielding accuracy comparable to consistent mass formulations. For the applied dual basis functions, they allow discontinuities at the element boundaries. In contrast, our study on dual lumping in IGA dynamics makes use of inverse Gramian and approximate dual basis functions, already introduced by \citet{Dornisch.2017} for the coupling of patches. These easily to compute functions are continuous and thus, the formulation is not limited to specific quadrature schemes. Parallel to our research on the use of approximate dual basis functions as test functions for mass lumping, a similar method has been proposed in \cite{Nguyen.2023}. There, the general applicability of the idea of using approximate dual mass lumping to higher dimensions and $4^{th}$ order partial differential equations (PDE), namely 2D plates and Kirchhoff-Love shells, is shown. Our research differs in the application to $2^{nd}$ order PDEs and a more precise treatment, regarding duality, of Dirichlet boundary conditions and the idea of the transformation operator, which simplifies and speeds up the numerical integration significantly in comparison to \cite{Nguyen.2023}. In contrast to other approaches in the literature, we also present the use of approximate dual basis functions for NURBS.

The paper is organized as follows: At first, a brief introduction to NURBS is given in Section \ref{sec:NURBS}, followed by the basics on dual functionals in Section \ref{sec:dual}. As B\'{e}zier dual basis functions were already discussed in detail in other contributions \cite{Dornisch.2017,Zou.2018,Zou.2020,Anitescu.2019}, we will only focus on the inverse Gramian and approximate dual basis functions. In Section \ref{sec:formulation}, we derive the element formulation for our investigations. Here, the dual basis functions are applied as test functions in the context of a Petrov-Galerkin method. We also provide the transformation of the corresponding Bubnov-Galerkin discretized formulation to our dual approach to demonstrate the straightforward implementation of the new lumping scheme. At this point, we want to stress that a conventional Bubnov-Galerkin formulation can be easily converted to the proposed Petrov-Galerkin formulation by means of a suitable transformation operator. That is to say, a simple pre-multiplication of a transformation matrix, which we rigorously derive in this contribution, suffices to implement the novel mass lumping scheme. Thus, it is easy to add the novel technique to existing codes and simply treat it as a black box approach. Numerical accuracy and costs are considered in Section \ref{sec:examples}. An insight on static convergence is shown, to first validate the element formulation and study the influence of the different methods to treat Dirichlet boundary conditions. For dynamic applications, we study the accuracy regarding spectral curves of the eigenvalue problem and the response to an external loading. In addition to contrasting the two examined dual basis functions, the accuracy is also compared to the diagonal mass formulation of \citet{Anitescu.2019}. Finally, we assess the computational costs and numerical accuracy to measure the overall efficiency of our lumping scheme. All gained results are reviewed in the conclusions provided in Section \ref{sec:conclusion}.

\section{Basic NURBS terminology}
\label{sec:NURBS}
Within this work, the application of dual basis functions is studied for one-dimensional NURBS basis functions. Therefore, an overview on all necessary definitions and notation for NURBS curves is given.\\

Piecewise polynomial B-spline curves $\mathbf{C}(\xi)$ of order $p$
\begin{equation}
\mathbf{C}(\xi)=\sum_{i=1}^n{N_i^p(\xi)~\mathbf{B}_i}
\label{eq:curve}
\end{equation}
are linear combinations of $n$ B-spline basis functions $N_i^p(\xi)$ with coefficients gained from the coordinates of the corresponding control points $\mathbf{B}_i$. The basis functions and thus, also the curve are defined across its parametric direction $\xi$. The coordinates of the spanned parametric space are arranged non-decreasingly in the knot vector $\mathbf{\Xi}=\{\xi_1,\xi_2,...,\xi_{n+p+1}\}$, such that $\xi_k\leq\xi_{k+1}$ holds true for $k=1,...,n+p$. These knots divide the underlying geometry -- not necessarily uniform -- into elements.\\

\begin{figure}[t]
\centering
\includegraphics[width=0.5\textwidth]{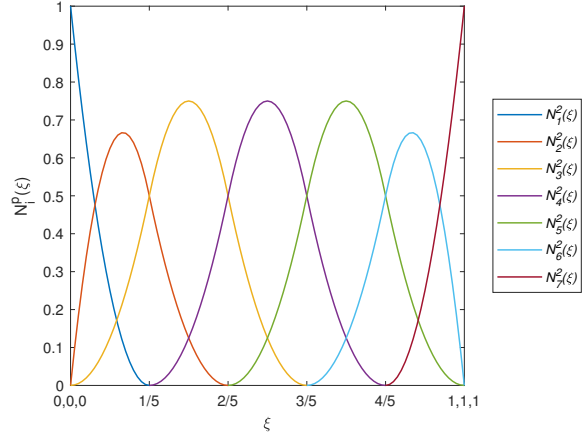}
\caption{B-spline basis functions $N_i(\xi)$ of order $p=2$ for an open knot vector $\mathbf{\Xi}=[0,0,0,\frac{1}{5},\frac{2}{5},\frac{3}{5},\frac{4}{5},1,1,1]$.}
\label{fig:basisfun}
\end{figure}

Bernstein polynomials are chosen as basis functions for B-spline curves. For an initial order of $p=0$, they are defined as 
\begin{equation}
N_i^0(\xi) = \begin{cases} 1\quad\text{if}\quad \xi_i\leq\xi\leq\xi_{i+1}\\0\quad\text{otherwise} \end{cases}~.
\end{equation}
For every higher polynomial order $p>0$, the univariate B-spline basis functions are defined recursively by
\begin{equation}
N_i^p(\xi) = \frac{\xi-\xi_i}{\xi_{i+p}-\xi_i}N_i^{p-1}(\xi)+\frac{\xi_{i+p+1}-\xi}{\xi_{i+p+1}-\xi_{i+1}}N_{i+1}^{p-1}(\xi)~.
\end{equation}
Thus, the $1^{\mathrm{st}}$ derivative of B-spline basis functions is computed by
\begin{equation}
\frac{d}{d\xi}N_i^p(\xi)=\frac{p}{\xi_{i+p}-\xi_i}N_i^{p-1}(\xi)-\frac{p}{\xi_{i+p+1}-\xi_{i+1}}N_{i+1}^{p-1}(\xi)~.
\end{equation}
Derivatives of higher order can be obtained following a similar scheme, but within this work no higher order differentiation is required. An example for quadratic basis functions is shown in Fig.~\ref{fig:basisfun}. There it can be seen that the support of each basis function equals $p+1$ knot spans.

The continuity of the curve at a specific knot is determined by its multiplicity $m$ and is $C^{p-m}$. A $p+1$ times repetition of the first and last entry in the knot vector refers to a so-called open knot vector. Commonly open knot vectors are used within IGA, as they form basis functions which are interpolatory at the outer knots, but not in general at the interior ones. Maximum continuity of $C^{p-1}$ is achieved by unique interior knots. Applying $k$-refinement instead of $p$-refinement to elevate the order of the B-spline curves ensures keeping the maximum continuity.

Besides the physical coordinates $\mathbf{X}_i$, an additional weight $w_i$ is introduced for each control point $\mathbf{B}_i$ to extend B-spline curves to NURBS curves 
\begin{equation}
\mathbf{B}_i = \left[\mathbf{X}_i^{\mathrm{T}},w_i\right]^{\mathrm{T}}~.
\end{equation}
The $p^{\mathrm{th}}$ order NURBS basis functions are then defined by 
\begin{equation}
R_i^p(\xi)=\frac{N_i^p(\xi)~w_i}{W(\xi)}
\end{equation}
with the weighting function
\begin{equation}
W(\xi)=\sum_{i=1}^n{N_i^p(\xi)~w_i}~.
\end{equation}
In analogy to Eq.~(\ref{eq:curve}), a NURBS curve is now computed as 
\begin{equation}
\mathbf{C}(\xi)=\sum_{i=1}^n{R_i^p(\xi)~\mathbf{B}_i}~,
\end{equation}
where $\mathbf{B}_i$ refers actually only to the physical coordinates $\mathbf{X}_i$. For two- or three-dimensional cases, NURBS are set up with a tensor product structure. Note that all concepts introduced within this work can be generalized to higher dimensional cases, recalling this tensor product structure.
 
\section{Dual functionals}
\label{sec:dual}
Let $\mathbf{N}=\{N_i\}_n$ be a set of $n$ one-dimensional spline basis functions and $\boldsymbol{\lambda}=\{\lambda_i\}_n$ its associated dual basis functions, both defined on the interval $I$. Than, the fundamental definition of the dual functional $f_i$ is
\begin{equation}
f_i(N_j)=\int\limits_{I_{j}}N_j\lambda_i~\mathrm{d}s~.
\label{eq:dual_def}
\end{equation}
The integral on the product of the spline $N_j$ with support on the interval $I_j$ and its dual basis function $\lambda_i$ is evaluated over the interval $I_j$. The main characteristic of a dual functional $f_i$ is the so-called bi-orthogonality
\begin{equation}
f_i(N_j)=\delta_{ij}~.
\label{eq:dual_prop}
\end{equation}
For an arbitrary spline basis function $N_j$ the dual functional equals the Kronecker delta 
\begin{equation}
\delta_{ij}= \begin{cases} 1\quad\text{if}\quad i=j\\0\quad\text{otherwise} \end{cases}~.
\end{equation}
The definition provided by Eq.~(\ref{eq:dual_def}) is quite similar to the computational approach of consistent mass formulations and naturally leads to the idea of dual lumping. Dual basis functions are applied as test functions and the corresponding B-Splines are kept as shape functions. Taking into account the definition of the Kronecker delta and the expression for the consistent mass matrix in the one-dimensional case
\begin{equation}
m_{ij}=\int\limits_{\Omega}{\rho~\lambda_iN_j~\mathrm{d}s}~,
\label{eq:m_Bspline}
\end{equation}
the entries of the mass matrix $m_{ij}$ are zero if $i \neq j$ and the density $\rho$ is a constant. A diagonal mass matrix is received immediately, and therefore, no further lumping is necessary.

Combining Eqs.~(\ref{eq:dual_def}) and (\ref{eq:dual_prop}), the criterion for constructing dual basis functions can be rewritten as
\begin{equation}
\int\limits_{I_j}{N_j\lambda_i~\mathrm{d}s=\delta_{ij}}~.
\end{equation}
Of course, this requirement could be met by a variety of functions, but for isogeometric purposes only functions are of interest which suit the already established routines and algorithms. Therefore, in many contributions an explicit formula \cite{Anitescu.2019, Zou.2020, Zou.2018, Dornisch.2017} based on Bézier extraction or the de Boor algorithm is proposed, keeping the initial B-spline support of $p+1$ knot spans. Unfortunately, these dual bases lead to discontinuities at element boundaries. In case of applying the functions from \cite{Schumaker.2007}, discontinuities appear even within knot spans. That is what \citet{Cottrell.2009} already denoted as downside of this approach, as it is not compatible with standard quadrature schemes.

As a consequence of keeping the initial support, the polynomial reproduction
\begin{equation}
x^r=\sum\limits_{j=1}^n{c_j^r\lambda_j} ~~\text{with}~~ c_j^r=\int\limits_{I}{x^rN_j~\mathrm{d}s}
\label{eq:reproduction}
\end{equation}
is usually limited to a degree $r<p$, where $p$ is the initial order of the underlying B-splines.

Below, we want to provide the basics of two kinds of duals, already introduced in \cite{Dornisch.2017}, relying on the idea of extending the support to ensure overall continuity and enable a full reproduction of $r=p$. The resulting dual functions 
$\boldsymbol{\lambda}(\xi)=[\lambda_1(\xi),\lambda_2(\xi),\cdots,\lambda_n(\xi)]^{\mathrm{T}}$ 
can be expressed as a linear transformation of the corresponding B-splines 
$\mathbf{N}(\xi)=[N_1(\xi),N_2(\xi),\cdots,N_n(\xi)]^{\mathrm{T}}$ 
by
\begin{equation}
\boldsymbol{\lambda}(\xi)=\mathbf{S}(\mathbf{\Xi})\mathbf{N}(\xi)~,
\label{eq:transformation}
\end{equation}
where $\mathbf{S}(\mathbf{\Xi})$ denotes the transformation operator, which is an essential feature of our implementation, as we will see in the following.

\subsection{Inverse Gramian dual basis functions}
\begin{figure}[t]
\centering
\includegraphics[width=0.5\textwidth]{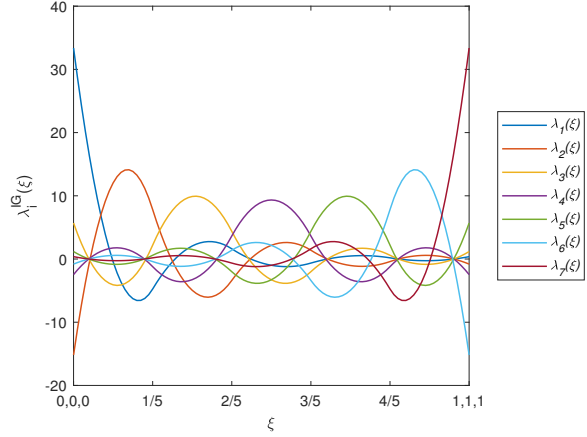}
\caption{Inverse Gramian dual basis functions $\lambda_i(\xi)$ corresponding to the B-spline basis functions shown in Fig.~\ref{fig:basisfun}.}
\label{fig:IG_basisfun}
\end{figure}

First, we consider to extend the support to its maximum, i.e., to the global one. In this case, the inverse of the Gramian $\mathbf{G}^{-1}$ is employed for the transformation. Thus, they are denoted as inverse Gramian (IG) dual basis functions. Here, the Gram matrix $\mathbf{G}_{N,N}$ is computed for the space spanned by the B-splines itself
\begin{equation}
\mathbf{G}_{N,N}=\int\limits_{\Omega}{\mathbf{N}_i^{\mathrm{T}}\mathbf{N}_j~\mathrm{d}s}~.
\end{equation} 
The computation yields a symmetric, block-diagonal, and always invertible matrix \cite{Dornisch.2017,Dornisch.2021,Lawson.1995,Dornisch.2015}. The inverse of the Gramian is fully populated and leads to dual basis functions
\begin{equation}
\boldsymbol{\lambda}(\xi)=\mathbf{G}_{N,N}^{-1}\mathbf{N}(\xi)
\end{equation}
with global support as can be seen in Fig.~\ref{fig:IG_basisfun}.
The computation of a mass matrix using $\boldsymbol{\lambda}(\xi)$ as test functions yields a consistent, but also diagonal mass matrix. Furthermore, also the requirement of reproduction (\ref{eq:reproduction}) is fulfilled up to the initial degree $p$.

\subsection{Approximate dual basis functions}
\begin{figure}[t]
\centering
\includegraphics[width=0.5\textwidth]{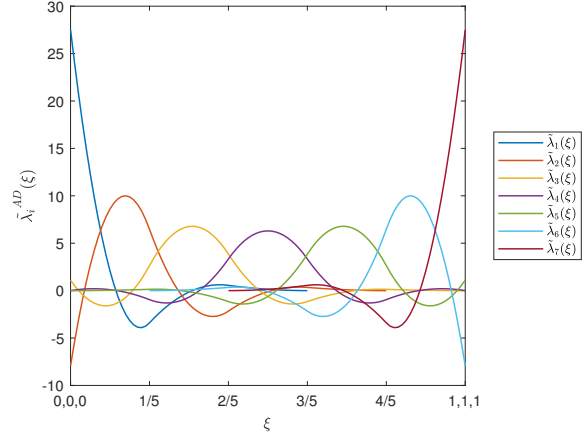}
\caption{Approximate dual basis functions $\tilde{\lambda}_i(\xi)$ corresponding to the B-spline basis functions shown in Fig.~\ref{fig:basisfun}.}
\label{fig:AD_basisfun}
\end{figure}

The second type of dual basis functions being considered are approximate dual basis functions, which also offer a maximum reproduction degree of $p$. Initially introduced in \cite{Chui.2004} in the context of harmonic analysis, they have been used in \cite{Dornisch.2017} for an isogeometric mortar method. In contrast to the IG duals, the initial B-spline support of $p+1$ is extended, but remains local. It is enlarged only minimally to a support, that guarantees continuity along the whole domain. Unfortunately, the bi-orthogonality criterion (\ref{eq:dual_prop}) cannot be fulfilled exactly, but approximately. Hence, this type of duals is denoted as \textit{approximate} dual (AD) basis functions. This circumstance is also marked through a tilde above the dual function vector as defined in Eq.~(\ref{eq:AD}). Considering a freely selectable reproduction degree $q\leq p$, the AD duals are computed by the following transformation of the initial spline space
\begin{equation}
\widetilde{\boldsymbol{\lambda}}(\xi)=\mathbf{S}_q(\mathbf{\Xi})\mathbf{N}(\xi)~.
\label{eq:AD}
\end{equation}
The transformation matrix $\mathbf{S}_q$ is given by
\begin{equation}
\mathbf{S}_q(\mathbf{\Xi})=\mathbf{U}_{\boldsymbol{\Xi};0} + \sum\limits_{v=1}^{q} \left[ \left( \prod\limits_{k=1}^{v} \mathbf{D}_{\boldsymbol{\Xi};p+k} \right) \mathbf{U}_{\boldsymbol{\Xi};v}  \left( \prod\limits_{k=1}^{v} \mathbf{D}_{\boldsymbol{\Xi};p+k} \right)^\mathrm{T} \right]~,
\end{equation}
see \citet{Chui.2004}.
Further definitions for the computation of the required terms will be presented below. First, the diagonal matrices $\mathbf{U}_{\mathbf{\Xi};v}$ have to be computed for $v=0,...,q$
\begin{equation}
\mathbf{U}_{\mathbf{\Xi};v}=\mathrm{diag}\left[u_{\mathbf{\Xi};p,1}^{(v)},\cdots ,u_{\mathbf{\Xi};p,n-v}^{(v)}\right]~.
\end{equation}
Their entries
\begin{equation}
u_{\mathbf{\Xi};p,j}^{(v)}=\frac{p+v+1}{\xi_{j+p+v+1}-\xi_j} \beta_{\mathbf{\Xi};p,j}^{(v)}
\end{equation}
with
\begin{equation}
\beta_{\mathbf{\Xi};p,j}=\left\{\begin{array}{ll}
	1 & ,~v=0 \\
	\frac{(p+1)!~(p-v)!}{(p+v+1)!~(p+v)!} F_v(\xi_{j+1},\cdots ,\xi_{j+p+v}) & ,~1\leq v\leq q
\end{array}\right.
\end{equation}
are all positive and necessitate an evaluation of the polynomial $F_v$ for $v>0$
\begin{equation}
F_v(x_1,\cdots ,x_r)=\frac{2^{-v}}{v!}\sum_{1\leq i_1,\cdots ,i_{2v}\leq r}{~\prod_{j=1}^v{\left(x_{i_{2j-1}}-x_{i_{2j}}\right)^2}}~.
\end{equation}
Next, for $r=p+1,...,p+q$ the matrices
\begin{equation}
\mathbf{D}_{\mathbf{\Xi};r}=\mathrm{diag}\left[d_{\mathbf{\Xi};r,1},\cdots ,d_{\mathbf{\Xi};r,n+p+1-r}\right]\mathbf{\Delta}_{n+p+1-r}
\end{equation}
have to be set up as the product of a diagonal matrix with the entries
\begin{equation}
d_{\mathbf{\Xi};r,j}=\frac{r}{\xi_{j+r}-\xi_j}
\end{equation}
and the rectangular banded matrix
\begin{equation}
\mathbf{\Delta}_{s}=\left[\begin{array}{rrrr}
	1 & & & \\
	-1 & 1 & & \\
	 & \ddots & \ddots & \\
	 & & -1 & 1 \\
	 & & & -1
\end{array}\right]_{s\times (s-1)}~.
\end{equation}
At the end, we obtain a symmetric and positive definite transformation matrix $\mathbf{S}_q(\mathbf{\Xi})$. In Fig.~\ref{fig:AD_basisfun}, an example of AD duals is shown. It can be clearly observed that they constitute continuous basis functions with a larger, but still local support. The support is enlarged to $p+2q+1$ basis functions per knot span. 
As these kinds of duals slightly violate the basic criteria of dual basis functions, we already denoted them as approximate ones. They will lead to banded, but diagonally-dominant mass matrices with significantly decreasing values of the off-diagonal terms. The computational effort of assembling the transformation matrix $\mathbf{S}_q$, its bandwidth and therefore also the resulting mass matrix, depends on the chosen reproduction degree $q$. Full reproduction is possible, but not necessarily needed as the received banded mass matrices require further lumping for an efficient computation, independent of $q$. Thus, in Sec.~\ref{sec:examples} the impact of reducing the reproduction degree on computational accuracy and efficiency will be studied to be able to weigh costs and benefits of this method.
\\Here, we only present the explicit construction of AD duals. For the mathematical background on approximate duals and a more detailed computation, especially of $F_v$, see \cite{Chui.2004}. It is to be noted that since $\mathbf{S}_q$ is constructed explicitly and needs to be computed only once per knot vector, the computational effort of computing $\mathbf{S}_q$ is very low and entirely negligible in comparison to the solution routines of the SOE.

\section{Isogeometric element formulation based on dual test functions}
\label{sec:formulation}
In this contribution, the main focus is on the application of duals as test functions. Therefore, we will stick with a simple one-dimensional element formulation with only one DOF per control point to investigate the fundamental properties as a basis for future studies. 

\subsection{Formulation for one-dimensional linear elasticity}
\begin{figure}[b]
\centering
\includegraphics[width=0.35\textwidth]{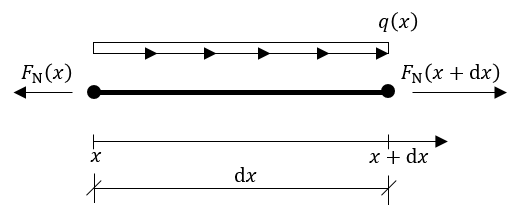}
\caption{Equilibrium on element level.}
\label{fig:PDE}
\end{figure}
The underlying partial differential equation to solve for this one-dimensional problem can be easily derived from Fig.~\ref{fig:PDE}, where $F_\mathrm{N}(x)$ and $q(x)$ are the distributions of normal forces and loading along the axial direction $x$, respectively. Considering the inertial force $F_\mathrm{I}=\rho(x)\mathrm{d}x~a$, where $a$ is the acceleration and $\rho(x)$ the mass density, this yields 
\begin{equation}
\frac{\mathrm{d}}{\mathrm{d}x}F_\mathrm{N}(x)+q(x)-\rho(x)a=0~.
\label{eq:PDE}
\end{equation}
The material law and kinematics \cite{Sadd.2014} are reduced to their components in $x$-direction 
\begin{equation}
\sigma = E\varepsilon
\label{eq:material}
\end{equation}
and
\begin{equation}
\varepsilon = \frac{\mathrm{d}}{\mathrm{d}x}u~,
\label{eq:kinematics}
\end{equation}
where $E$ is Young's modulus, $\sigma$ the stress, $\varepsilon$ the strain, and $u$ the displacement, each in axial direction. Following the Petrov-Galerkin routine, Equation~(\ref{eq:PDE}) is transferred into the weak form 
\begin{equation}
\int\limits_\mathcal{L}\left(\frac{\mathrm{d}}{\mathrm{d}x}F_\mathrm{N}(x)+q(x)-\rho(x)a\right)\delta u~\mathrm{d}s=0~.
\label{eq:PDE_var}
\end{equation}
We assume $u$ to be a scalar field $u \in \mathcal{V}$ with $\mathcal{V}=\left\{u\in\left(H^1(\mathcal{L})\right)^1 |u=0~\mathrm{on}~\partial\mathcal{L}_\mathrm{D}\right\}$ on the one-dimensional domain $\mathcal{L}$. The corresponding test function is chosen from a subspace $\delta u\in \mathcal{W}\supseteq \mathcal{V}$. 
Integration by parts is performed once, using the notation $(\cdots)_{,i}=\frac{\mathrm{d}}{\mathrm{d}i}(\cdots)$. This yields
\begin{equation}
\int\limits_\mathcal{L}\left(F_\mathrm{N}~\delta u\right)_{,x}~\mathrm{d}s-\int\limits_\mathcal{L}F_\mathrm{N}~\delta u_{,x}~\mathrm{d}s+\int\limits_\mathcal{L}q~\delta u~\mathrm{d}s-\int\limits_\mathcal{L}\rho a~\delta u~\mathrm{d}s=0~.
\label{eq:PDE_pInt}
\end{equation}
Introducing the normal vector $\mathbf{n}$ along the boundary $\partial\mathcal{L}$, the first term of Eq.~(\ref{eq:PDE_pInt}) can be also expressed as 
\begin{equation}
\int\limits_\mathcal{L}\left(F_\mathrm{N}~\delta u\right)_{,x}~\mathrm{d}s=n_x F_\mathrm{N}\delta u|_{\partial\mathcal{L}}~.
\label{eq:bound}
\end{equation}
For the one-dimensional case, the normal vector is reduced to its component $n_x$ in $x$-direction. We assume homogeneous Dirichlet boundary conditions for the test function $\delta u=0$ on $\partial\mathcal{L}_\mathrm{D}$ with $\partial\mathcal{L}=\partial\mathcal{L}_\mathrm{D}\cup\partial\mathcal{L}_\mathrm{N}$. Thus, only Neumann boundary conditions on $\partial\mathcal{L}_\mathrm{N}$ remain in Eq.~(\ref{eq:bound}), covering the occurrence of axial point loads $F$. With $F_\mathrm{N}=F$ on $\partial\mathcal{L}_\mathrm{N}$, the relation $F_\mathrm{N}=\sigma A$ and including Eqs.~(\ref{eq:material}) and (\ref{eq:kinematics}) into Eq.~(\ref{eq:PDE_pInt}) leads to the formulation
\begin{equation}
\int\limits_\mathcal{L}\delta u~\rho~a~\mathrm{d}s+\int\limits_\mathcal{L}\delta\varepsilon~EA~\varepsilon~\mathrm{d}s=F\delta u|_{\partial\mathcal{L}_\mathrm{N}}+\int\limits_\mathcal{L}q~\delta u~\mathrm{d}s~.
\label{eq:formulation}
\end{equation}

\subsection{NURBS-based isogeometric Petrov-Galerkin discretization}
\label{sec:discretization}
The discretization of the test functions $\delta u^{\mathrm{h}}$ and $\delta\varepsilon^{\mathrm{h}}$ and the unknowns $u^{\mathrm{h}}$ and $\varepsilon^{\mathrm{h}}$ follows the scheme of the geometry interpolation in Eq.~(\ref{eq:curve}). The NURBS shape functions $R^{\mathrm{s}}$ are taken from $\mathcal{V}$, while the test functions are discretized by \linebreak$R^{\mathrm{t}}\in\mathcal{W}$
\begin{subequations}
\begin{align}
\delta u^{\mathrm{h}} &=\sum_{I=1}^{n_{\mathrm{np}}^{\mathrm{t}}}{R^{\mathrm{t}}_I~\delta u_I}~,
\label{eq:varu}\\
\delta\varepsilon^{\mathrm{h}} &=\sum_{I=1}^{n_{\mathrm{np}}^{\mathrm{t}}}{R^{\mathrm{t}}_{I,x}~\delta u_I}~,
\label{eq:vare}\\
u^{\mathrm{h}} &=\sum_{I=1}^{n_{\mathrm{np}}^{\mathrm{s}}}{R^{\mathrm{s}}_I~u_I}~,
\label{eq:u}\\
\varepsilon^{\mathrm{h}} &=\sum_{I=1}^{n_{\mathrm{np}}^{\mathrm{s}}}{R^{\mathrm{s}}_{I,x}~u_I}~.
\end{align}
\label{eq:discretization}
\end{subequations}
With $a^{\mathrm{h}}=\frac{\mathrm{d}^2}{\mathrm{d}t^2}u^{\mathrm{h}}$, we derive
\begin{equation}
a^{\mathrm{h}} =\sum_{I=1}^{n_{\mathrm{np}}^{\mathrm{s}}}{R^{\mathrm{s}}_I~\frac{\mathrm{d}^2}{\mathrm{d}t^2}u_I}~
\end{equation}
from Eq.~(\ref{eq:u}). 
Establishing $B^{\mathrm{s}}_I=R^{\mathrm{s}}_{I,x}$, $B^{\mathrm{t}}_I=R^{\mathrm{t}}_{I,x}$ and $\ddot{u}_I=\frac{\mathrm{d}^2}{\mathrm{d}t^2}u_I$, Eq.~(\ref{eq:formulation}) is discretized to
\begin{equation}
\int\limits_{\mathcal{L}}{\left(\sum_{I=1}^{n_{\mathrm{np}}^{\mathrm{t}}}{{R}^{\mathrm{t}}_I\delta {u}_I} \rho\sum_{J=1}^{n_{\mathrm{np}}^{\mathrm{s}}}{{R}^{\mathrm{s}}_J {\ddot{u}}_J}\right)\mathrm{d}s}+\int\limits_{\mathcal{L}}{\left(\sum_{I=1}^{n_{\mathrm{np}}^{\mathrm{t}}}{{B}^{\mathrm{t}}_I\delta {u}_I} EA\sum_{J=1}^{n_{\mathrm{np}}^{\mathrm{s}}}{{B}^{\mathrm{s}}_J {u}_J}\right)\mathrm{d}s}=\sum_{I=1}^{n_{\mathrm{np}}^{\mathrm{t}}}{{R}^{\mathrm{t}}_I\delta {u}_I ~F|_{\partial\mathcal{L}_N}}+\int\limits_{\mathcal{L}}{\left(\sum_{I=1}^{n_{\mathrm{np}}^{\mathrm{t}}}{R}^{\mathrm{t}}_I\delta {u}_I ~q\right)\mathrm{d}s}~,
\label{eq:formulation_disc}
\end{equation}
where $n_{\mathrm{np}}$ is the number of control points. Moving the discrete values out of the integrals, the final formulation
\begin{equation}
\sum_{I=1}^{n_{\mathrm{np}}^{\mathrm{t}}}{\sum_{J=1}^{n_{\mathrm{np}}^{\mathrm{s}}}{\delta {u}_I\int\limits_{\mathcal{L}}{{R}^{\mathrm{t}}_I \rho {R}^{\mathrm{s}}_J ~\mathrm{d}s}~{\ddot{u}}_J}}+\sum_{I=1}^{n_{\mathrm{np}}^{\mathrm{t}}}{\sum_{J=1}^{n_{\mathrm{np}}^{\mathrm{s}}}{\delta {u}_I\int\limits_{\mathcal{L}}{{B}^{\mathrm{t}}_I EA {B}^{\mathrm{s}}_J ~\mathrm{d}s}~{u}_J}}=\sum_{I=1}^{n_{\mathrm{np}}^{\mathrm{t}}}{\delta {u}_I\left({R}^{\mathrm{t}}_I~F|_{\partial\mathcal{L}_N}+\int\limits_{\mathcal{L}}{{R}^{\mathrm{t}}_I~q~\mathrm{d}s}\right)}
\label{eq:formulation_final}
\end{equation}
is derived, where the left part of the equation yields the computation of the entries $m_{IJ}$ of the mass and $k_{IJ}$ of the stiffness matrix, and the load entries $f_{I}$ are provided on the right-hand side of the equation
\begin{equation}
m_{IJ}=\int\limits_{\mathcal{L}}{{R}^{\mathrm{t}}_I \rho {R}^{\mathrm{s}}_J ~\mathrm{d}s}~,
\label{eq:mij}
\end{equation}
\begin{equation}
k_{IJ}=\int\limits_{\mathcal{L}}{{B}^{\mathrm{t}}_I EA {B}^{\mathrm{s}}_J ~\mathrm{d}s}~,
\label{eq:kij}
\end{equation}
\begin{equation}
f_{I}={R}^{\mathrm{t}}_I~F|_{\partial\mathcal{L}_N}+\int\limits_{\mathcal{L}}{{R}^{\mathrm{t}}_I~q~\mathrm{d}s}~.
\label{eq:fi}
\end{equation}
According to these equations, the mass matrix $\mathbf{M}^e$, stiffness matrix $\mathbf{K}^e$ and load vector $\mathbf{F}^e$ are computed on element level first, by considering only the $n_{\mathrm{en}}$ associated control points
\begin{equation}
\mathbf{M}_{IJ}^e=\left[
\begin{array}{ccc}
	m^e_{11} & \cdots & m^e_{1n_{\mathrm{en}}}\\
	\vdots & \ddots & \vdots\\
	m^e_{n_{\mathrm{en}}1} & \cdots & m^e_{n_{\mathrm{en}}n_{\mathrm{en}}}
\end{array}\right]
~,\qquad
\mathbf{K}_{IJ}^e=\left[
\begin{array}{ccc}
	k^e_{11} & \cdots & k^e_{1n_{\mathrm{en}}}\\
	\vdots & \ddots & \vdots\\
	k^e_{n_{\mathrm{en}}1} & \cdots & k^e_{n_{\mathrm{en}}n_{\mathrm{en}}}
\end{array}\right]
~,\qquad
\mathbf{F}_I^e=\left[
\begin{array}{c}
	f^e_1\\
	\vdots\\
	f^e_{n_{\mathrm{en}}}
\end{array}\right]~.
\end{equation}
Subsequently, an assembly of all $n_{\mathrm{el}}$ element matrices $\mathbf{K}^e$ and $\mathbf{F}^e$ to the global SOE yields
\begin{equation}
\bigcup_{e=1}^{n_{\mathrm{el}}}{\sum_{I=1}^{n_{\mathrm{np}}^{\mathrm{t}}}{\sum_{J=1}^{n_{\mathrm{np}}^{\mathrm{s}}}{\delta\mathbf{u}^e~\mathbf{M}_{IJ}^e~\ddot{\mathbf{u}}^e}}}+\bigcup_{e=1}^{n_{\mathrm{el}}}{\sum_{I=1}^{n_{\mathrm{np}}^{\mathrm{t}}}{\sum_{J=1}^{n_{\mathrm{np}}^{\mathrm{s}}}{\delta\mathbf{u}^e~\mathbf{K}_{IJ}^e~\mathbf{u}^e}}}=\bigcup_{e=1}^{n_{\mathrm{el}}}{\delta\mathbf{u}^e \mathbf{F}^e_I}~,
\label{eq:loop}
\end{equation}
where the unknowns are arranged element-wise as
\begin{subequations}
\begin{align}
\ddot{\mathbf{u}}^e&=\left[\ddot{u}_I\right]_{I=1,...,n_{\mathrm{en}}}~,\\
\mathbf{u}^e&=\left[u_I\right]_{I=1,...,n_{\mathrm{en}}}~,\\
\delta\mathbf{u}^e&=\left[\delta u_I\right]_{I=1,...,n_{\mathrm{en}}}~.
\end{align}
\end{subequations}
Equation~(\ref{eq:loop}) can be rewritten in an assembled form as 
\begin{equation}
\delta\hat{\mathbf{u}}^{\mathrm{T}}\mathbf{M}\hat{\ddot{\mathbf{u}}}+\delta\hat{\mathbf{u}}^{\mathrm{T}}\mathbf{K}\hat{\mathbf{u}}=\delta\hat{\mathbf{u}}^{\mathrm{T}} \mathbf{F}~.
\end{equation}
Neglecting the trivial solution $\delta\hat{\mathbf{u}}=\mathbf{0}$, the final system of equations 
\begin{equation}
\mathbf{M}\hat{\ddot{\mathbf{u}}}+\mathbf{K}\hat{\mathbf{u}}=\mathbf{F}
\label{eq:LGS}
\end{equation}
is obtained. For the static case $\ddot{\mathbf{u}}=\mathbf{0}$ it has the solution $\hat{\mathbf{u}}=\mathbf{K}^{-1}\mathbf{F}$.

At first glance, there seems to be no difference between the presented Petrov-Galerkin and a standard Bubnov-Galerkin formulation, but in contrast to Bubnov-Galerkin formulations with $R^{\mathrm{s}}\equiv R^{\mathrm{t}}$, the chosen test functions differ from the shape functions. Hence, the stiffness matrices are non-symmetric and will raise the computational effort of matrix inversions and almost double the required memory amount for the storage of mass and stiffness matrices compared to symmetric systems of the same size. Since we employ a matrix inversion-free explicit time integration method, this drawback reduces to the doubled memory requirement. We will show in Sec.~\ref{sec:transformation} that the proposed usage of dual test functions can be expressed as a simple
transformation of the NURBS shape functions. Subsequently, the entire discretized SOE based on dual test functions can be shown to be a simple transformation of the standard Bubnov-Galerkin SOE. Furthermore, in Sec.~\ref{sec:examples_static} we will show that the transformation does not impact stability and accuracy of the static solution.

\subsection{Transfering NURBS-NURBS discretizations into dual formulations}
%% Herausziehen der Transformationsmatrix S, zeigen dass möglich
\label{sec:transformation}
In this section, the formulas derived in Sec.~\ref{sec:discretization} are adapted to our dual formulation. Therefore, the dual function is applied as $R^{\mathrm{t}}:=\lambda$, so that the test function (\ref{eq:varu}) is now discretized by $\delta u^{\mathrm{h}} =\sum_{I=1}^{n_{\mathrm{np}}}{\lambda_I~\delta u_I}$. This yields the entries of the mass matrix 
\begin{equation}
m_{IJ}^{\mathrm{dual}}=\int\limits_{\mathcal{L}}{{\lambda}_{I} \rho {R}^{\mathrm{s}}_J~\mathrm{d}s}~,
\label{eq:m_dual0}
\end{equation}
entries of the stiffness matrix 
\begin{equation}
k_{IJ}^{\mathrm{dual}}=\int\limits_{\mathcal{L}}{{\lambda}_{I,x} EA {B}^{\mathrm{s}}_J~\mathrm{d}s}~,
\label{eq:k_dual0}
\end{equation}
and the components of the load vector
\begin{equation}
f_{I}^{\mathrm{dual}}={\lambda}_I~F|_{\partial\mathcal{L}_N}+\int\limits_{\mathcal{L}}{{\lambda}_I~q~\mathrm{d}s}~.
\end{equation}
Test functions are chosen as dual functions $\lambda_i$, but shape functions remain discretized by NURBS. To denote that fact, Eq.~(\ref{eq:m_dual0}) is rewritten
\begin{equation}
m_{IJ}^{\mathrm{dual}}=\int\limits_{\mathcal{L}}{{\lambda}_{I}^{\mathrm{dual}} \rho {R}_J^{\mathrm{NURBS}}~\mathrm{d}s}~.
\end{equation}
We recall the transformation matrix $\mathbf{S}=\left[S_{IK}\right]$ with $I,K=1,...,n_{\mathrm{np}}^{\mathrm{s}}$ as the connection between B-Spline shape functions and the corresponding dual test functions as defined in Eq.~(\ref{eq:transformation}). Expressing the dual NURBS test functions $\lambda(\xi)=\mathbf{S}^\mathrm{R}R(\xi)$ by a transformation of NURBS basis functions, and writing the NURBS as weighted B-Splines, we can see from the mass matrix entries
\begin{equation}
m_{IJ}^{\mathrm{dual}}=\int\limits_{\mathcal{L}}{\left(\sum_{K=1}^{n_{\mathrm{np}}^{\mathrm{s}}}{{S}_{IK}^\mathrm{R}\frac{N_K^{\mathrm{t}}\cdot w_K}{W}}\right) \rho \frac{N_J^{\mathrm{s}}\cdot w_J}{W}~\mathrm{d}s}~,
\label{eq:m_dual_NURBS}
\end{equation}
that the criterion of bi-orthogonality, compare to Eq.~(\ref{eq:m_Bspline}), can only be preserved by neglecting the influence of the non-constant weightung function $W$. Thus, the transformation matrix for the use of NURBS basis functions is defined on element level as 
\begin{equation}
\mathbf{S}^\mathrm{R}(\xi) = \mathbf{S}W^2(\xi)~.
\label{eq:transformation_NURBS}
\end{equation}
We insert transformation (\ref{eq:transformation_NURBS}) into Eq.~(\ref{eq:m_dual_NURBS}) and rewrite the shape and test functions as NURBS
\begin{equation}
m_{IJ}^{\mathrm{dual}}=\int\limits_{\mathcal{L}}{\left(\sum_{K=1}^{n_{\mathrm{np}}^{\mathrm{s}}}{{S}_{IK}W^2{R}_{K}^{\mathrm{NURBS}}}\right) \rho {R}_J^{\mathrm{NURBS}}~\mathrm{d}s}~.
\label{eq:m_dual_NURBS2}
\end{equation}
Moving the transformation operator $\mathbf{S}$ out of the integral, as it is a constant matrix, and the applied weights to the right, it can be shown, that the formulation based on dual test functions is just a transformation of the standard Bubnov-Galerkin formulation using NURBS as test and shape functions and involving the weighting function within the residual
\begin{equation}
m_{IJ}^{\mathrm{dual}}=\sum_{K=1}^{n_{\mathrm{np}}^{\mathrm{s}}}{{S}_{IK}\underbrace{\int\limits_{\mathcal{L}}{{R}_{K}^{\mathrm{NURBS}} \rho {R}_J^{\mathrm{NURBS}}W^2}~\mathrm{d}s}_{m_{KJ}^{\mathrm{NURBS}}}}~\leftrightarrow~\mathbf{M}^{\mathrm{dual}}=\mathbf{S}\mathbf{M}^{\mathrm{NURBS}}~.
\label{eq:m_dual}
\end{equation}
Following the same routine, it can be also shown that
\begin{equation}
\mathbf{K}^{\mathrm{dual}}=\mathbf{S}\mathbf{K}^{\mathrm{NURBS}}
\label{eq:k_dual}
\end{equation}
and
\begin{equation}
\mathbf{F}^{\mathrm{dual}}=\mathbf{S}\mathbf{F}^{\mathrm{NURBS}}
\label{eq:f_dual}
\end{equation}
hold true and therefore, the whole system of equations (\ref{eq:LGS}) can be rewritten as
\begin{equation}
\mathbf{S}\mathbf{M}\hat{\ddot{\mathbf{u}}}+\mathbf{S}\mathbf{K}\hat{\mathbf{u}}=\mathbf{S}\mathbf{F}~.
\label{eq:SOE_dual}
\end{equation}
Thus, the solution $\hat{\mathbf{u}}$ or $\hat{\ddot{\mathbf{u}}}$ is independent from the type of chosen test functions, as they can be expressed as a combination of the chosen shape functions. The entries of the stiffness matrix and the load vector are also derived through slightly modified integrals as
\begin{equation}
k_{IJ}^{\mathrm{NURBS}}=\int\limits_{\mathcal{L}}{\left({B}_{I}^{\mathrm{NURBS}}+2R_I\frac{W_{,x}}{W}\right) EA {B}_J^{\mathrm{NURBS}}W^2}~\mathrm{d}s~,
\label{eq:k_dual_NURBS}
\end{equation}
and
\begin{equation}
f_{I}^{\mathrm{NURBS}}={R}_I^{\mathrm{NURBS}}W^2~F|_{\partial\mathcal{L}_N}+\int\limits_{\mathcal{L}}{{R}_I^{\mathrm{NURBS}}~q~W^2~\mathrm{d}s}~.
\label{eq:f_dual_NURBS}
\end{equation}
In the remainder of this article, the NURBS basis functions are reduced to its special case of B-Splines, as all weights $w_i$ of the control points are chosen as $w_i=1~\forall i \in \left\lbrace 1,\ldots,n_{\mathrm{np}}^s\right\rbrace$. Thus, all modifications in the integrals in Eqs.~(\ref{eq:m_dual_NURBS2}), (\ref{eq:k_dual_NURBS}) and (\ref{eq:f_dual_NURBS}) related to the weighting function $W$ can be ignored, as a constant $W$ results in the standard Bubnov-Galerkin formulation. Nevertheless, the presented dual approach is not limited to B-Splines and the implementation for NURBS in general is straightforward, as was already demonstrated for the isogeometric mortar method \cite{Dornisch.2017}. For clarification, in Sec.~\ref{sec:examples_NURBS} an example for varying weights is given.

\subsection{Treatment of boundary conditions}
The implementation of boundary conditions is quite similar to standard IGA formulations. We only have to take special care with regard to strongly enforced homogeneous Dirichlet boundary conditions if the transformation matrix $\mathbf{S}$ is applied after assembly to the SOE~(\ref{eq:SOE_dual}), which describes the common situation in most numerical codes. The size of the system matrices, e.g., the stiffness matrix $\overline{\mathbf{K}}$ is reduced to $(n_{\mathrm{np}}-k)\times (n_{\mathrm{np}}-k)$, where $k$ is the number of fixed control points. As the transformation matrix is related to the full spline space, $\mathbf{S}$ is of dimension $n_{\mathrm{np}}\times n_{\mathrm{np}}$. Recalling Eq.~(\ref{eq:transformation}), a single value of the $i^{\mathrm{th}}$ dual function at an arbitrary evaluation point can be calculated as
\begin{equation}
\lambda_i = S_{i1} N_1 + ... + S_{im} N_m + ... + S_{in_{\mathrm{np}}} N_{n_{\mathrm{np}}}~.
\end{equation}
Note that at this points no weights for NURBS are included, as the transformation operator extracted from the integrals is only related to the B-Splines spanned by the underlying knot vector. Nonetheless, the presented idea of treating the Dirichlet boundary conditions is also valid for the more general case of NURBS.

To be able to process the matrix multiplication given in Eq.~(\ref{eq:SOE_dual}), a transformation matrix $\overline{\mathbf{S}}$ of reduced size has to be computed in advance. Comparing the IG and AD functions in Figs.~\ref{fig:IG_basisfun} and \ref{fig:AD_basisfun} with the corresponding B-spline basis functions in Fig.~\ref{fig:basisfun}, it is immediately clear that the dual basis functions are not interpolatory at the external knots, typically considered for the application of this kind of boundary conditions. Simply neglecting a single control point $m$ and its corresponding entries in $\mathbf{S}$ 
\begin{equation}
\lambda_i = S_{i1} N_1 + ... + S_{i(m-1)} N_{(m-1)} + S_{i(m+1)} N_{(m+1)} + ... + S_{in_{\mathrm{np}}} N_{n_{\mathrm{np}}}~,
\end{equation}
causes a loss of accuracy as the term $S_{im} N_m$ is not equal to zero by definition. In order to apply the duality principle correctly without significant losses of accuracy through the dimensionality reduction, we remove the $k$ basis functions concerned by the implementation of homogeneous Dirichlet conditions in an analogous fashion as known from a static condensation technique. Therefore, the linear system of equations describing the transformation from B-Splines to dual basis functions
\begin{equation}
\boldsymbol{\lambda}=\mathbf{S}\mathbf{N}
\label{eq:SOE_full}
\end{equation}
has to be rearranged, such that 
\begin{equation}
\mathbf{S}=\left[\begin{array}{cc}
	\mathbf{A} &\mathbf{B}\\
	\mathbf{B}^\top &\mathbf{C}
\end{array}\right]
\end{equation}
with 
\begin{subequations}
\begin{equation}
\boldsymbol{A}=
\left[\begin{array}{ccc}
	S_{11} & \cdots & S_{1k}\\
	\vdots & \ddots & \vdots\\
	S_{k1} & \cdots & S_{kk}\\
\end{array}\right]~,
\end{equation}\\
\begin{equation}
\boldsymbol{B}=
\left[\begin{array}{ccc}
	S_{1~k+1} & \cdots & S_{1n}\\
	\vdots & \ddots & \vdots\\
	S_{k~k+1} & \cdots & S_{kn}\\
\end{array}\right]~,
\end{equation}\\
\begin{equation}
\boldsymbol{C}=
\left[\begin{array}{ccc}
	S_{k+1~k+1} &\cdots &S_{k+1~n}\\
	\vdots &\ddots &\vdots\\
	S_{n~k+1} &\cdots &S_{nn}\\
\end{array}\right]~.
\end{equation}
\end{subequations}
To remove the $k$ basis functions corresponding to the fixed control points, we now separate the equations to be retained
\begin{equation}
\left[\begin{array}{c}
	\lambda_{k+1}\\	\vdots\\ \lambda_n
\end{array}\right]=\boldsymbol{B}^\top
\left[\begin{array}{c}
	N_1\\	\vdots\\ N_k
\end{array}\right]+
\boldsymbol{C}
\left[\begin{array}{c}
	N_{k+1}\\	\vdots\\ N_n
\end{array}\right]
\label{eq:SOE_k+1}
\end{equation}
from the first $k$ equations, which are given by
\begin{equation}
\left[\begin{array}{c}
	\lambda_1\\	\vdots\\ \lambda_k
\end{array}\right]=\boldsymbol{A}
\left[\begin{array}{c}
	N_1\\	\vdots\\ N_k
\end{array}\right]+
\boldsymbol{B}
\left[\begin{array}{c}
	N_{k+1}\\	\vdots\\ N_n
\end{array}\right]~.
\label{eq:SOE_k}
\end{equation}
Equation~(\ref{eq:SOE_k}) is transformed to
\begin{equation}
\left[\begin{array}{c}
	N_1\\	\vdots\\ N_k
\end{array}\right]=
\boldsymbol{A}^{-1}\left(
\left[\begin{array}{c}
	\lambda_1\\	\vdots\\ \lambda_k
\end{array}\right]
-\boldsymbol{B}
\left[\begin{array}{c}
	N_{k+1}\\	\vdots\\ N_n
\end{array}\right]
\right)
\end{equation}
and inserted into Eq.~(\ref{eq:SOE_k+1})
\begin{equation}
\left[\begin{array}{c}
	\lambda_{k+1}\\	\vdots\\ \lambda_n
\end{array}\right]=\boldsymbol{B}^\top
\boldsymbol{A}^{-1}\left(
\left[\begin{array}{c}
	\lambda_1\\	\vdots\\ \lambda_k
\end{array}\right]
-\boldsymbol{B}
\left[\begin{array}{c}
	N_{k+1}\\	\vdots\\ N_n
\end{array}\right]
\right)+
\boldsymbol{C}
\left[\begin{array}{c}
	N_{k+1}\\	\vdots\\ N_n
\end{array}\right]~.
\label{eq:SOE_new}
\end{equation}

Now the interior dual functions $\lambda_{k+1},\ldots\lambda_n$ do not depend on the boundary basis functions $N_1,\ldots N_k$ anymore. Thus, the second and third term in Eq.~(\ref{eq:SOE_new}) are zero at the boundary for all dual functions $\lambda_{k+1},\ldots\lambda_n$. Next, we want the interior dual functions to be independent of the boundary dual functions $\lambda_{1},\ldots\lambda_k$. This is the only way to ensure that the interior dual functions are zero at the boundary, which is required for the enforcement of strong Dirichlet boundary conditions. Without loss of generality, we set 
\begin{equation}
\left[\begin{array}{c}
\lambda_1\\
\vdots\\
\lambda_k
\end{array}\right]=\mathbf{0}
\label{eq:static_con}
\end{equation}
and thus obtain the dimensionally reduced transformation matrix
\begin{equation}
\overline{\mathbf{S}}=\mathbf{C}-\mathbf{B}^\top\mathbf{A}^{-1}\mathbf{B}\,,
\end{equation}
which can be directly used to transform the Bubnov-Galerkin system matrices and vector with build-in Dirichlet boundary conditions. It is to be noted that the result is the same as if a standard static condensation with the condition of Eq.~(\ref{eq:static_con}) is applied to Eq.~(\ref{eq:SOE_full}).

Of course, it is also possible to compute the dual system matrices on element level, using the corresponding row of the full transformation matrix $\mathbf{S}$ to calculate the entries associated with a specific control point. Usually the fixed DOFs will be simply skipped within these element-wise assembly routines and thus the influence of non-participating shape and test functions is correctly prevented. But as the dual lumping scheme should be easily adaptable to existing codes, we presented this reduction procedure based on standard Bubnov-Galerkin system matrices of reduced size and the transformation matrix of full size, which further has to be reduced to the transformation operator of reduced size. The more simplistic approach of neglecting all entries related to fixed DOFs within $\mathbf{K}_{\mathrm{dual}}^{\mathrm{full}}=\mathbf{S}\mathbf{K}_{\mathrm{dual}}^{\mathrm{full}}$ was not chosen, as the resulting $L_2$-error norm of normal forces in Fig.~\ref{fig:e0} is not convincing. The accuracy is limited through resulting disturbances at the fixed boundaries. For our one-dimensional examples, we did not observe any problems concerning duality, stability, or accuracy using our boundary treatment approach. As $\overline{\mathbf{S}}$ is always applied to both sides of the equations, the solution remains unchanged for consistent mass matrices and will just be affected by the additional mass lumping. 

Note, that a similar treatment of Dirichlet boundary conditions is also considered within the latest study of \citet{Hiemstra.2023}, where the mathematical foundation is exemplified for a single homogeneous boundary condition.

\section{Numerical examples}
\label{sec:examples}
\begin{figure}[b]
\centering
\includegraphics[width=0.3\textwidth]{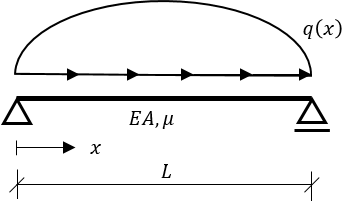}
\caption{System sketch for truss under uniaxial loading $q(x)$.}
\label{fig:sysplot}
\end{figure}

\begin{figure}[t]
\centering
\begin{subfigure}[h]{0.4\textwidth}
	\centering
	\includegraphics[width=\textwidth]{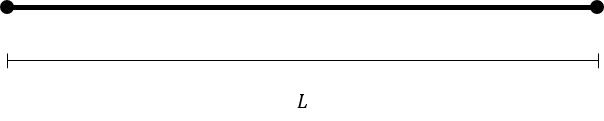}
	\caption{mesh~A}
	\label{fig:mesh1}
\end{subfigure}\\
\vspace*{3mm}
\begin{subfigure}[h]{0.4\textwidth}
	\centering
	\includegraphics[width=\textwidth]{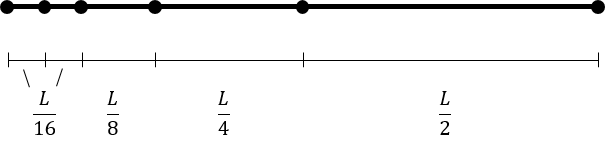}
	\caption{mesh~B}
	\label{fig:mesh2}
\end{subfigure}\\
\vspace*{3mm}
\begin{subfigure}[h]{0.4\textwidth}
	\centering
	\includegraphics[width=\textwidth]{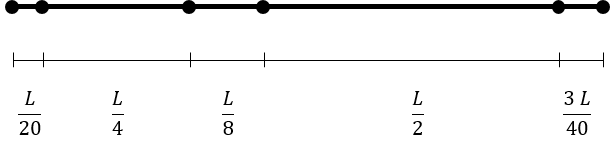}
	\caption{mesh~C}
	\label{fig:mesh3}
\end{subfigure}
\caption{Types of meshes used for numerical examples. Dots ($\medbullet$) denote knots of the coarsest mesh, which coincide with the control points for polynomial order $p=1$. Meshes for polynomial orders $p>1$ are constructed using $k$-refinement. Further $h$-refinement is applied uniformly within each initial knot span.}
\label{fig:mesh}
\end{figure}

For the following numerical examples a simple truss is considered to study the performance of the derived element formulation. A system of length $L$, extensional stiffness $EA$ and distributed mass $\mu$ is shown in Fig.~\ref{fig:sysplot}. As we want to study the basic behavior of dual test functions within explicit dynamics, only academic examples are chosen. The implementation for higher-dimensional problems is straightforward, taking into account the tensor product structure of NURBS. Note that a similar technique for B-Splines has been developed by \citet{Nguyen.2023}, where the focus is on fourth order PDEs describing beam, plate, or shell models.
To test the formulation not only on uniform knot vectors, also two non-uniform meshes are considered, see Fig.~\ref{fig:mesh}.

\subsection{Static problems: Conditioning and convergence properties}
\label{sec:examples_static}
Our first numerical example is a general validation of our formulation, especially the chosen type of condensation within $\mathbf{S}$, and a comparison of the convergence properties for the different test functions. The standard formulation with NURBS as shape and test functions is compared to three kinds of dual test functions: the IG duals, the AD duals with minimal reproduction degree of $q_{\mathrm{min}}=1$, and the AD duals with maximum reproduction degree of $q_{\mathrm{max}}=p$, which is equal to the original polynomial order $p$ of the corresponding NURBS.

For this static application $q(x) = \frac{P_0}{L}\sin(\pi\frac{x}{L})$ is used as load distribution with $P_0=100{,}000\,$kN. The system parameters are chosen as length $L=10\,$m and extensional stiffness $EA=1{,}649{,}335\,$kN, corresponding to a Young's modulus $E=2.1\cdot 10^8\,$kN/m² of a circular cross section with $\o=0.1\,$m.

\begin{figure}[t]
\centering
\begin{subfigure}[h]{0.40\textwidth}
	\centering
	\includegraphics[width=\textwidth]{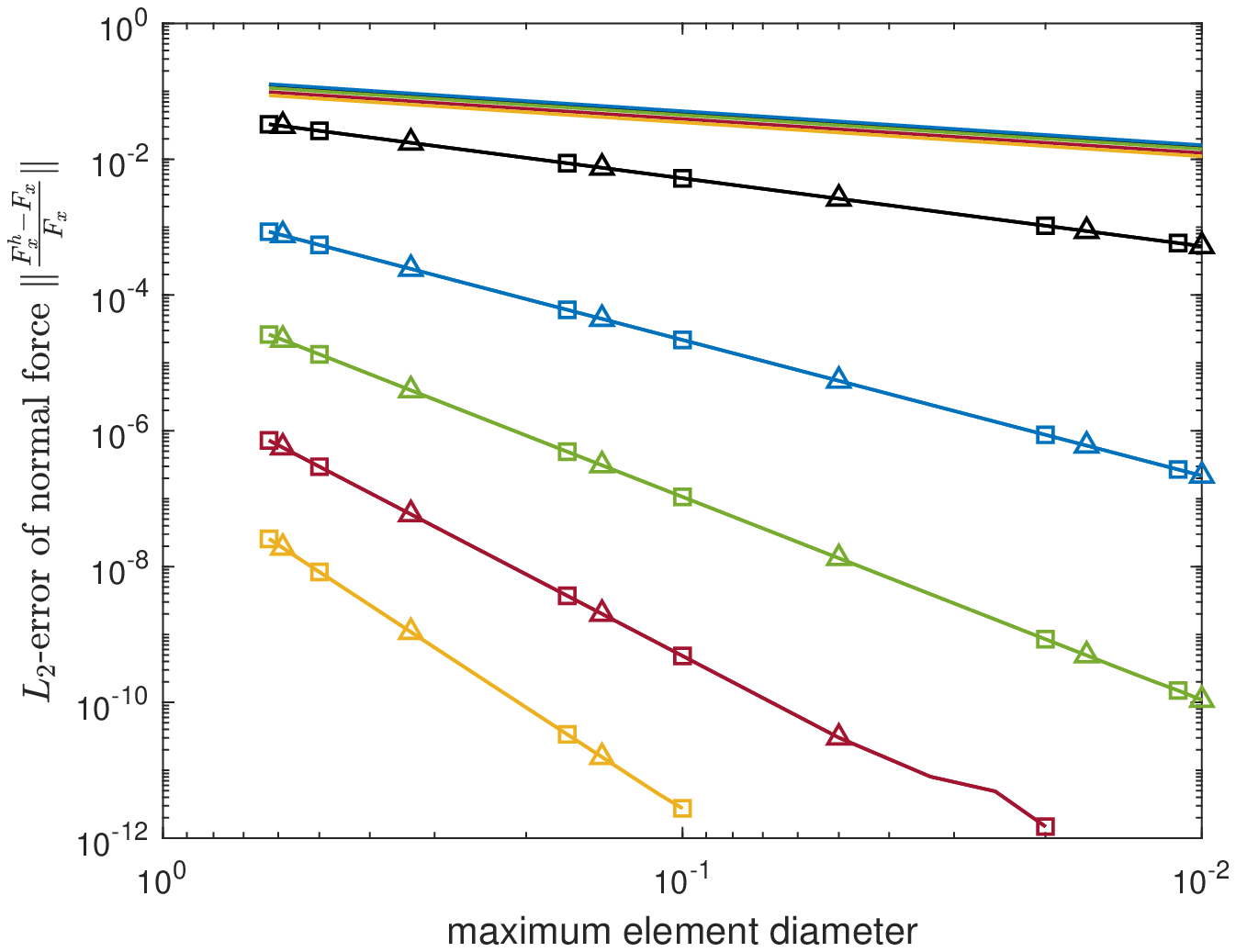}
\end{subfigure}\\
\begin{subfigure}[h]{0.4\textwidth}
	\centering
	\includegraphics[width=\textwidth]{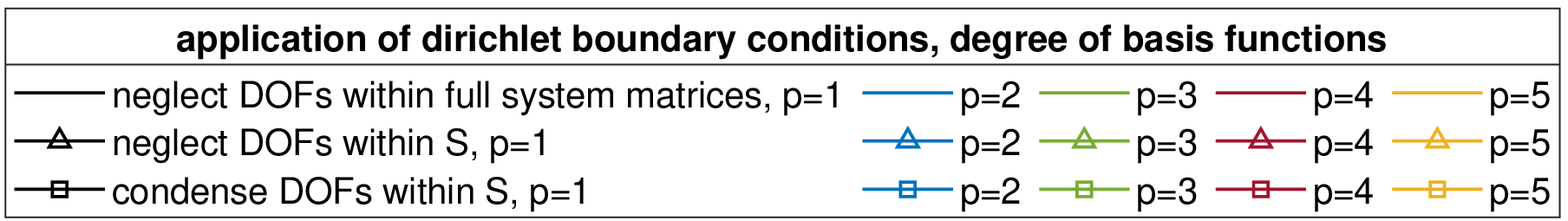}
\end{subfigure}
\caption{$L_2$-error norm of normal forces for three different methods to apply homogeneous Dirichlet boundary conditions, NURBS of order $p=1,...,5$ are used as shape functions, AD with $q_{\mathrm{max}}$ are used as test functions. Computation was done on uniform mesh~A (Fig.~\ref{fig:mesh1}).}
\label{fig:e0}
\end{figure}

In Fig.~\ref{fig:e0}, the $L_2$-error norm of normal forces is compared for three different ways of applying the homogeneous Dirichlet boundary conditions. For the simplistic approach, denoted as `neglect DOFs within full system' in Fig.~\ref{fig:e0}, the rows and columns related to fixed DOFs are neglected within the full assembled dual stiffness matrix $\mathbf{K}_{\mathrm{dual}}^{\mathrm{full}}$. This intervention violates the former equilibrium as the additional contribution of inner control points to that boundary is denied. In the vicinity of the fixed boundaries, the stresses deviate strongly; proper convergence behavior is lost. In \cite{Nguyen.2023}, the test functions related to these boundary conditions are replaced by B-Splines. Thus, the spanned space of test functions is modified and not dual anymore at specific points, but allows to treat the boundary conditions in a standard manner. This idea is equivalent to obtaining the transformation operator by neglecting fixed DOFs related entries within the full transformation matrix $\mathbf{S}$ and only than compute the system matrices through multiplication with the remaining part of the transformation operator. For common IGA formulations expected `$p+1$' convergence is obtained. The results do not differ from the proposed condensation method, because both approaches only operate on the transformation matrix and not the system matrices. The assembly on element level is not investigated, as the results are equal to the second and third approach as effect of the possible extraction of the transformation operator.

\begin{figure}[t]
\centering
\begin{subfigure}[h]{0.40\textwidth}
	\centering
	\includegraphics[width=\textwidth]{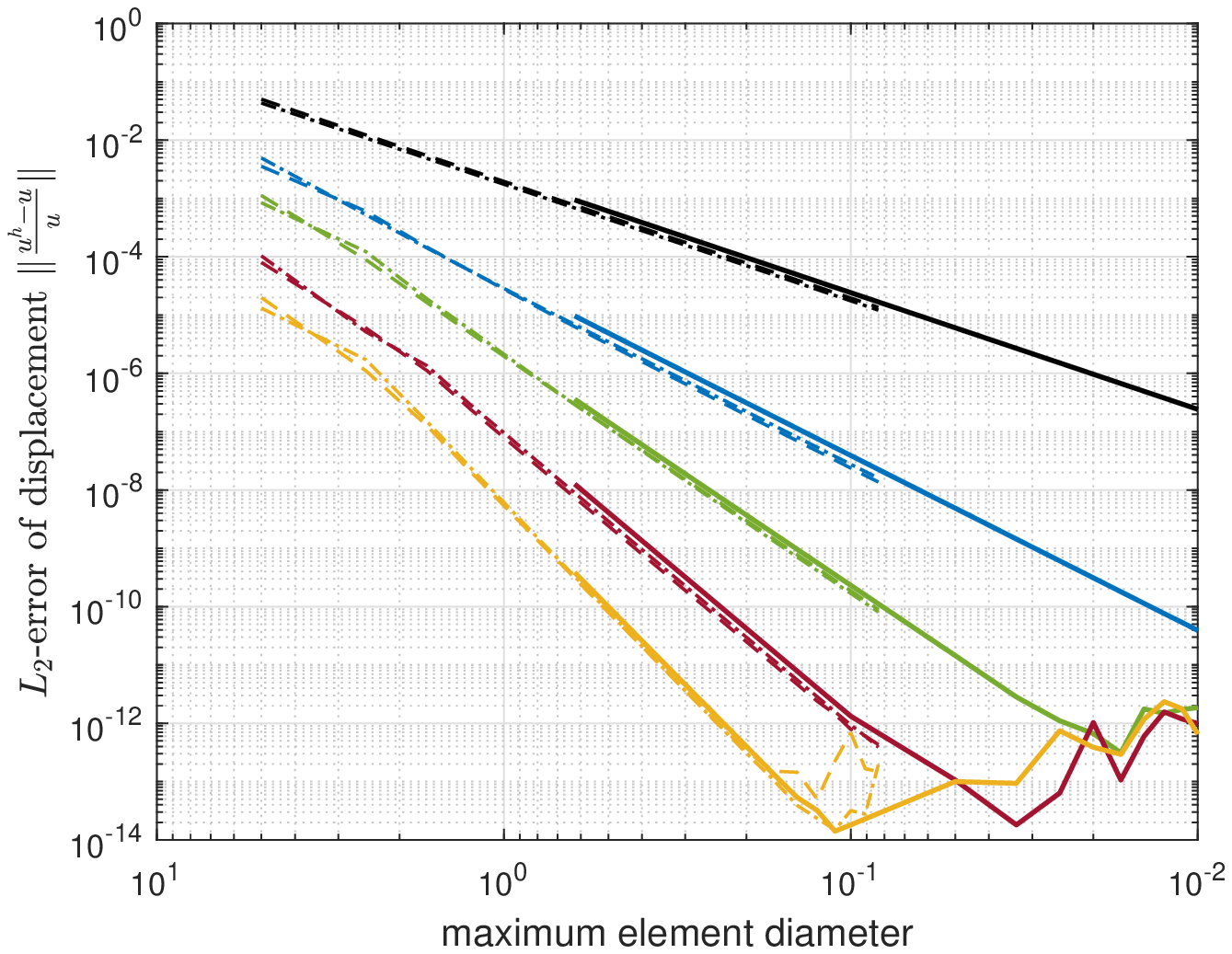}
	\caption{NURBS}
	\label{fig:e1_L2_NURBS}
\end{subfigure}
\quad
\begin{subfigure}[h]{0.40\textwidth}
	\centering
	\includegraphics[width=\textwidth]{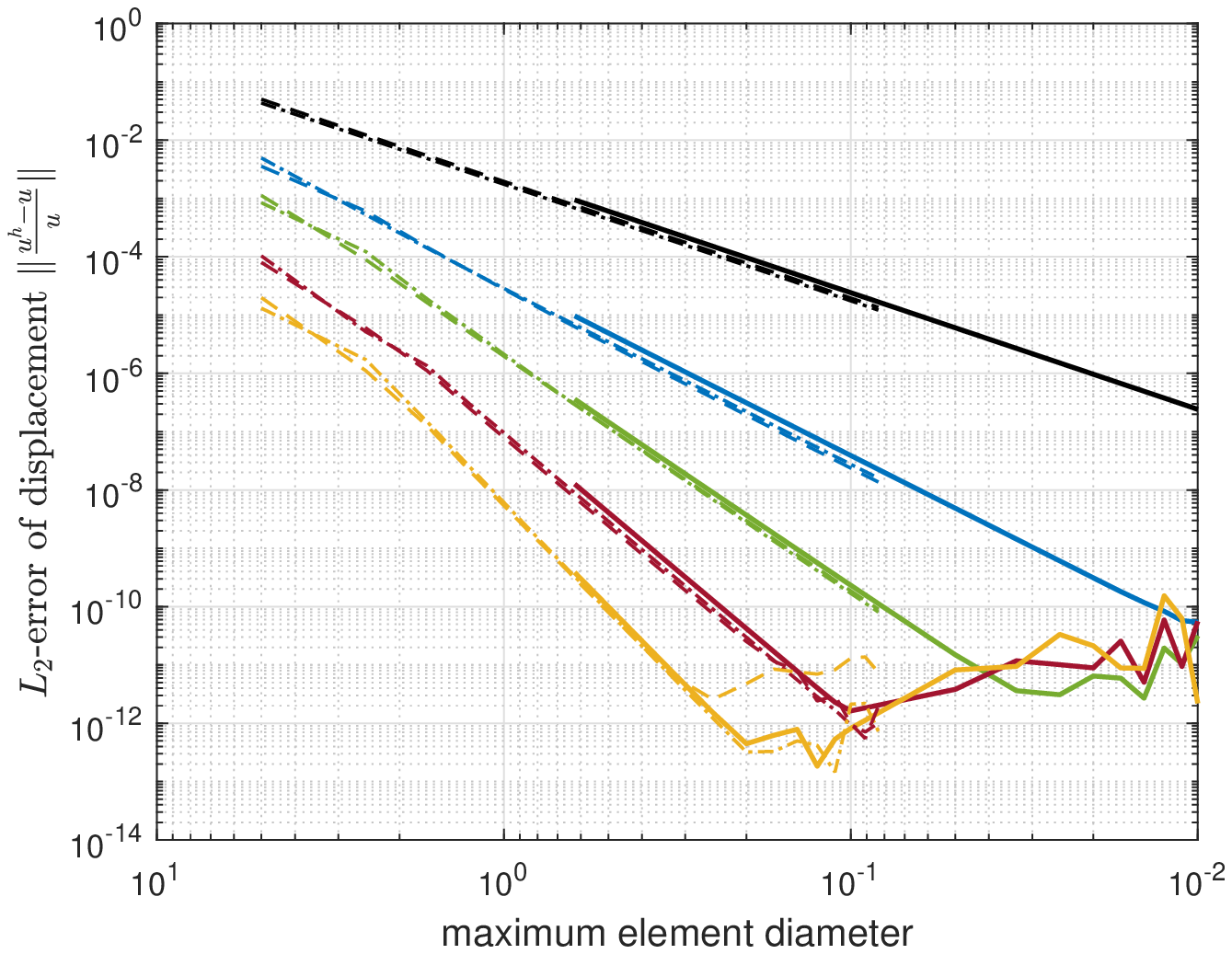}
	\caption{IG}
	\label{fig:e1_L2_IG}
\end{subfigure}\\
\begin{subfigure}[h]{0.40\textwidth}
	\centering
	\includegraphics[width=\textwidth]{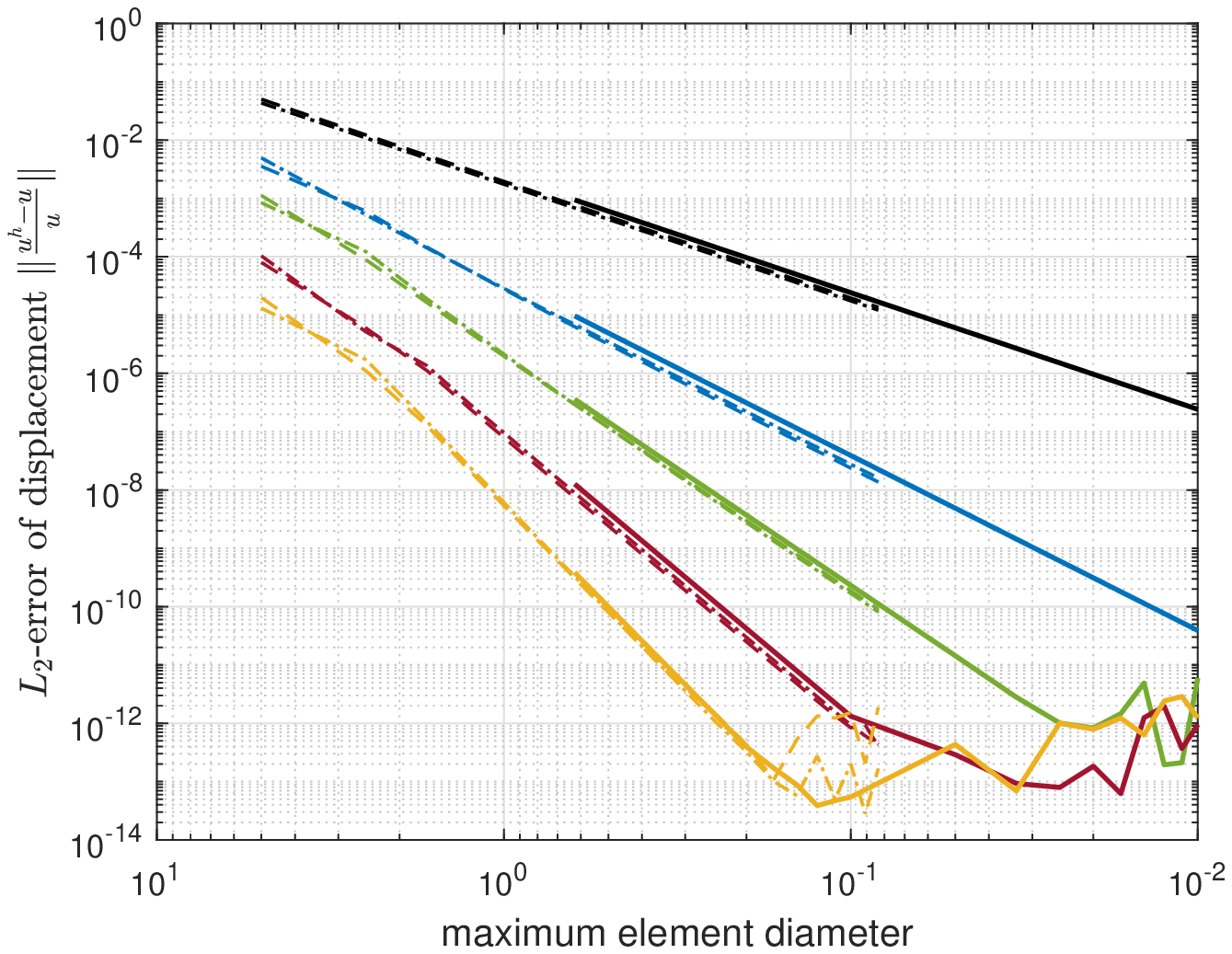}
	\caption{AD $q_{\mathrm{min}}$}
	\label{fig:e1_L2_ADmin}
\end{subfigure}
\quad
\begin{subfigure}[h]{0.40\textwidth}
	\centering
	\includegraphics[width=\textwidth]{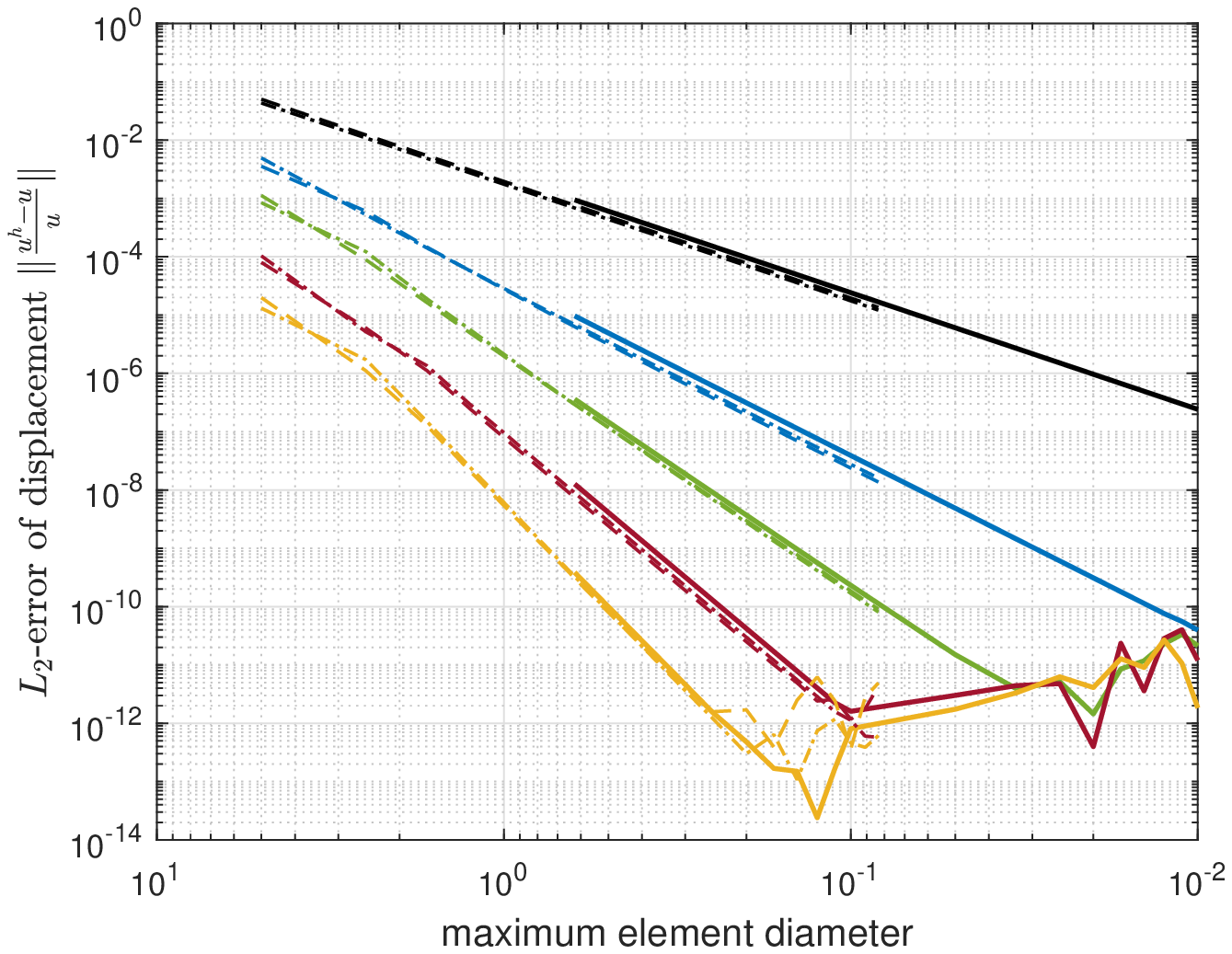}
	\caption{AD $q_{\mathrm{max}}$}
	\label{fig:e1_L2_ADmax}
\end{subfigure}\\
\begin{subfigure}[h]{0.25\textwidth}
	\centering
	\includegraphics[width = \textwidth]{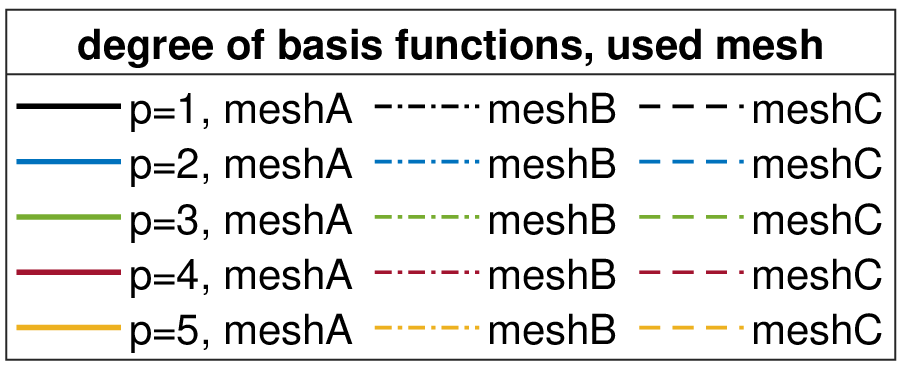}
\end{subfigure}
\caption{$L_2$-error norm of displacements for three different meshes (Fig.~\ref{fig:mesh}), NURBS of order $p=1,...,5$ are used as shape functions, type of test functions varies.}
\label{fig:e1_L2}
\end{figure}
We have shown in Sec.~\ref{sec:transformation} that using dual test functions can be interpreted as a simple transformation of the standard NURBS formulation. Hence, it promises the same results and should not affect the accuracy, if the SOE remains well-conditioned. This fact can be observed in Fig.~\ref{fig:e1_L2}. The $L_2$-error norm is equal for all examined test functions. Only minor deviations can be spotted in the case of rather small elements caused by numerical reasons during the computation process, but the general convergence behavior and the rates do not depend on the chosen test functions. As the error norm is plotted against the maximum element diameter, the non-uniform meshes B and C seem to perform slightly better than the uniform mesh~A. This is due to the higher amount of also smaller elements accompanied by that refinement.

\begin{figure}[b]
\centering
\begin{subfigure}[ht]{0.230\textwidth}
	\centering
	\includegraphics[width=\textwidth]{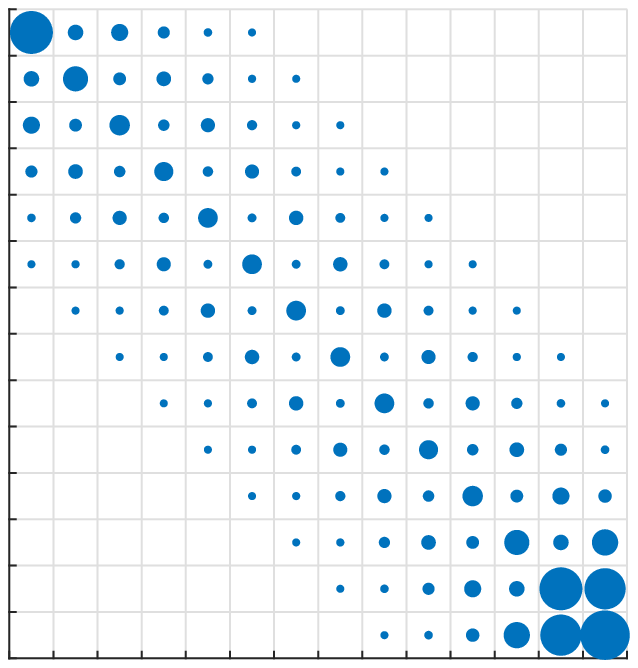}
	\caption{NURBS}
\end{subfigure}
\begin{subfigure}[ht]{0.230\textwidth}
	\centering
	\includegraphics[width=\textwidth]{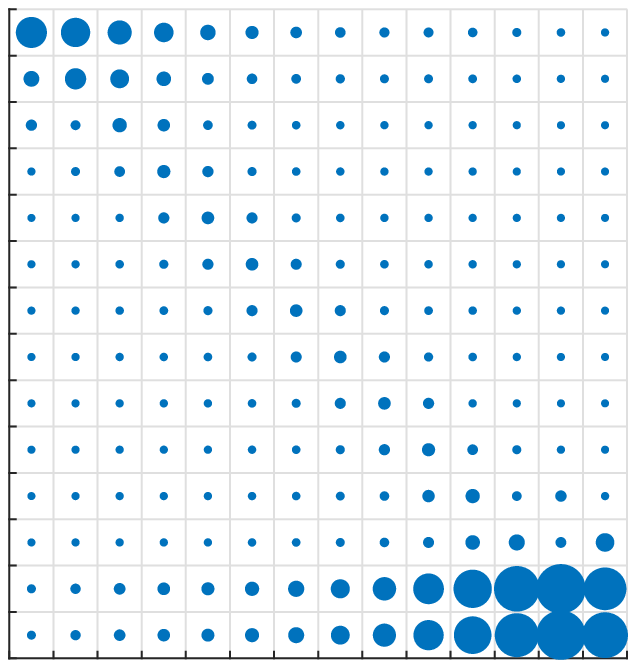}
	\caption{IG}
\end{subfigure}
\begin{subfigure}[ht]{0.230\textwidth}
	\centering
	\includegraphics[width=\textwidth]{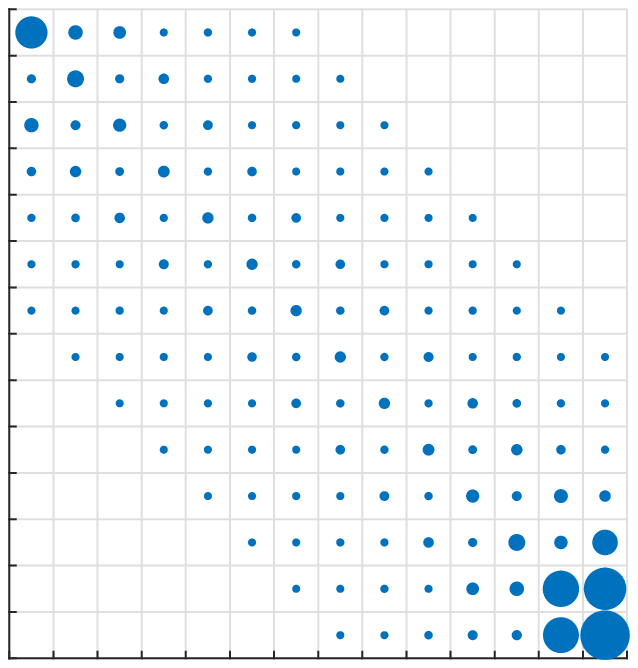}
	\caption{AD $q_{\mathrm{min}}$}
\end{subfigure}
\begin{subfigure}[ht]{0.230\textwidth}
	\centering
	\includegraphics[width=\textwidth]{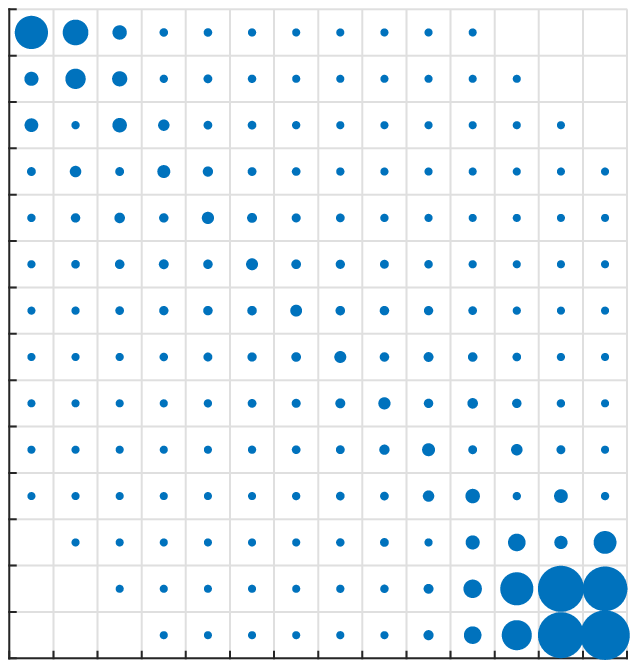}
	\caption{AD $q_{\mathrm{max}}$}
\end{subfigure}
\caption{Nonzero entries of stiffness matrix $\mathbf{K}$ with applied Dirichlet boundary conditions for the system shown in Fig.~\ref{fig:sysplot} discretized by 10 elements, size of the dots refers to the magnitude of the entries, NURBS of order $p=5$ are used as shape functions, type of test functions varies}
\label{fig:e1_K0}
\end{figure}

Another fact worth mentioning is the (non)impact of the reproduction degree $q$ for AD test functions. Even the minimal degree $q_{\mathrm{min}}$ provides the same accuracy and convergence rates as the standard NURBS formulation. But the computational efficiency is another factor that needs to be thoroughly assessed. Using dual test functions has an affect on the structure of the 
stiffness matrix $\mathbf{K}$, as shown in Fig.~\ref{fig:e1_K0}, since the support of the original NURBS shape functions is enlarged as described in Sec.~\ref{sec:dual}. Thus, AD duals cause slightly higher computational costs in explicit dynamic computations than the standard NURBS formulation. As a lower $q$ results in a less densely populated $\mathbf{K}$, reducing the reproduction degree also lowers the costs for the calculations to be comparable to the standard NURBS formulation. IG duals are most costly with no further option to improve the efficiency.

\begin{figure}[t]
\centering
\begin{subfigure}[ht]{0.40\textwidth}
	\centering
	\includegraphics[width=\textwidth]{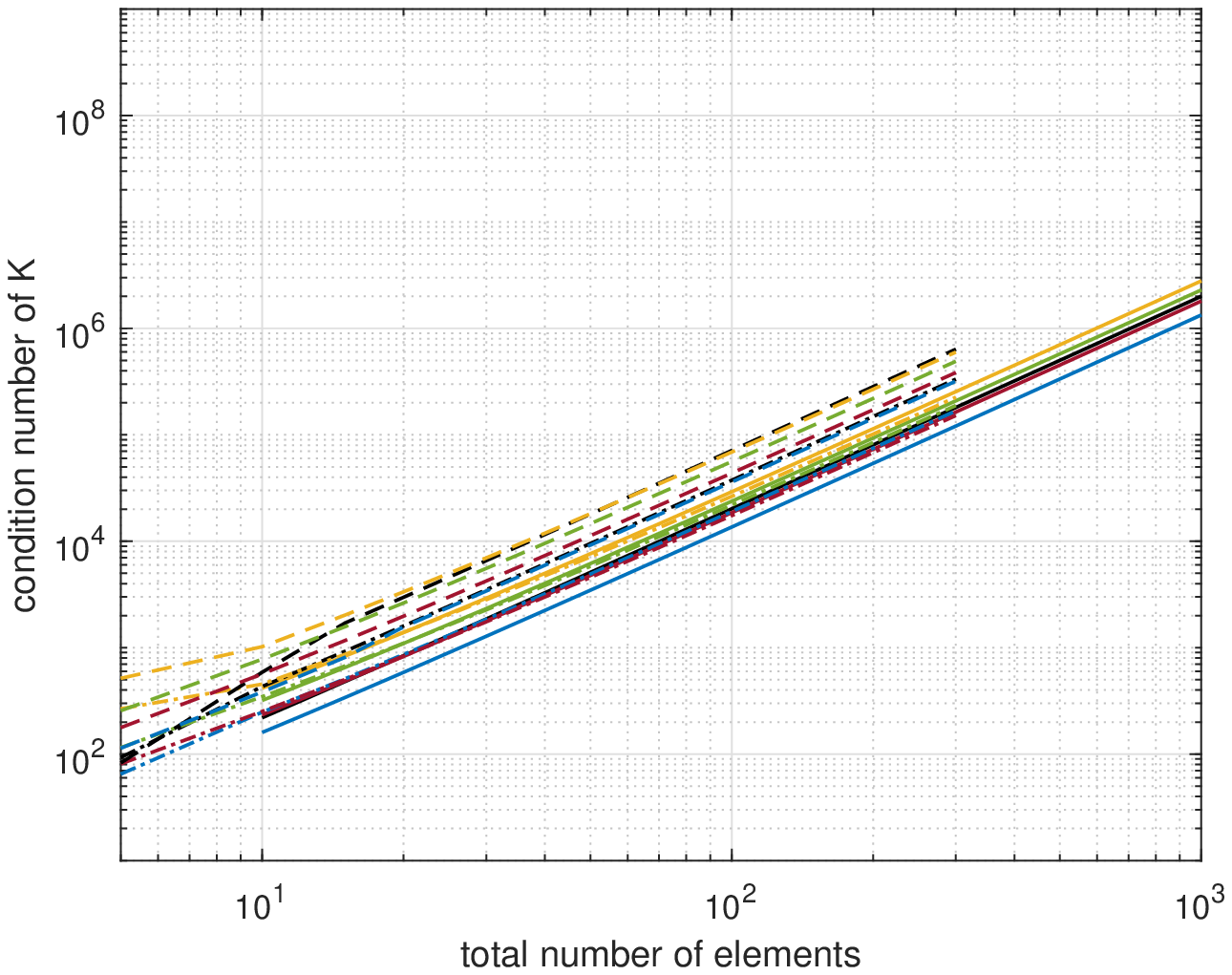}
	\caption{NURBS}
	\label{fig:e1_K_NURBS}
\end{subfigure}
\quad
\begin{subfigure}[ht]{0.40\textwidth}
	\centering
	\includegraphics[width=\textwidth]{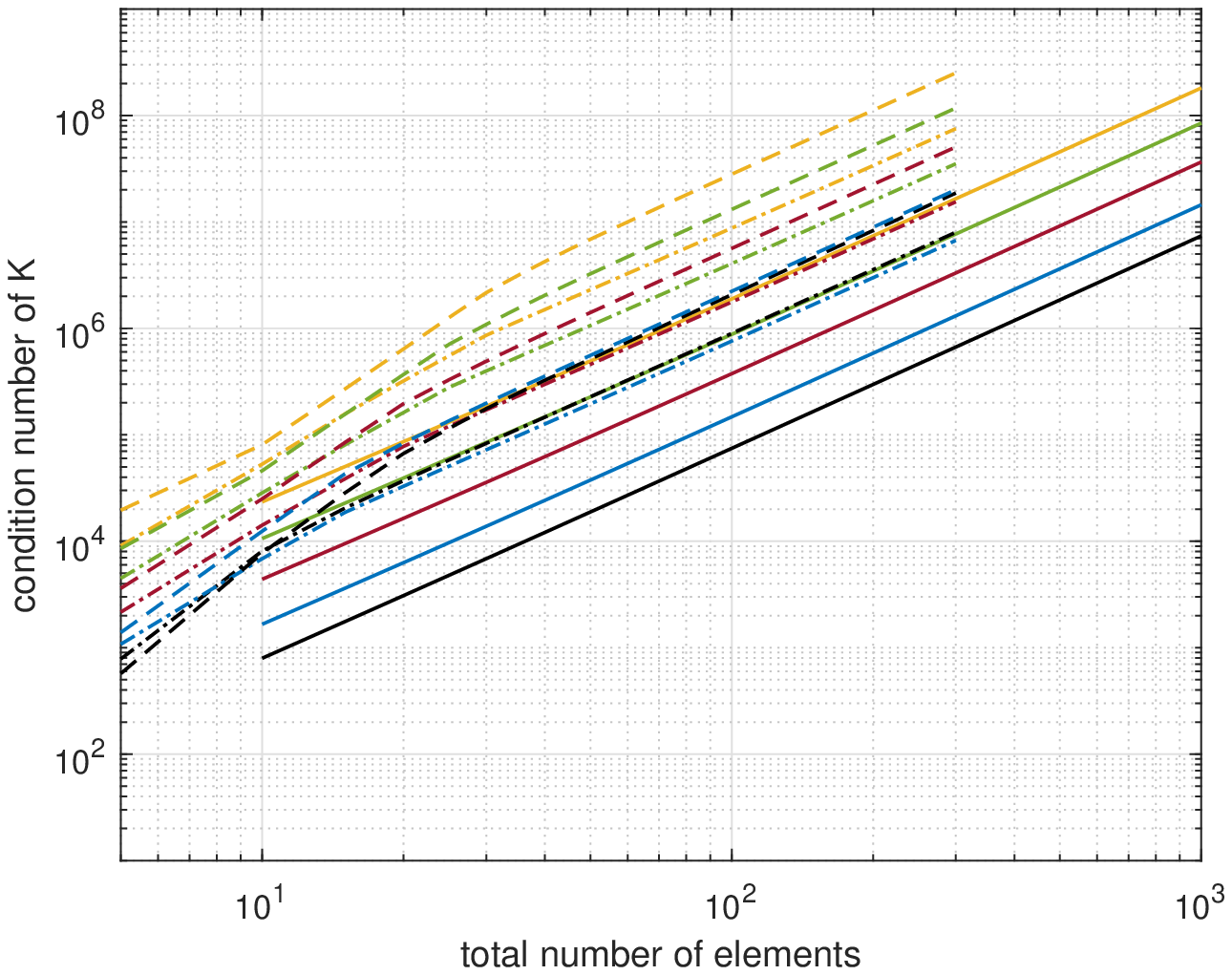}
	\caption{IG}
	\label{fig:e1_K_IG}
\end{subfigure}\\
\begin{subfigure}[ht]{0.40\textwidth}
	\centering
	\includegraphics[width=\textwidth]{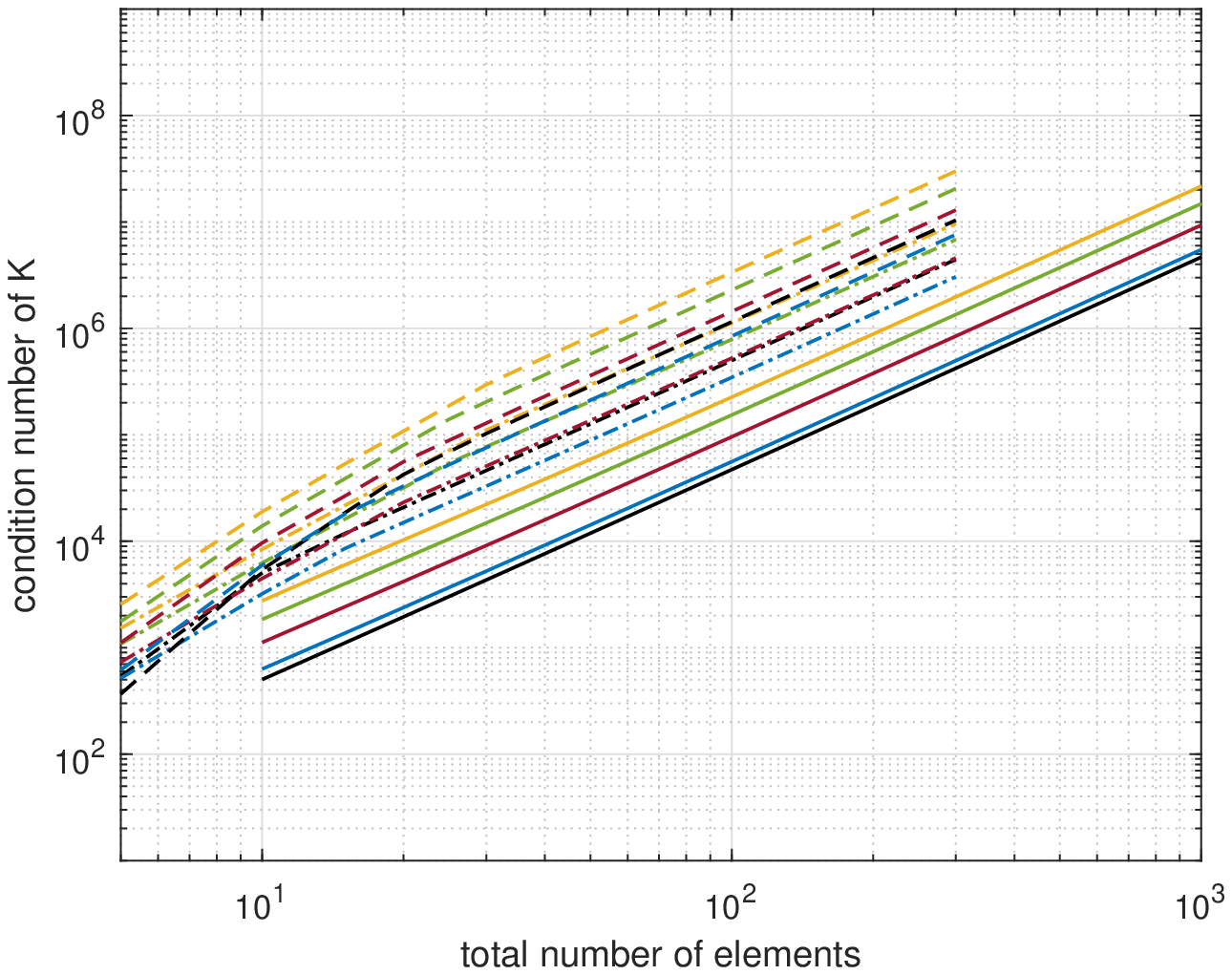}
	\caption{AD $q_{\mathrm{min}}$}
	\label{fig:e1_K_ADmin}
\end{subfigure}
\quad
\begin{subfigure}[ht]{0.40\textwidth}
	\centering
	\includegraphics[width=\textwidth]{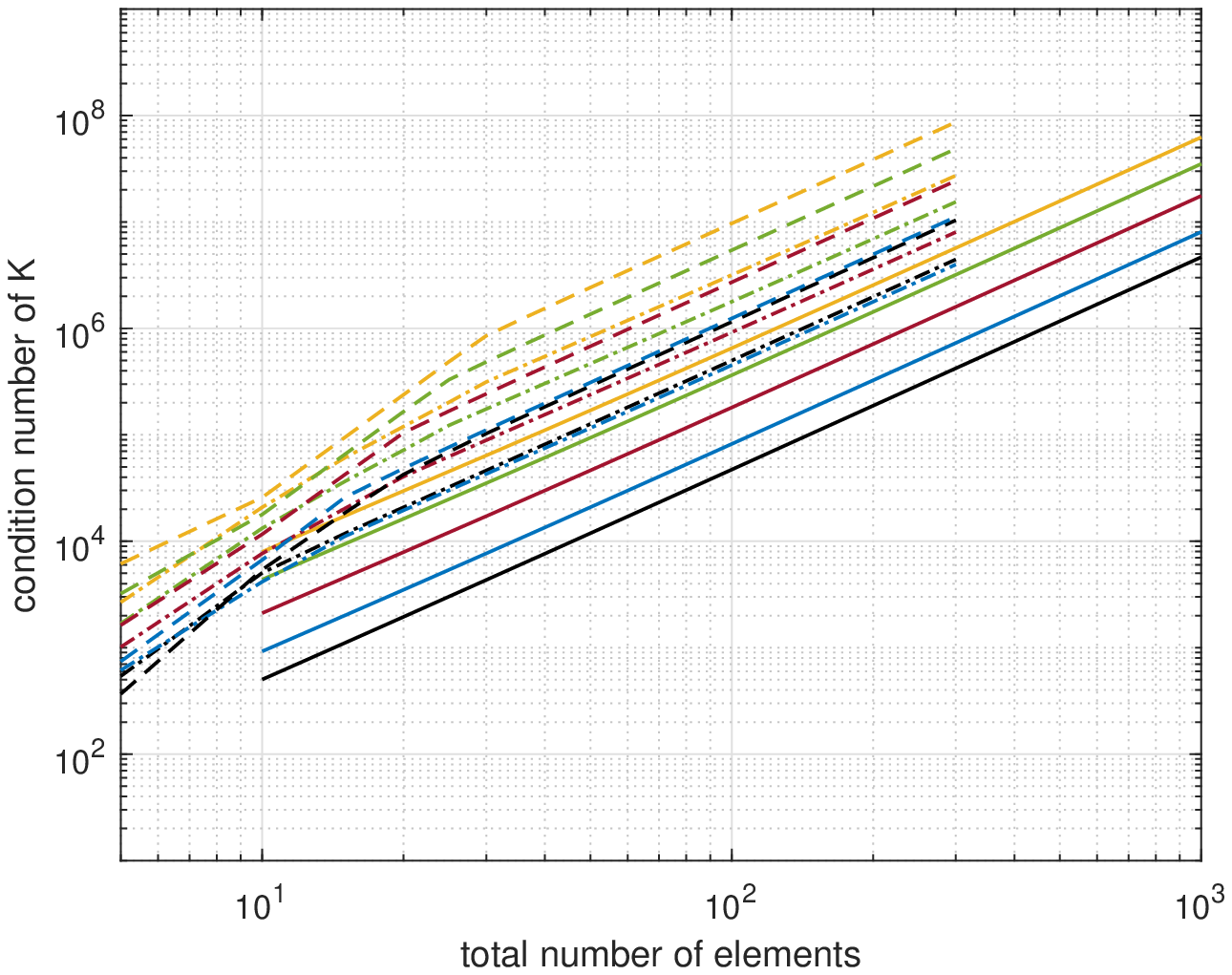}
	\caption{AD $q_{\mathrm{max}}$}
	\label{fig:e1_K_ADmax}
\end{subfigure}\\
\begin{subfigure}[h]{0.25\textwidth}
	\centering
	\includegraphics[width = \textwidth]{legend_example1.eps}
\end{subfigure}
\caption{Condition number of $\mathbf{K}$ for three different meshes (Fig.~\ref{fig:mesh}), NURBS of order $p=1,...,5$ are used as shape functions, type of test functions varies.}
\label{fig:e1_K}
\end{figure}

The choice of test functions also effects the conditioning of $\mathbf{K}$. Due to the fact that the dual formulation is no longer a Bubnov-Galerkin formulation, the stiffness matrices are non-symmetric, and their inversion in general yields higher computational costs. We can notice from Fig.~\ref{fig:e1_K} that NURBS test functions, which solely yield a symmetric $\mathbf{K}$, show the lowest condition number. Additionally, a relation between the increasing support and the growth of the condition number can be detected. Raising the support from NURBS to AD duals with $q_{\mathrm{min}}$ to AD duals with $q_{\mathrm{max}}$ to IG duals results in a gradual increase of the condition number of $\mathbf{K}$. The condition number is directly connected to the inversion of $\mathbf{K}$ and therefore, a measure for the accuracy level of the solution of the system of equations for static and implicit dynamic analyses.

The general behavior of an increasing condition number for raising polynomial degree $p$ or increasing number of elements does not differ between using NURBS and duals as test functions, which is comprehensively documented for IGA and other higher-order methods in \cite{Eisentraeger.2020}. By applying duals, the condition becomes more sensitive to a higher number of elements or polynomial degree: Figures~\ref{fig:e1_K_IG} to \ref{fig:e1_K_ADmax} show a slightly larger slope and  wider shift between different polynomial orders compared to Fig.~\ref{fig:e1_K_NURBS}. A larger shift from the uniform mesh to non-uniform meshes can be noticed, too. Using a non-uniform mesh deteriorates the condition number of the stiffness matrix as it is usually expected.

\subsection{Approximation of mode shapes applying B-Splines}
\label{sec:examples_eig}
\begin{figure}[t]
\centering
\begin{subfigure}[ht]{0.230\textwidth}
	\centering
	\includegraphics[width=\textwidth]{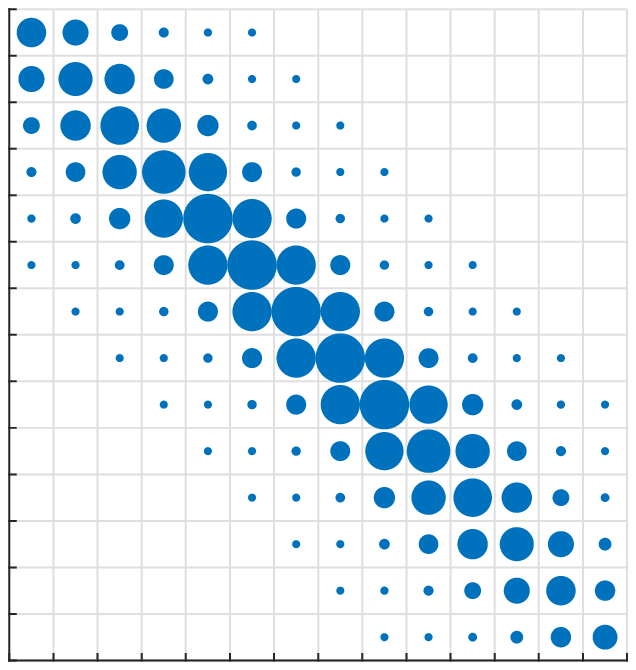}
	\caption{NURBS}
\end{subfigure}
\begin{subfigure}[ht]{0.230\textwidth}
	\centering
	\includegraphics[width=\textwidth]{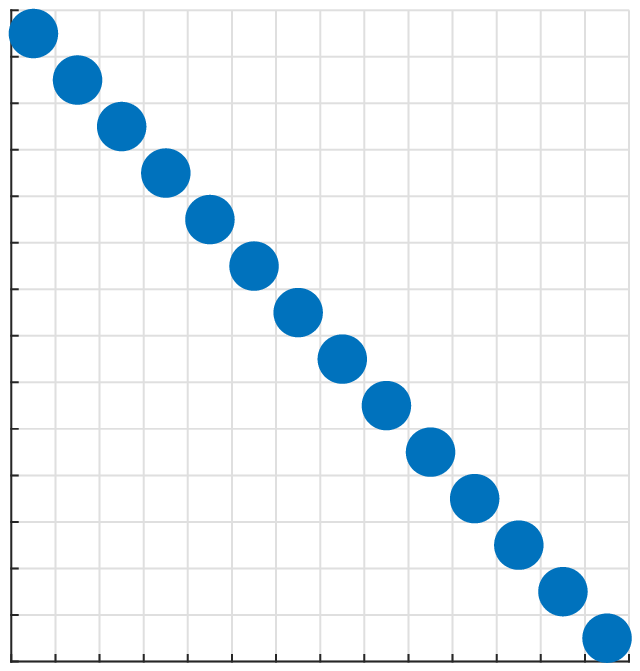}
	\caption{IG}
\end{subfigure}
\begin{subfigure}[ht]{0.230\textwidth}
	\centering
	\includegraphics[width=\textwidth]{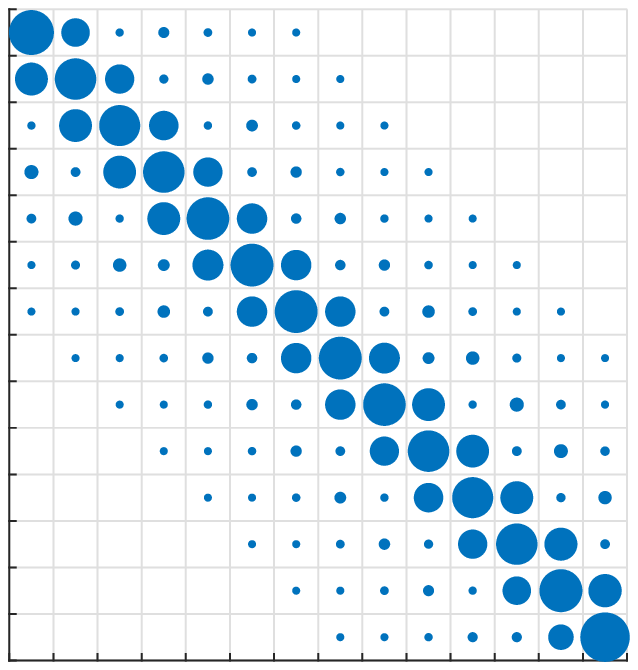}
	\caption{AD $q_{\mathrm{min}}$}
	\label{fig:e1_M0_ADmin}
\end{subfigure}
\begin{subfigure}[ht]{0.230\textwidth}
	\centering
	\includegraphics[width=\textwidth]{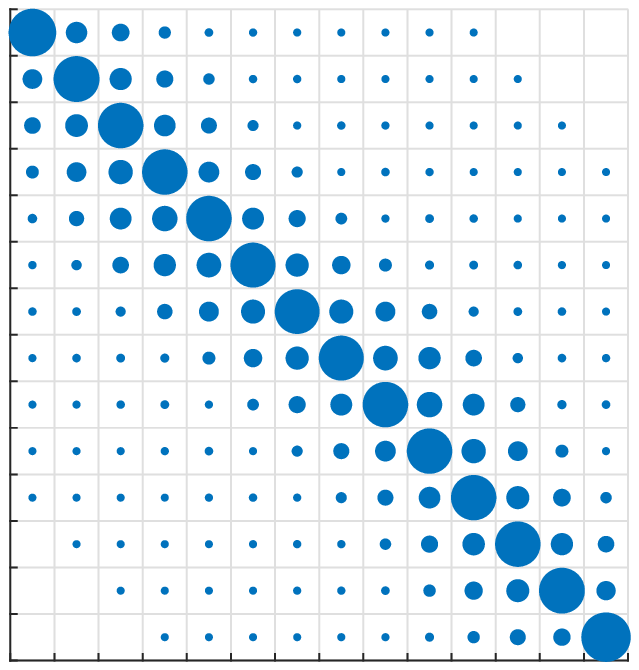}
	\caption{AD $q_{\mathrm{max}}$}
	\label{fig:e1_M0_ADmax}
\end{subfigure}
\caption{Nonzero entries of mass matrix $\mathbf{M}$ with applied Dirichlet boundary conditions for the system shown in Fig.~\ref{fig:sysplot} discretized by 10 elements, size of the dots refers to the magnitude of the entries, NURBS of order $p=5$ are used as shape functions, type of test functions varies.}
\label{fig:e1_M0}
\end{figure}

\begin{figure}[t]
\centering
\begin{subfigure}[ht]{0.180\textwidth}
	\centering
	\includegraphics[width=\textwidth]{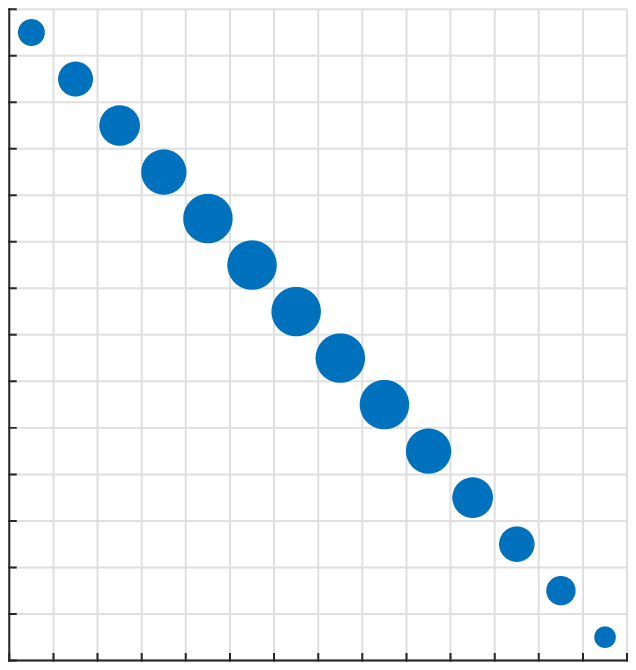}
	\caption{NURBS}
	\label{fig:e1_M1_NURBS}
\end{subfigure}
\qquad
\begin{subfigure}[ht]{0.180\textwidth}
	\centering
	\includegraphics[width=\textwidth]{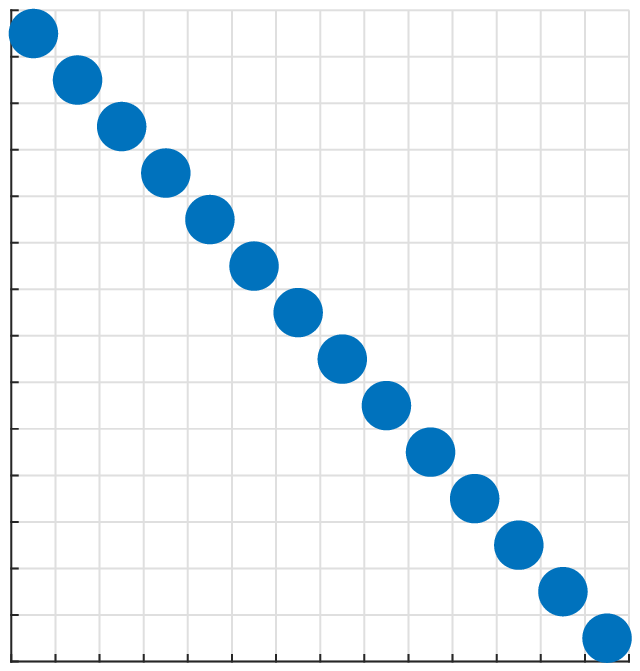}
	\caption{IG}
\end{subfigure}
\qquad
\begin{subfigure}[ht]{0.180\textwidth}
	\centering
	\includegraphics[width=\textwidth]{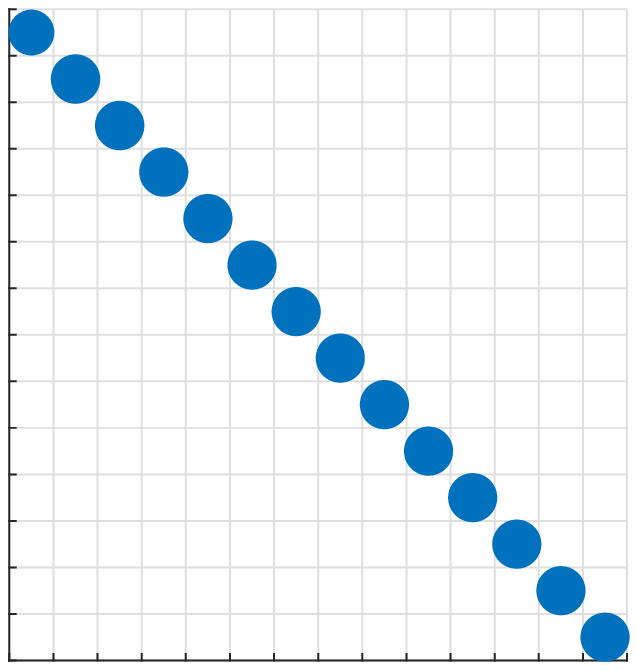}
	\caption{AD $q_{\mathrm{min}}$}
\end{subfigure}
\qquad
\begin{subfigure}[ht]{0.180\textwidth}
	\centering
	\includegraphics[width=\textwidth]{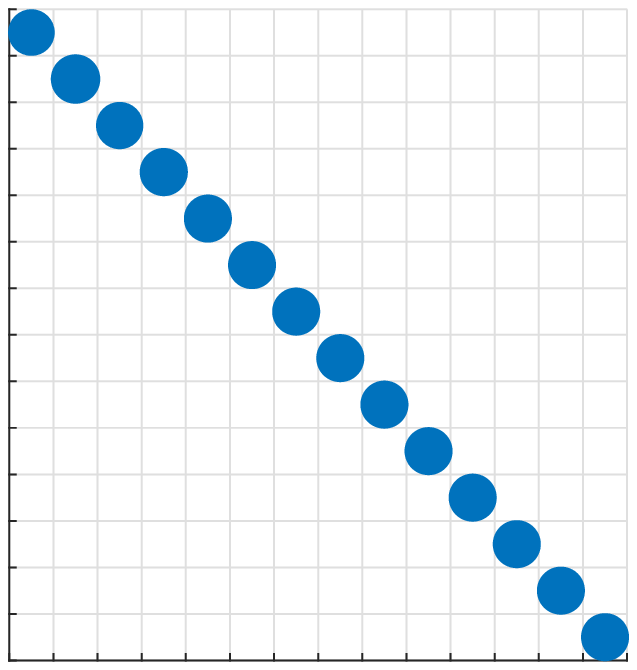}
	\caption{AD $q_{\mathrm{max}}$}
\end{subfigure}
\caption{Nonzero entries of the lumped mass matrices from Fig.~\ref{fig:e1_M0}.}
\label{fig:e1_M1}
\end{figure}

Applications in the field of dynamic problems additionally require the computation of the mass matrix $\mathbf{M}$. Figure~\ref{fig:e1_M0} shows the density pattern of $\mathbf{M}$ depending on the chosen kind of test function. For AD and IG test functions the mass matrix can be obtained by applying a global transformation, cf. Eqs.~(\ref{eq:m_dual}) and (\ref{eq:SOE_dual}). As already discussed in Sec.~\ref{sec:dual}, the IG test functions produce consistent diagonal mass matrices, while AD duals enlarge its bandwidth up to $p+q+1$. Hence, an additional row-sum lumping step will be applied to AD formulations within the following examples to compare the efficiency to standard NURBS formulations, where the mass matrices are also lumped by utilizing the row-sum technique. The size of the points in Figs.~\ref{fig:e1_M0_ADmin} and \ref{fig:e1_M0_ADmax} illustrates that for AD test functions the magnitude of the off-diagonal terms $|m_{ij}|_{i\neq j}<<|m_{ii}|$ is much smaller in comparison to the ones on the main diagonal. For this basic example consisting of 10 elements and employing a basis function degree $p=5$, about 80\% of the total mass is distributed along the main diagonal as effect of the diagonally-dominant mass matrix, whereas in the case of NURBS test functions less than 40\% are distributed along the main diagonal. The mass distribution obtained by the AD formulation with row sum lumping cannot be distinguished by the eye from the consistent diagonal mass matrix obtained by IG test functions, see Fig.~\ref{fig:e1_M1}. In contrast to that, for standard NURBS functions with applied row sum lumping, a clear distribution along the domain is noticeable. As for the AD approach the major part of the consistent mass is already concentrated on the main diagonal, the loss of accuracy through additional row-sum lumping should not be as huge as for the standard formulation.

Analogous to Sec.~\ref{sec:transformation}, it can be shown that 
\begin{equation}
(\mathbf{S}~\mathbf{K}-\omega^2\mathbf{S}~\mathbf{M})\mathbf{\Phi}=0~\leftrightarrow~(\mathbf{K}-\omega^2\mathbf{M})\mathbf{\Phi}=0
\end{equation}
holds true and thus, eigenfrequencies $\omega$ and mode shapes $\mathbf{\Phi}$ of the dual formulation should not differ from the symmetric NURBS formulation. Due to this fact the so called `outliers', already discussed in the early years of IGA by \cite{Cottrell.2006,Hughes.2008}, also appear within the proposed formulation. These additional frequencies without physical relevance, caused by the high polynomial order of NURBS, are of comparable huge magnitude. They appear for the last $p$ eigenfrequencies for even and $p-1$ for odd polynomial degrees $p$, if uniform meshes are used. Within the succeeding numerical examples the outlier frequencies are harmless like for many applications. As already stated by \citet{Cottrell.2006}, the outliers still affect on the computational costs as they determine the critical time step size of explicit time integration schemes. In general, for wave propagation problems the risk remains that the highest modes, which are the outlier ones, also participate in the solution and thus, deteriorate accuracy as well. Recent studies by \citet{Hiemstra.2021} propose a technique to remove the outlier frequencies and hence, how to avoid the negative impact on computational efficiency and accuracy. Fortunately, for the basic examples within this work, the outlier frequencies do not affect the solution and thus, we did not take further measures to remove them. Nevertheless, in the spectral plots, which are presented at a later point, they do not show up, as we decided to omit them for better clarity of the results of interest. An insight regarding the behavior of outliers for the dual in comparison to the standard formulation is given in Fig.~\ref{fig:e2_spec_out} and demonstrates that the general behavior is not affected. Thus, taking further measures as in \cite{Hiemstra.2021,Nguyen.2022} should be considered for efficient computations, but within this study we will solely focus on the investigation of the dual lumping scheme to avoid false interpretations of the results.
\begin{figure}[t]
\centering
\begin{subfigure}[ht]{0.40\textwidth}
	\centering
	\includegraphics[width=\textwidth]{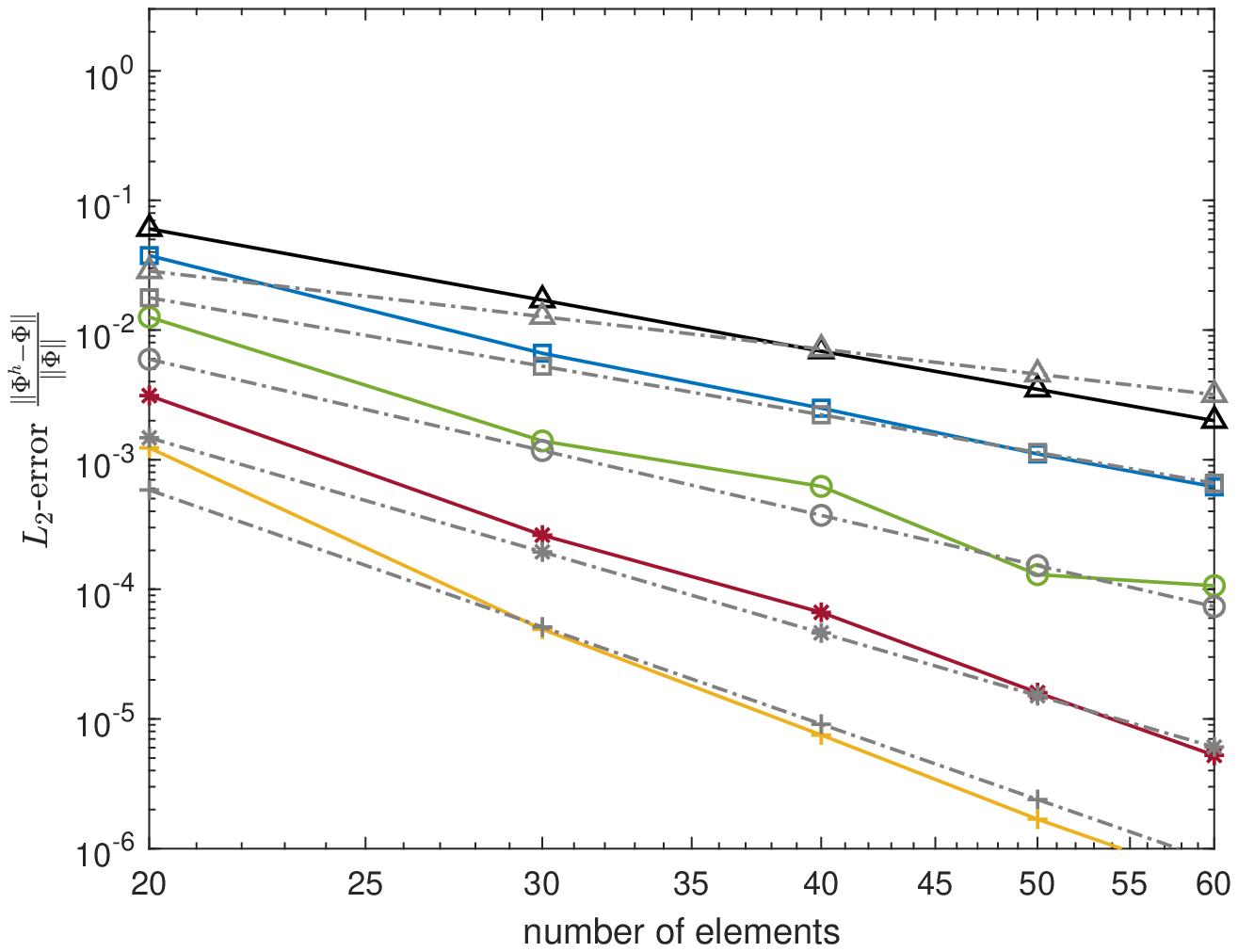}
	\caption{$M_{\mathrm{NURBS,AD,IG}}^{\mathrm{cons}}$}
	\label{fig:e2_L2_NURBSc}
\end{subfigure}
\quad
\begin{subfigure}[ht]{0.40\textwidth}
	\centering
	\includegraphics[width=\textwidth]{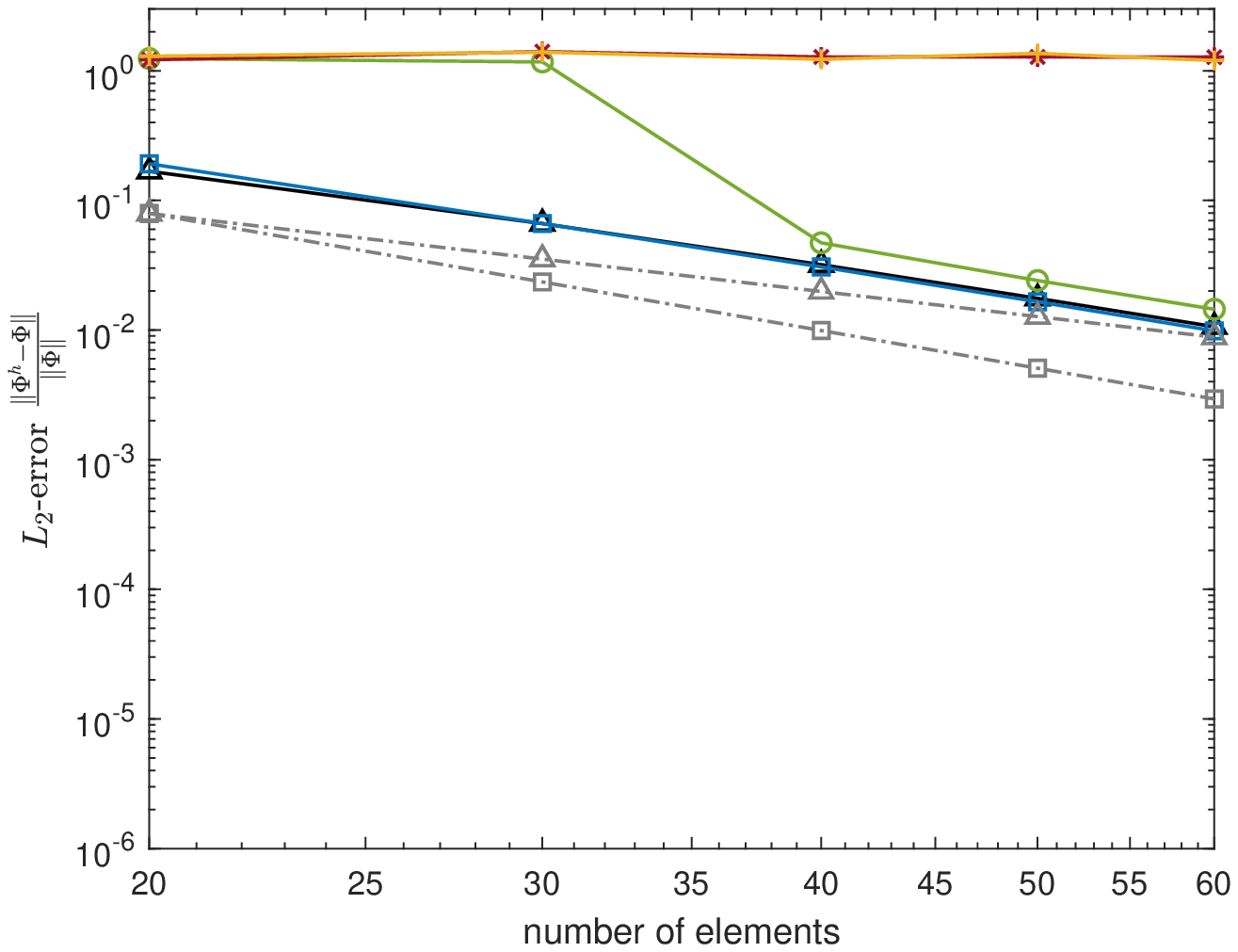}
	\caption{$M_{\mathrm{NURBS}}^{\mathrm{row-sum}}$}
	\label{fig:e2_L2_NURBSd}
\end{subfigure}\\
\begin{subfigure}[ht]{0.40\textwidth}
	\centering
	\includegraphics[width=\textwidth]{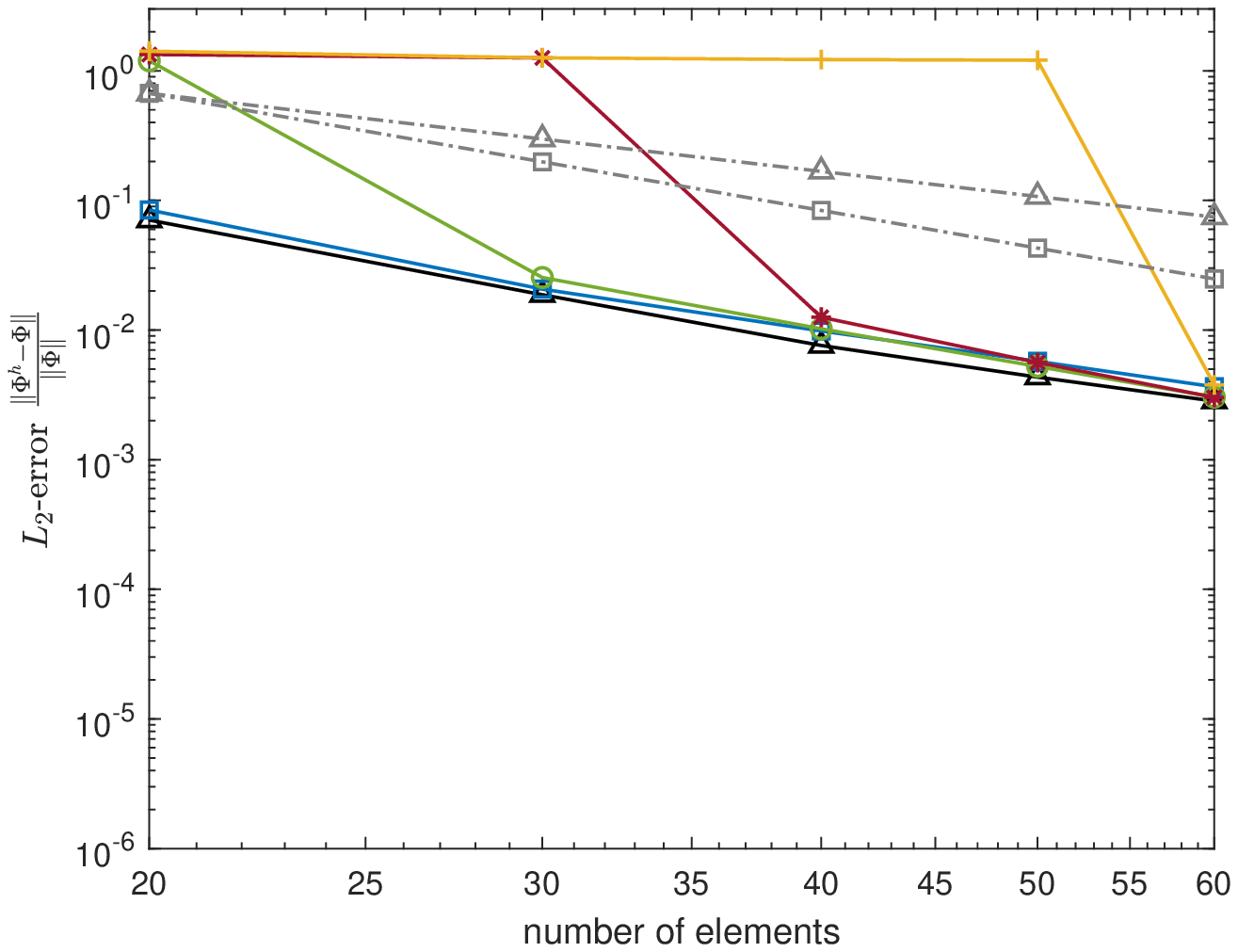}
	\caption{$M_{\mathrm{AD}~q_{\mathrm{min}}}^{\mathrm{row-sum}}$}
	\label{fig:e2_L2_ADmin}
\end{subfigure}
\quad
\begin{subfigure}[ht]{0.40\textwidth}
	\centering
	\includegraphics[width=\textwidth]{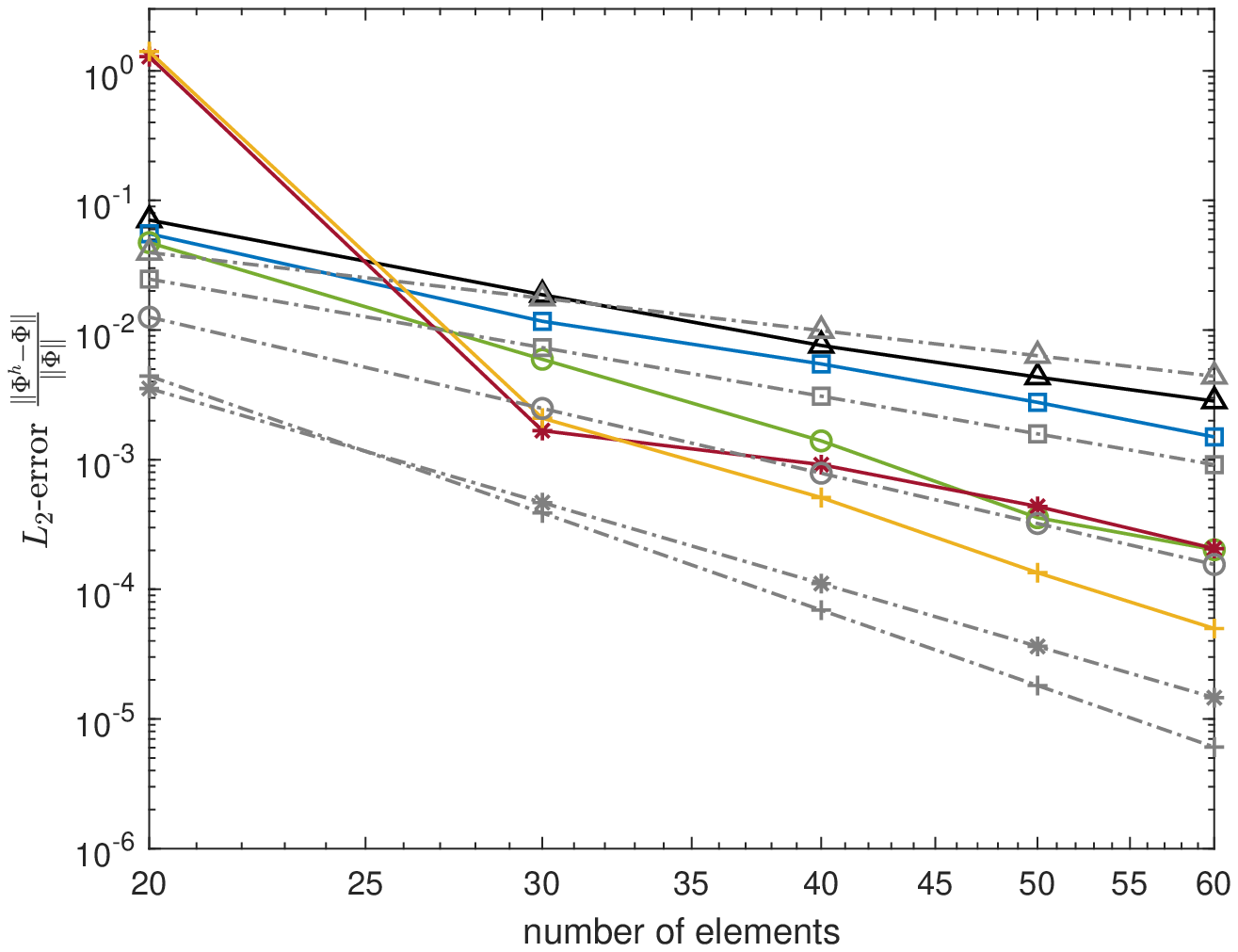}
	\caption{$M_{\mathrm{AD}~q_{\mathrm{max}}}^{\mathrm{row-sum}}$}
	\label{fig:e2_L2_ADmax}
\end{subfigure}\\
\begin{subfigure}[ht]{1.00\textwidth}
	\centering
	\includegraphics[width=\textwidth]{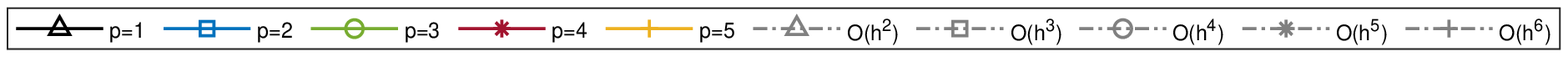}
\end{subfigure}
\caption{$L_2$-error norm of the $10^{\mathrm{th}}$ mode shape corresponding to $f_{10}$, NURBS of order $p$ are used as shape functions, type of test functions varies. Computation was done on uniform mesh~A (Fig.~\ref{fig:mesh1}).}
\label{fig:e2_L2}
\end{figure}

\begin{figure}[t]
\centering
\begin{subfigure}[ht]{0.40\textwidth}
	\centering
	\includegraphics[width=\textwidth]{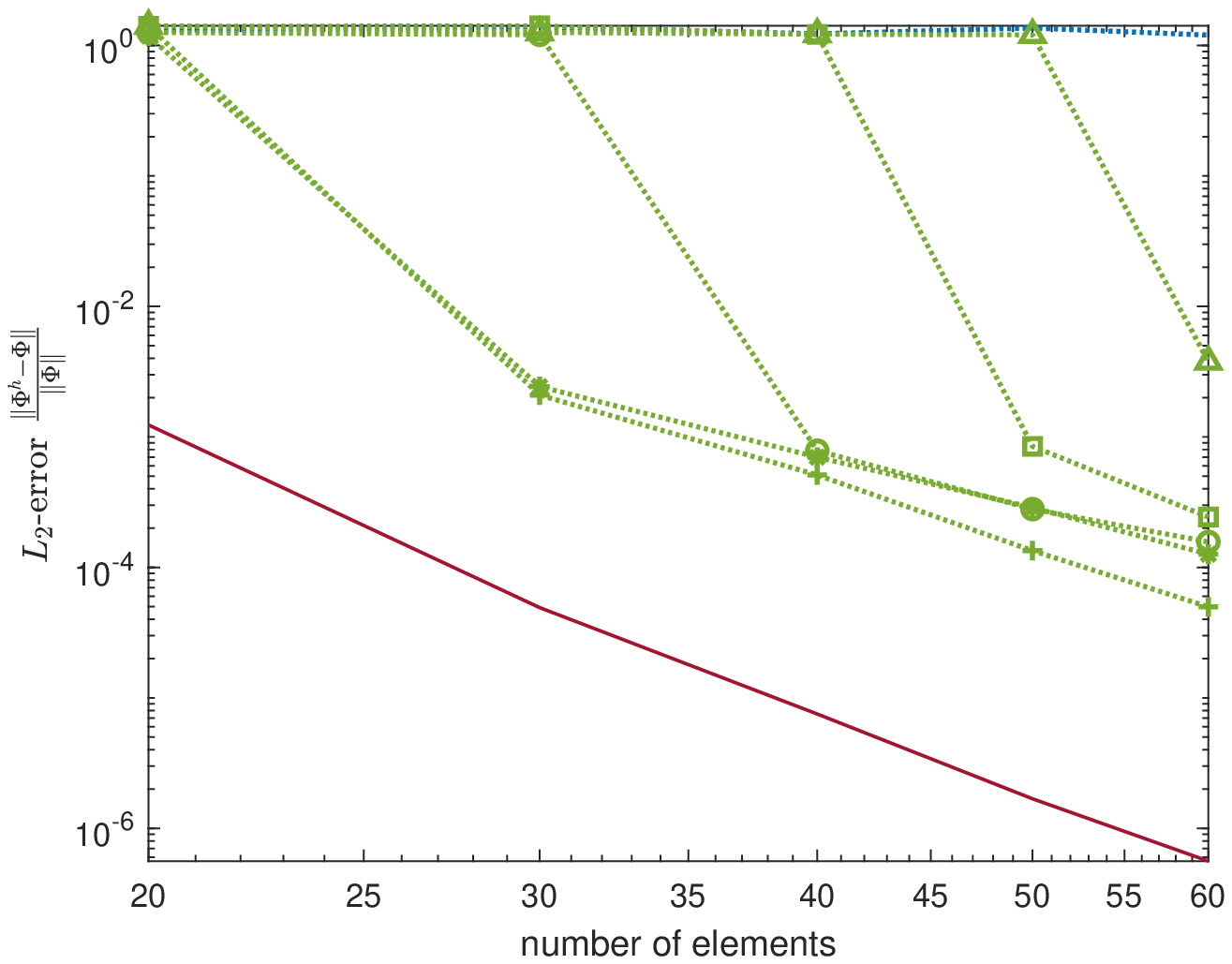}
\end{subfigure}
\quad
\begin{subfigure}[ht]{0.10\textwidth}
	\centering
	\includegraphics[width=\textwidth]{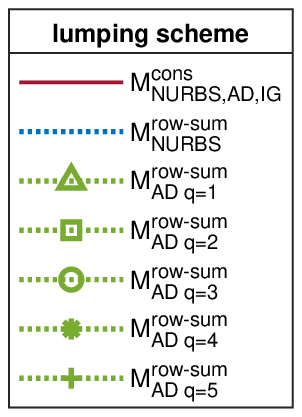}
\end{subfigure}
\caption{$L_2$-error norm of the $10^{\mathrm{th}}$ mode shape from Fig.~\ref{fig:e2_L2} comparing different test functions, NURBS of order $p=5$ are used as shape functions. Computation was done on uniform mesh~A (Fig.~\ref{fig:mesh1}).}
\label{fig:e2_L2_p5}
\end{figure}
Before the computational costs are measured, we will focus on the accuracy of the computed mode shapes and eigenfrequencies. The same set of test functions as in Sec.~\ref{sec:examples_static} is applied to the example from Fig.~\ref{fig:sysplot}, only using the uniform mesh~A. The results are compared to the analytical solution 
\begin{equation}
\omega_n=\frac{\pi}{2L}(2n-1)\sqrt{\frac{EA}{\mu}},~\text{with}~\mu=\rho A
\end{equation}
and
\begin{equation}
\Phi_n=\sin\left(\frac{\pi}{2L}(2n-1) x\right)~.
\end{equation}
The truss properties are chosen as 
$E=1$, $A=\pi/4$ 
and 
$\rho=10^{-4}$. 
Figure~\ref{fig:e2_L2} depicts the accuracy of the approximation of the $10^{\mathrm{th}}$ mode shape, applying NURBS shape functions with polynomial orders $p=1~\text{to}~5$. Since all consistent formulations provide the same outcomes by means of the global transformation, their results are grouped together to the lumping scheme "$\mathbf{M}^{\mathrm{cons}}_{\mathrm{NURBS,AD,IG}}$". The results in Fig.~\ref{fig:e2_L2_NURBSc} are compared to row-sum lumped approaches, using NURBS (Fig.~\ref{fig:e2_L2_NURBSd}) and AD with minimal (Fig.~\ref{fig:e2_L2_ADmin}) and maximum reproduction (Fig.~\ref{fig:e2_L2_ADmax}) as test functions. For the lowest possible order $p=1$ in Fig.~\ref{fig:e2_L2}, almost no differences between the consistent and lumped approaches can be spotted. This behavior is as expected, since for $p=1$ IGA is identical to standard FEM, wherefore row-sum lumping is known to provide very good results. The bandwidth of the mass matrices is close to one, not far from diagonal. Thus, the applied lumping scheme does not significantly affect the results. With the consistent formulation the expected convergence rate of `$p+1$' is achieved. Applying row-sum lumping within the Bubnov-Galerkin formulation leads to higher errors and limited convergence rates. A growth or stagnation of the error norm arises for higher polynomial orders $p>2$ until the amount of elements is large enough to pass into convergence. The results get even worse raising the NURBS order $p$. This is again due to the rising bandwith, which is than summed up on the diagonal.

Replacing the NURBS by AD test functions with $q_\mathrm{min}$ improves the  convergence behavior only in that way, that the graphs for orders $p>2$ pass into convergence for a lower amount of elements and the general accuracy level is slightly enhanced. For both cases, a maximum convergence rate of $O(h^3)$ is received for the studied refinements. Thus, the results based on lumped mass matrices have to compete with the consistent formulation of order $p=2$. From Fig.~\ref{fig:e2_L2_ADmax} it is noticeable, that for AD with maximum reproduction employing higher polynomial orders seems convincing, as further $k$-refinement raises the accuracy level and the convergence rate as well. In Fig.~\ref{fig:e2_L2_p5}, the improvement of convergence through raising the degree of reproduction is depicted for a polynomial order of $p=5$. The consistent formulation always provides the best convergence behavior and in particular the pure row-sum lumping technique the worst. A clear distinction between different degrees of reproduction, from minimal to maximum, can be recognized. With the increase of the reproduction degree the graph moves gradually towards the results of the consistent formulation. It can be noticed that the higher the degree of reproduction is chosen, the less elements are necessary to obtain convergence. If $q_\mathrm{max}$ is applied, the same convergence rate as for the consistent formulation is captured, but the same accuracy level cannot be achieved. Nevertheless, in comparison to the standard row-sum lumping, the dual lumping scheme with maximum reproduction can be entitled as high order convergent.

\begin{figure}[t]
\centering
\includegraphics[width = 0.5\textwidth]{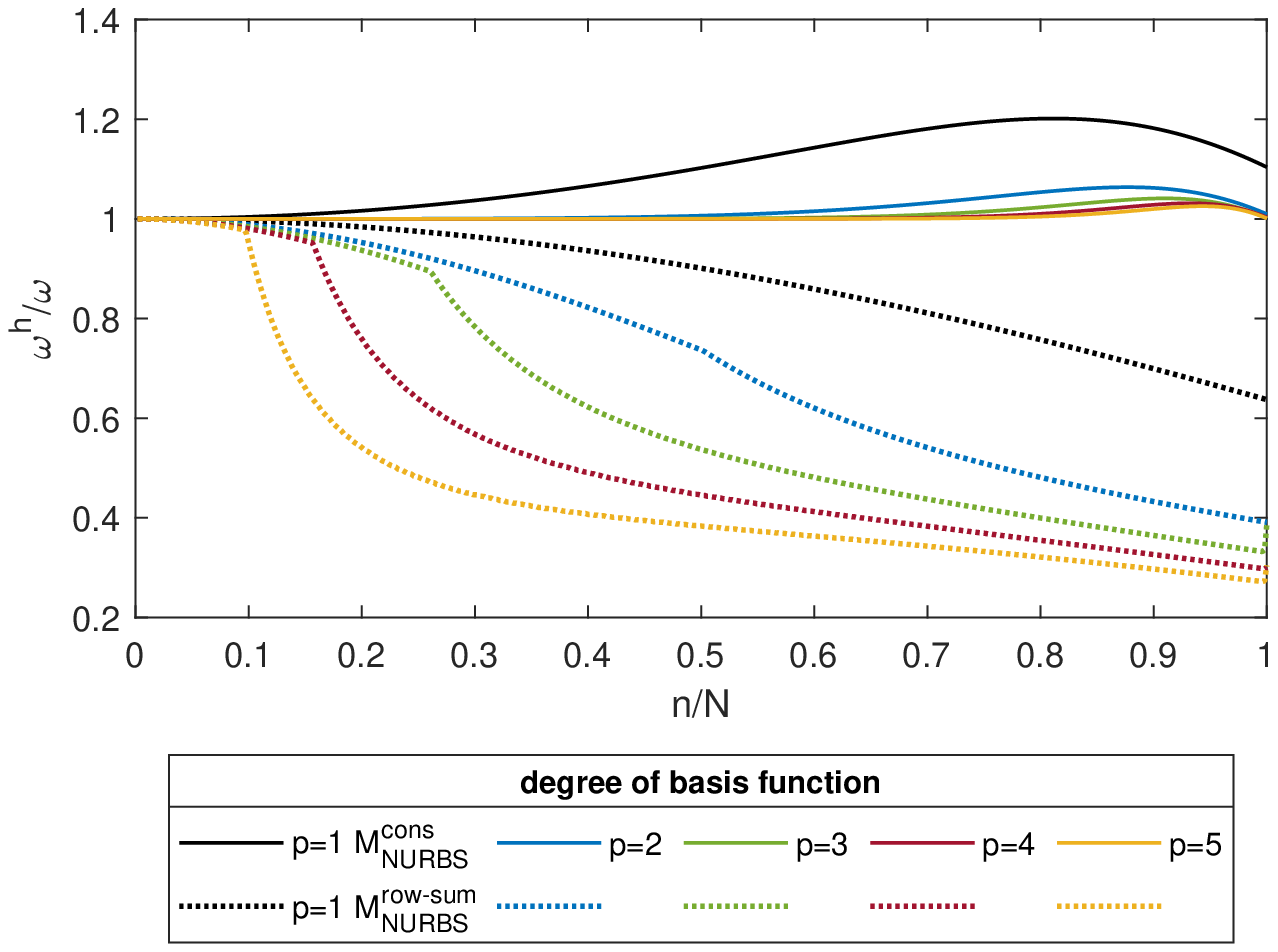}
\caption{Numerical spectra comparing Bubnov-Galerkin formulation with consistent $\mathbf{M}_\mathrm{NURBS}^\mathrm{cons}$ and row-sum lumped mass matrices $\mathbf{M}_\mathrm{NURBS}^\mathrm{row-sum}$ for different polynomial orders $p$. Computation was done on uniform mesh~A (Fig.~\ref{fig:mesh1}).}
\label{fig:e2_spec_NURBS}
\end{figure}

\begin{figure}[t]
\centering
\begin{subfigure}[ht]{0.40\textwidth}
	\centering
	\includegraphics[width=\textwidth]{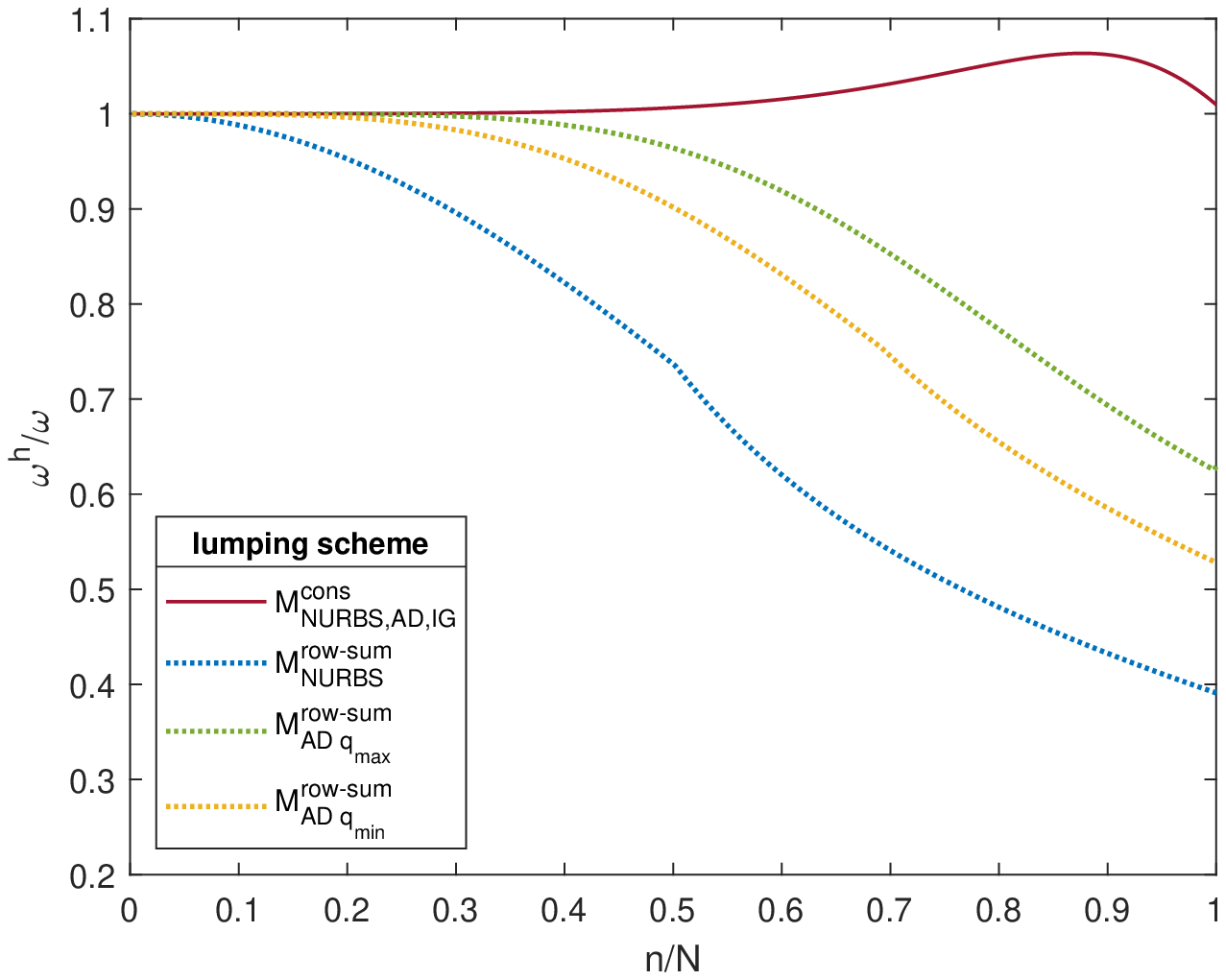}
	\caption{$p=2$}
	\label{fig:e2_spec_p2}
\end{subfigure}
\quad
\begin{subfigure}[ht]{0.40\textwidth}
	\centering
	\includegraphics[width=\textwidth]{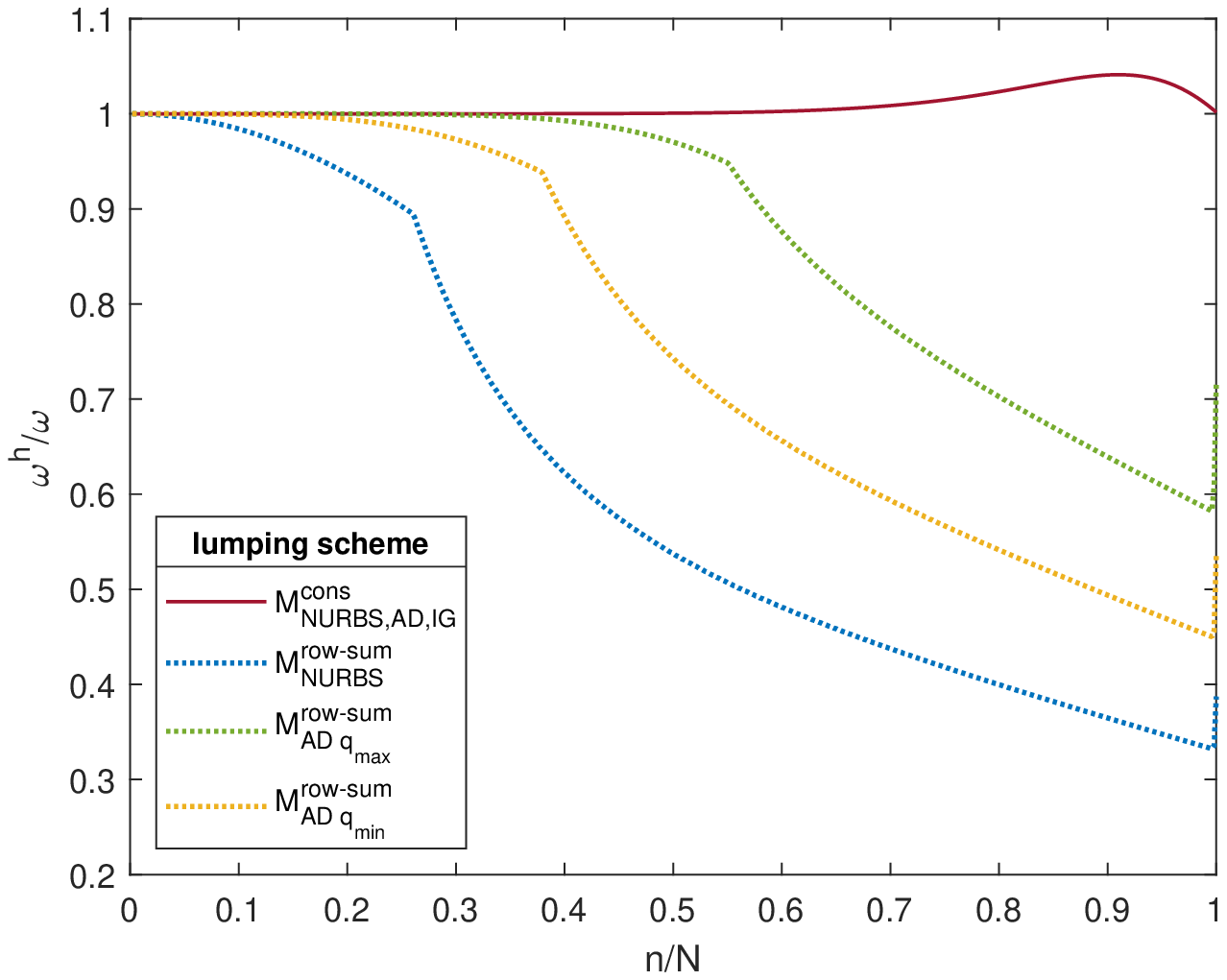}
	\caption{$p=3$}
	\label{fig:e2_spec_p3}
\end{subfigure}\\
\begin{subfigure}[ht]{0.40\textwidth}
	\centering
	\includegraphics[width=\textwidth]{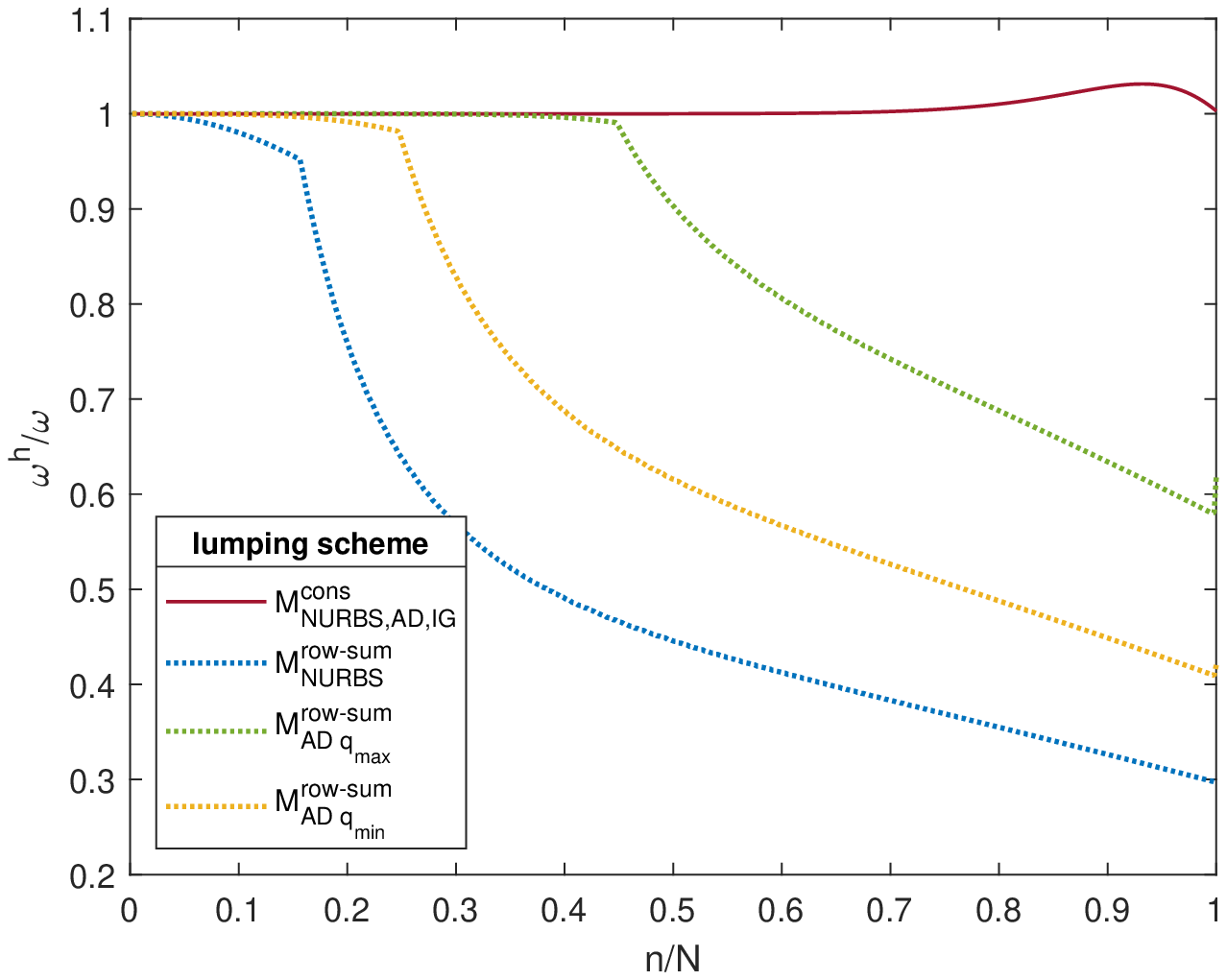}
	\caption{$p=4$}
	\label{fig:e2_spec_p4}
\end{subfigure}
\quad
\begin{subfigure}[ht]{0.40\textwidth}
	\centering
	\includegraphics[width=\textwidth]{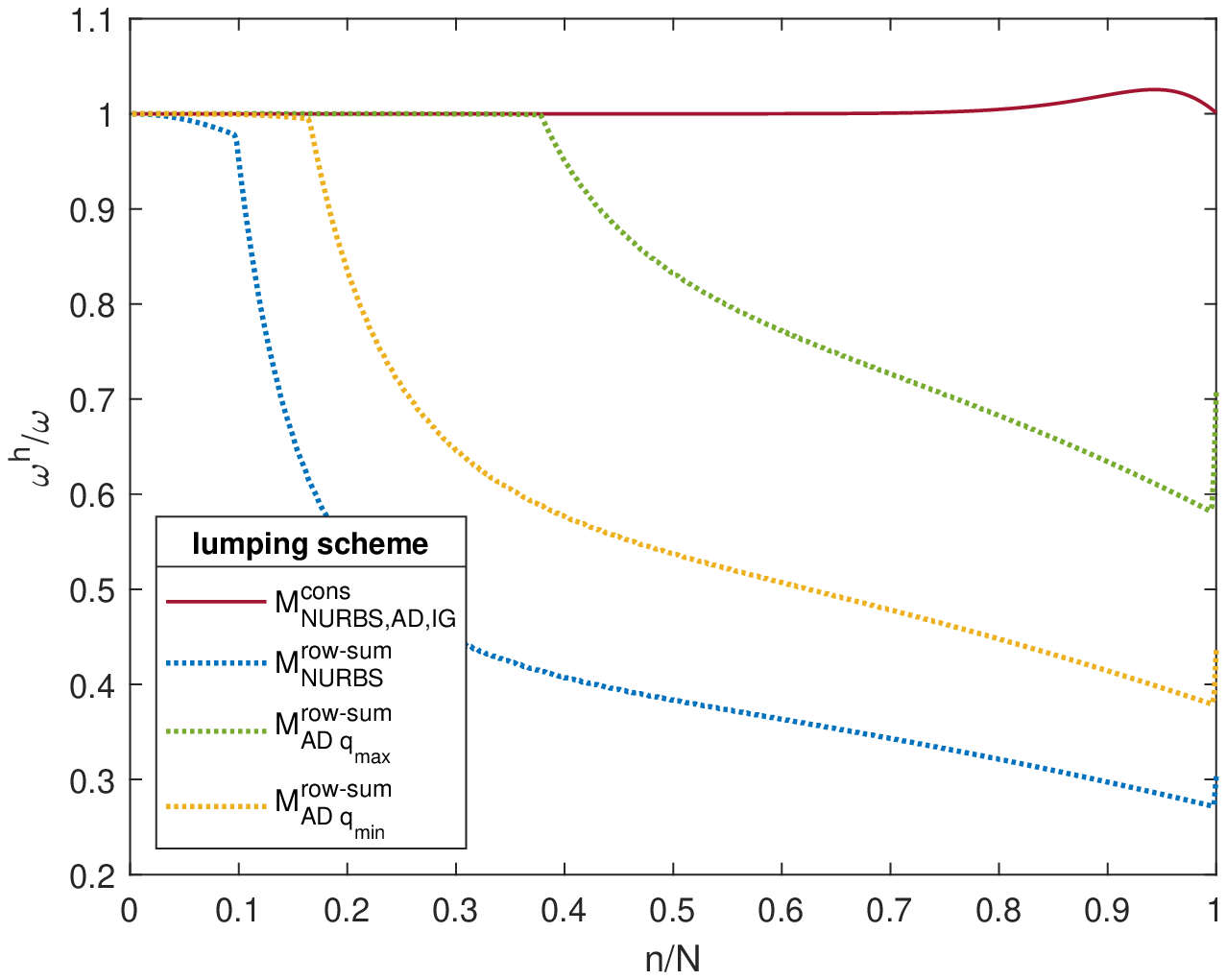}
	\caption{$p=5$}
	\label{fig:e2_spec_p5}
\end{subfigure}
\caption{Numerical spectra of the consistent formulation in comparison with the lumped Bubnov-Galerkin formulation and the dual lumping approach AD with minimal and maximum reproduction degree for polynomial orders $p=2$ to 5. Computation was done on uniform mesh~A (Fig.~\ref{fig:mesh1}).}
\label{fig:e2_spec_dual}
\end{figure}

\begin{figure}[t]
\centering
\includegraphics[width=0.40\textwidth]{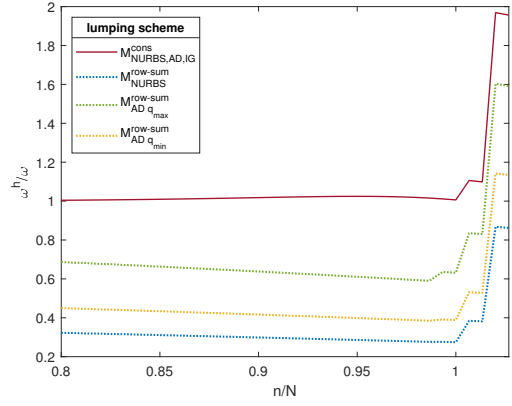}
\caption{Numerical spectra from Fig.~\ref{fig:e2_spec_p5}: last 20$\%$ in detail, extended by the occurring outlier frequencies.}
\label{fig:e2_spec_out}
\end{figure}

The observed convergence behavior for the $10^{th}$ mode shape can be noticed for the corresponding eigenvalue as well. Further, the findings of this specific mode can be transferred completely to all other modes. The approximation of all eigenfrequencies captured  by the system is best to examine within a plot of the overall spectrum. Using a consistent mass matrix, a low ratio $n/N$ of examined eigenfrequency $n$ and total number of eigenfrequencies $N$, determined by the number of elements, provides better accuracy than a higher value, but getting closer to the last captured eigenfrequencies, the ratio $\frac{\omega^\mathrm{h}}{\omega}$ is tending to 1 again. For common NURBS formulations and a fixed mesh with 200 elements, Fig.~\ref{fig:e2_spec_NURBS} compares the accuracy of eigenfrequencies computed by consistent and lumped mass matrices. Without lumping the findings are equal to the static applications: the accuray increases if the NURBS order $p$ is elevated. However, for applied row-sum lumping the results get worse while raising $p$ and they do not show a reversal to better approximation towards the ratio of $n/N=1$. In contrast to consistent mass formulations, where $\omega^h>\omega$ holds true, it is also noticed that due to lumping the eigenfrequencies are underestimated. Within an explicit time integration that fact will allow to choose a comparatively larger critical time step. Consequently, a computation based on poorly approximated eigenmodes will be run with less time steps to solve, possibly leading to even less accurate results. To improve the quality of the outcome, NURBS of low order or a larger amount of elements should be considered to keep the range of frequencies of interest at a low ratio $n/N$. Of course the usage of a high number of elements was not the intent of IGA and therefore, there is a huge demand for improved lumping schemes, such as the proposed dual lumping. Figure~\ref{fig:e2_spec_dual} shows the improvements through applying the dual lumping scheme. The eigenfrequencies are underestimated as well, but better accuracy is provided. For instance, for a polynomial order $p=4$ the standard row-sum lumping approach approximates the highest eigenfrequency by only 30$\%$. With AD test functions of maximum reproduction the accuracy can be raised to almost 60$\%$. Allowing 10$\%$ deviation they are accurate for half of the spectrum, where standard lumping only yields accuracy for less than a fifth. Even for the minimal degree of reproduction the dual lumping approach achieves always better results than standard lumping. The general behavior and the noticed facts remain untouched by the occurring outlier eigenfrequencies, as shown in Fig.~\ref{fig:e2_spec_out} for the last part of the discussed spectra. In that specific example, the participation of outliers will, independently from the chosen test functions, lead to around twice as many time steps as it would take for computations based on system matrices of same size, but without these spurious frequencies.

\begin{figure}[t]
\centering
\begin{subfigure}[ht]{0.40\textwidth}
	\centering
	\includegraphics[width=\textwidth]{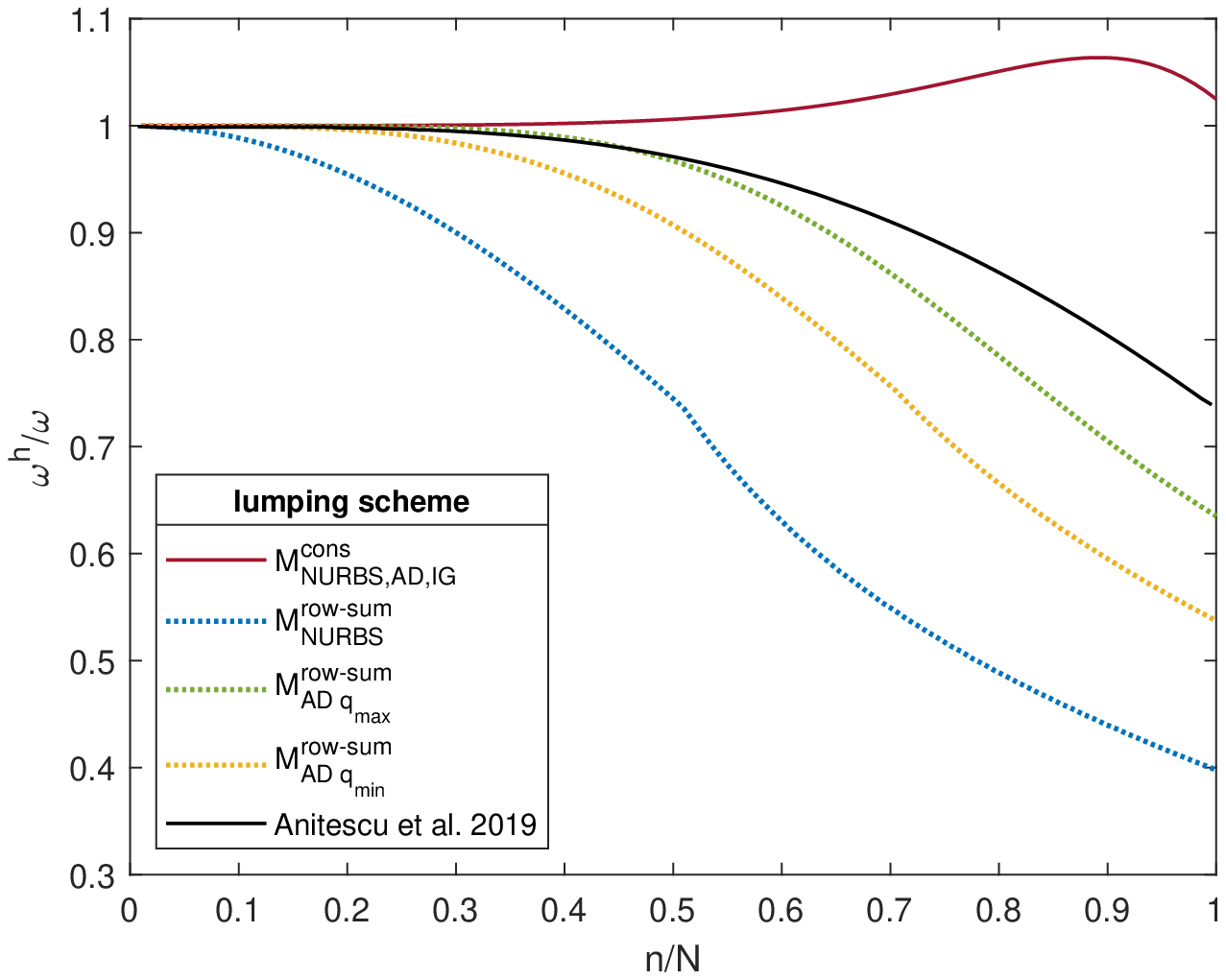}
	\caption{$p=2$}
	\label{fig:e2_spec_Anitescu_p2}
\end{subfigure}
\quad
\begin{subfigure}[ht]{0.40\textwidth}
	\centering
	\includegraphics[width=\textwidth]{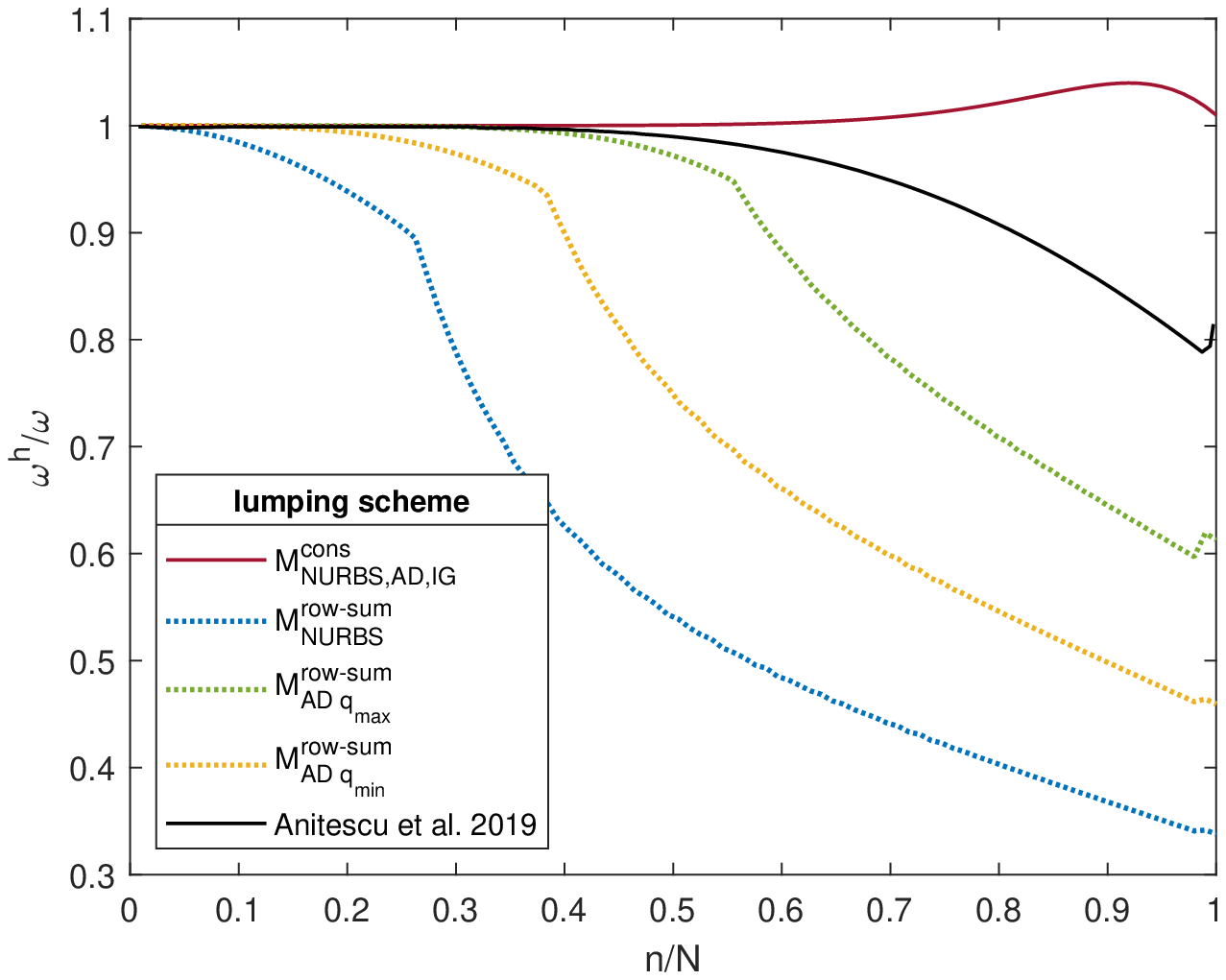}
	\caption{$p=3$}
	\label{fig:e2_spec_Anitescu_p3}
\end{subfigure}
\caption{Numerical spectra in comparison with the dual lumping approach of  \citet{Anitescu.2019} for polynomial orders $p=\{2,3\}$.}
\label{fig:e2_spec_Anitescu}
\end{figure}

\begin{figure}[b]
\centering
\includegraphics[width=0.3\textwidth]{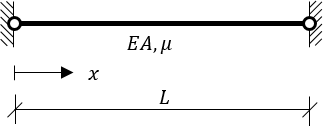}
\caption{System sketch for truss with fixed-fixed boundary conditions.}
\label{fig:sysplot_dyn}
\end{figure}

As mentioned in the introduction (Sec.~\ref{sec:intro}), a dual lumping scheme was already introduced by \citet{Anitescu.2019}, providing a consistent diagonal mass matrix without full reproduction. A comparison to the formulation conceived within this work is given in Fig.~\ref{fig:e2_spec_Anitescu}. The spectra correspond to a fixed-fixed uniaxial truss, see Fig.~\ref{fig:sysplot_dyn}, with the analytical solution 
\begin{equation}
\omega_n = \frac{\pi}{L} n\sqrt{\frac{E}{\rho}}~.
\end{equation} 

The proposed lumping scheme in \cite{Anitescu.2019} does not overestimate the analytical solution as the other consistent formulations do. Better accuracy is received than for the AD lumping scheme with additional row-sum lumping, irrespective of polynomial order $p$ or chosen degree of reproduction. In the case of $p=2$, see Fig.~\ref{fig:e2_spec_Anitescu_p2}, the gap between the formulations rises only slightly towards the largest eigenmode. Figure~\ref{fig:e2_spec_Anitescu_p3} shows clearly the influence of additional row-sum lumping. While the formulation from \cite{Anitescu.2019} leads to improved results through increased polynomial order, for the AD formulation a slight decline in accuracy for the high eigenmodes can be detected. However, for full reproduction, the results are close together for nearly 60$\%$ of the spectrum. Already mentioned before, it is also obvious that with the proposed IG lumping scheme, more accurate results compared to those obtained in \cite{Anitescu.2019} are achieved as full reproduction is secured, cf. red curves in Figs.~\ref{fig:e2_spec_dual} and \ref{fig:e2_spec_Anitescu}. What the spectral plots cannot depict is the effort of computing the system matrices. The derived formulations based either on IG or AD test functions seem to be much easier to implement in existing codes as the adaptation of dual basis functions can be shifted from element level to a simple transformation operation.

\begin{figure}[t]
\centering
\begin{subfigure}[ht]{0.40\textwidth}
	\centering
	\includegraphics[width=\textwidth]{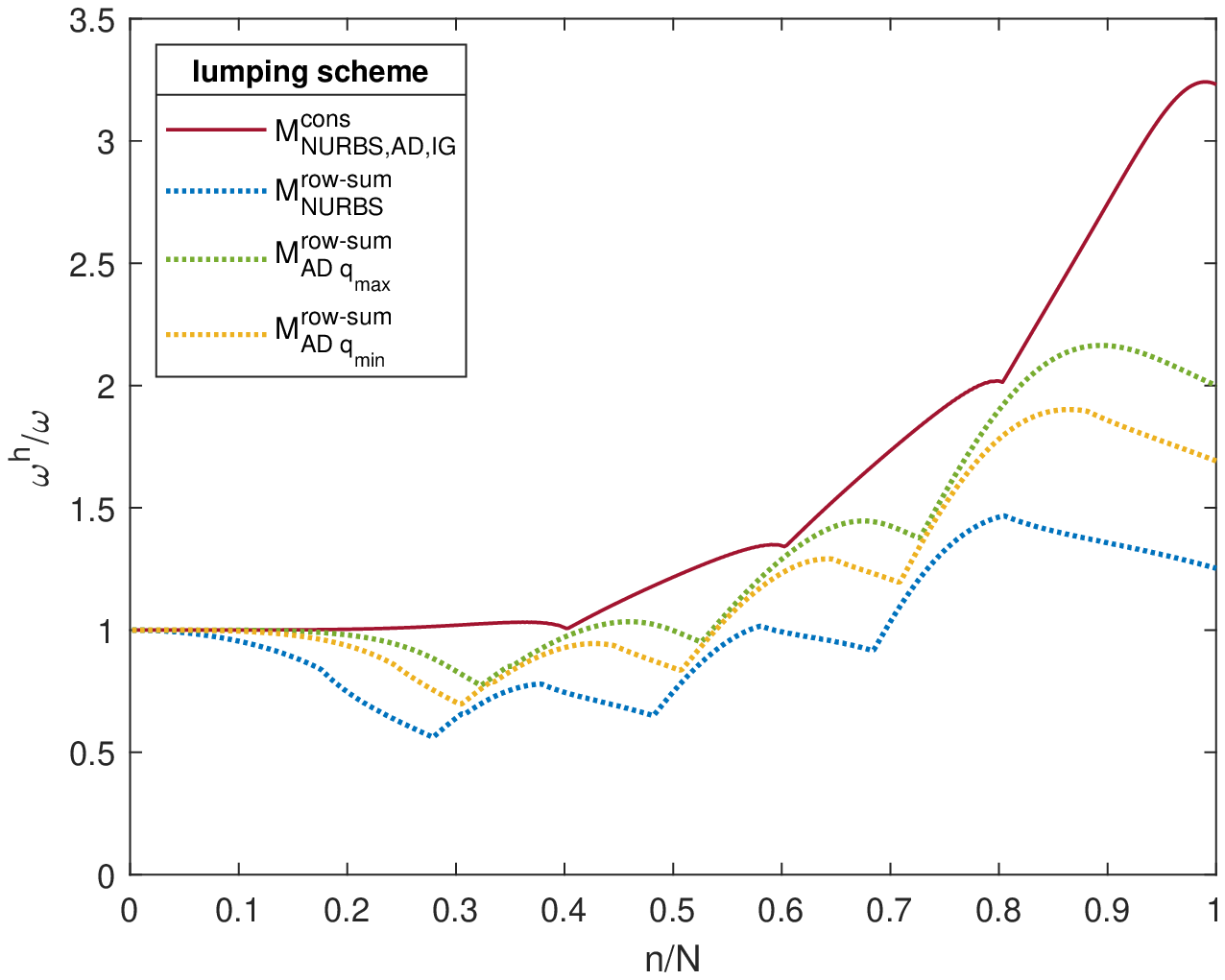}
	\caption{$p=2$}
	\label{fig:e2_spec_netB_p2}
\end{subfigure}
\quad
\begin{subfigure}[ht]{0.40\textwidth}
	\centering
	\includegraphics[width=\textwidth]{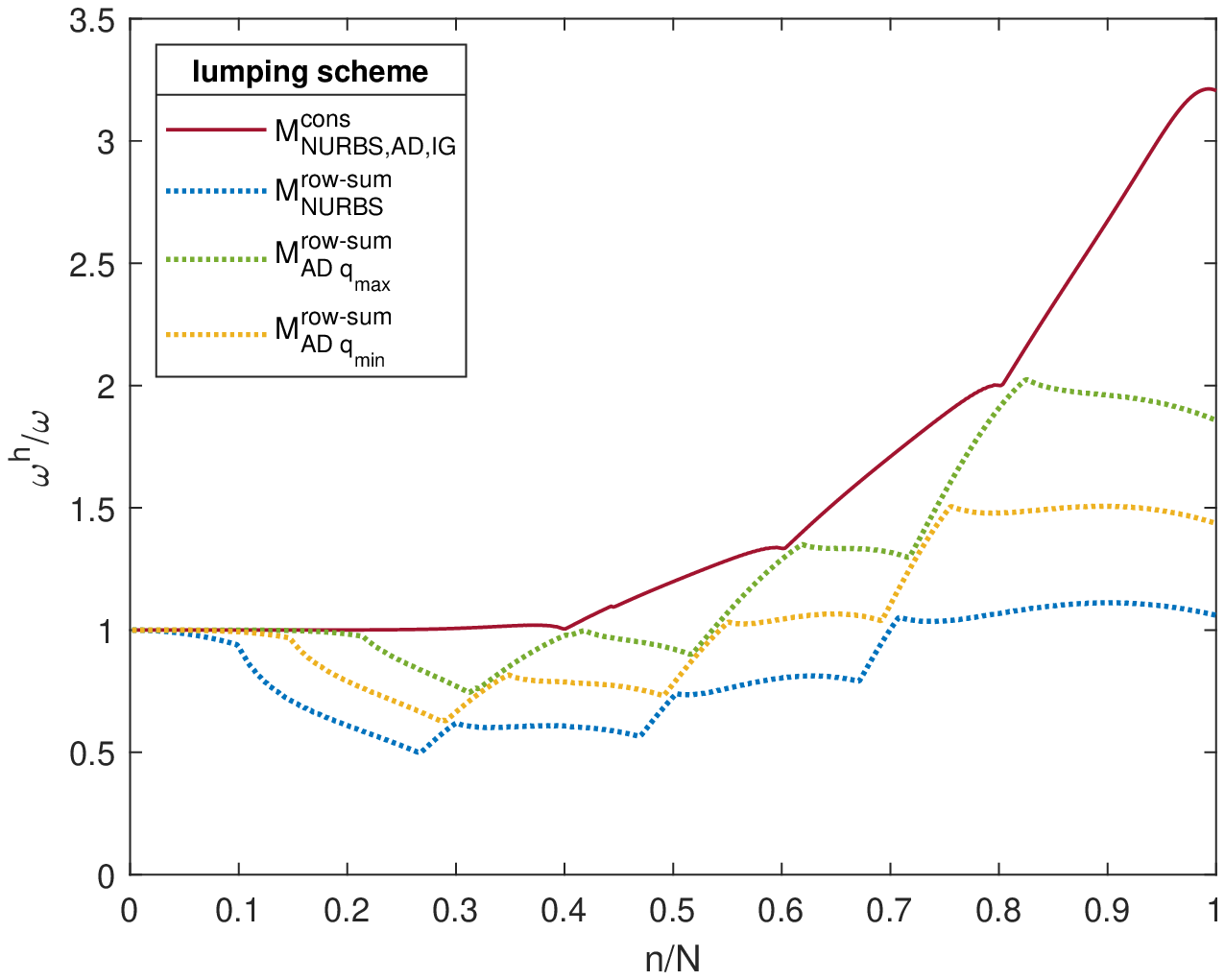}
	\caption{$p=3$}
	\label{fig:e2_spec_netB_p3}
\end{subfigure}\\
\begin{subfigure}[ht]{0.40\textwidth}
	\centering
	\includegraphics[width=\textwidth]{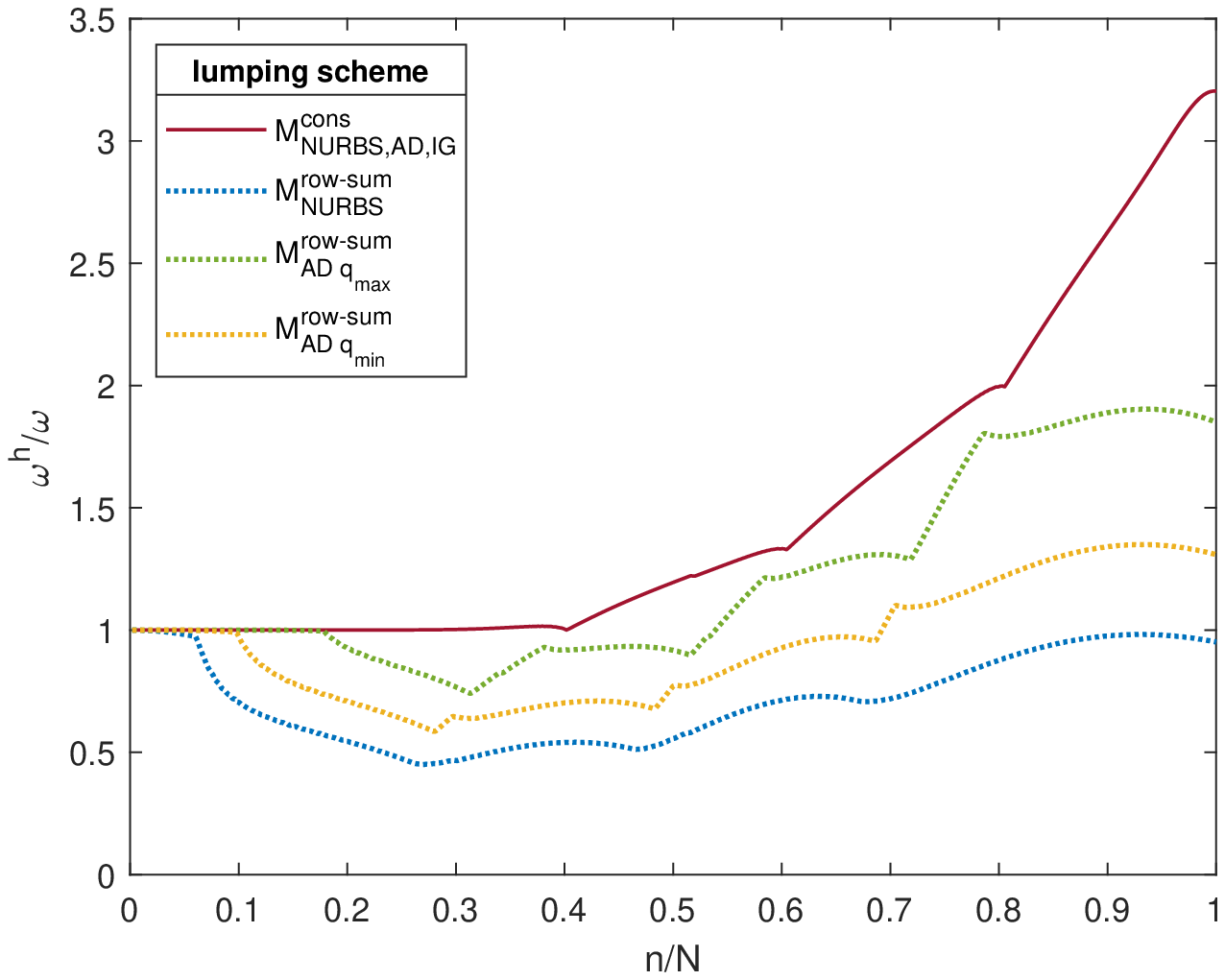}
	\caption{$p=4$}
	\label{fig:e2_spec_netB_p4}
\end{subfigure}
\quad
\begin{subfigure}[ht]{0.40\textwidth}
	\centering
	\includegraphics[width=\textwidth]{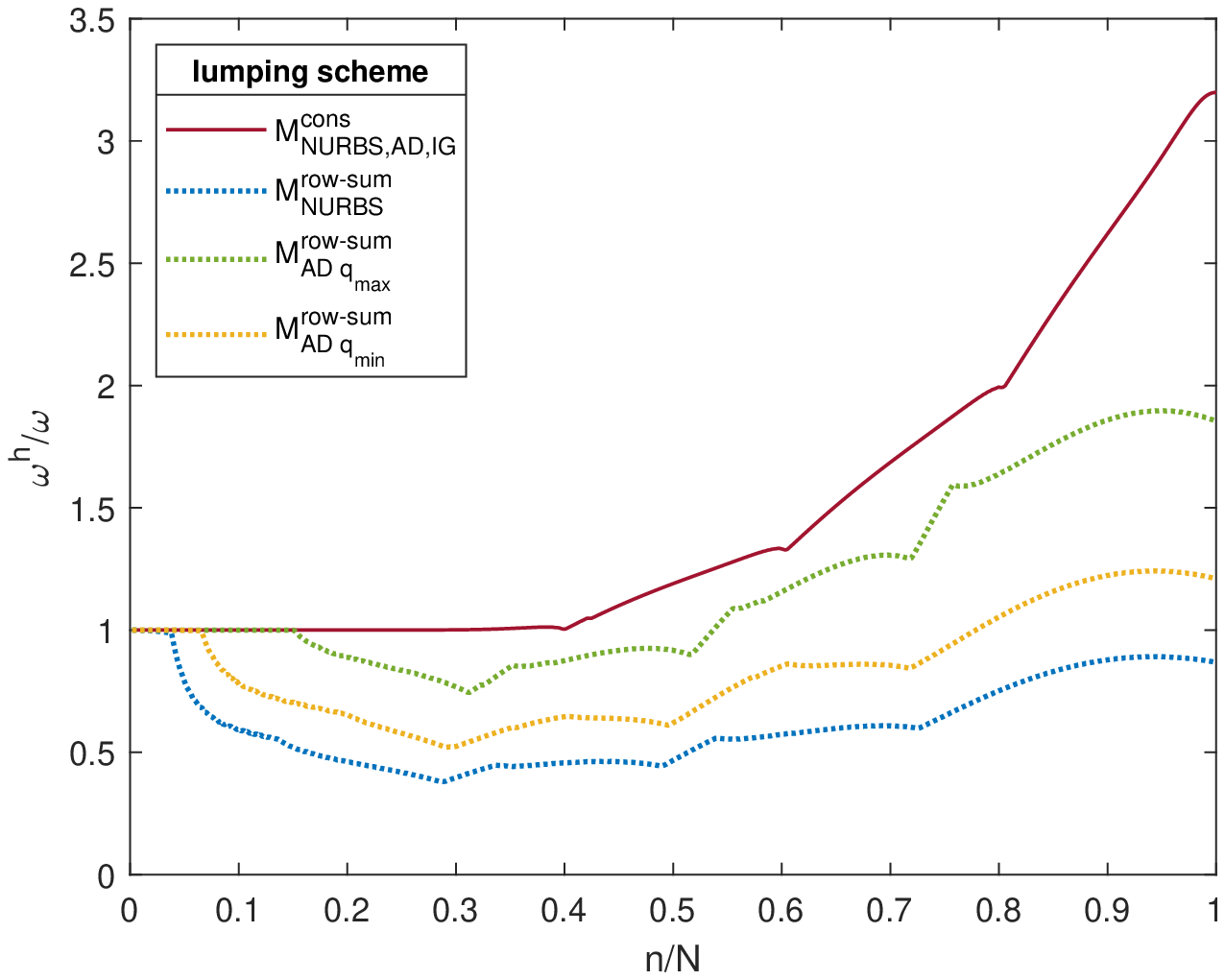}
	\caption{$p=5$}
	\label{fig:e2_spec_netB_p5}
\end{subfigure}
\caption{Numerical spectra of the consistent formulation in comparison with the lumped Bubnov-Galerkin formulation and the dual lumping approach AD with minimal and maximum reproduction degree for polynomial orders $p=2$ to 5. Computation was done on non-uniform mesh~B (Fig.~\ref{fig:mesh2}).}
\label{fig:e2_spec_netB}
\end{figure}

\begin{figure}[t]
\centering
\begin{subfigure}[ht]{0.40\textwidth}
	\centering
	\includegraphics[width=\textwidth]{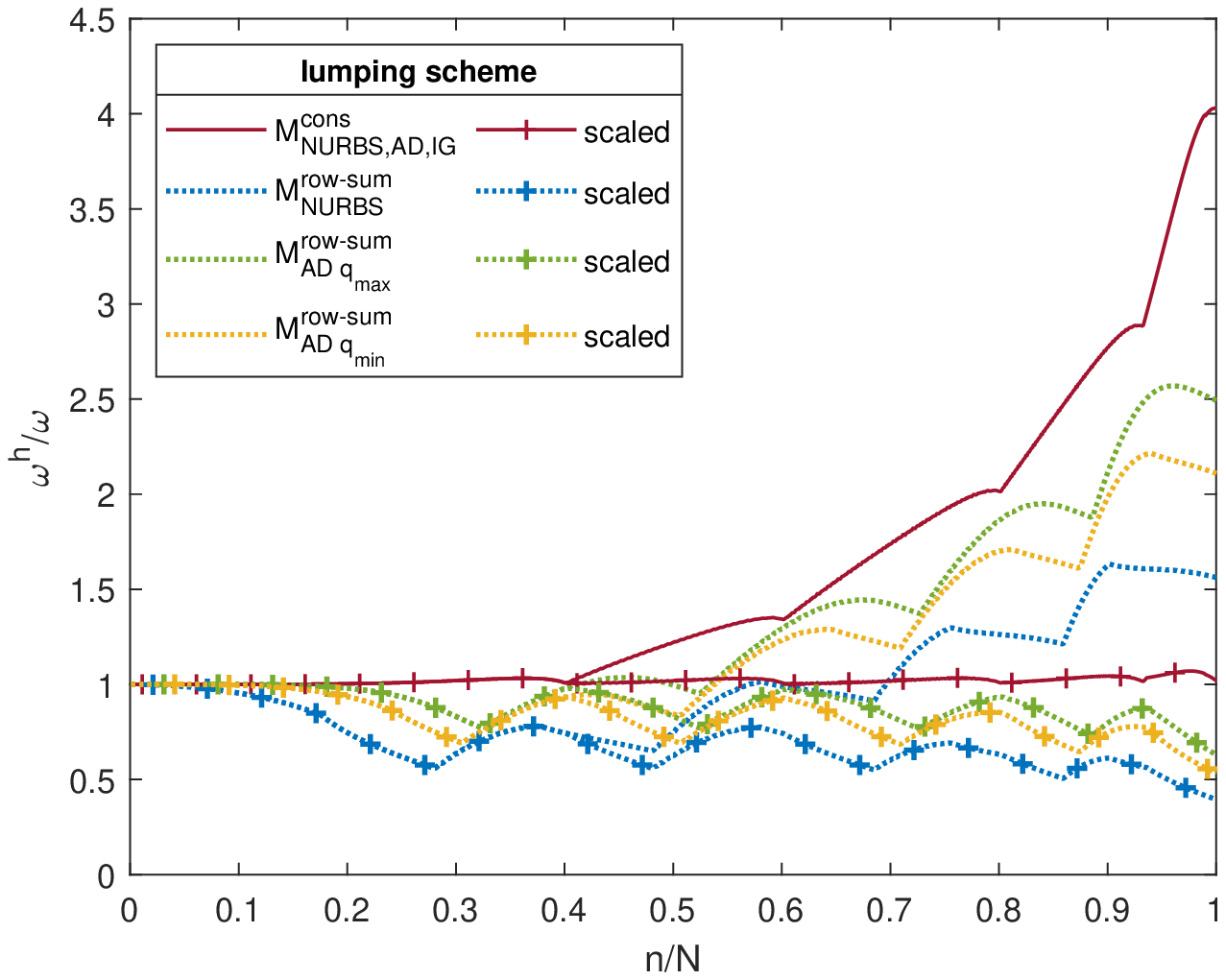}
	\caption{$p=2$}
\end{subfigure}
\quad
\begin{subfigure}[ht]{0.40\textwidth}
	\centering
	\includegraphics[width=\textwidth]{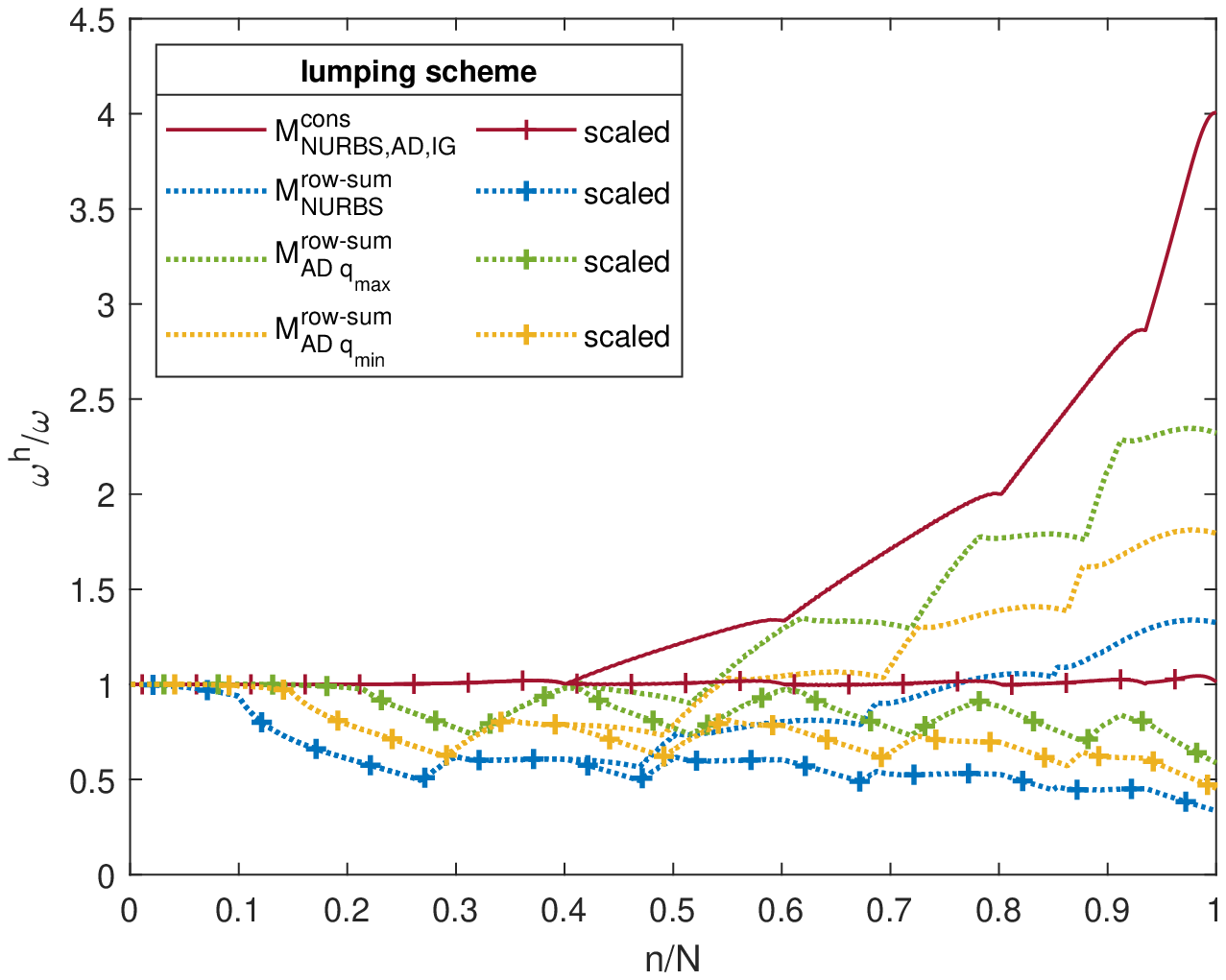}
	\caption{$p=3$}
\end{subfigure}\\
\begin{subfigure}[ht]{0.40\textwidth}
	\centering
	\includegraphics[width=\textwidth]{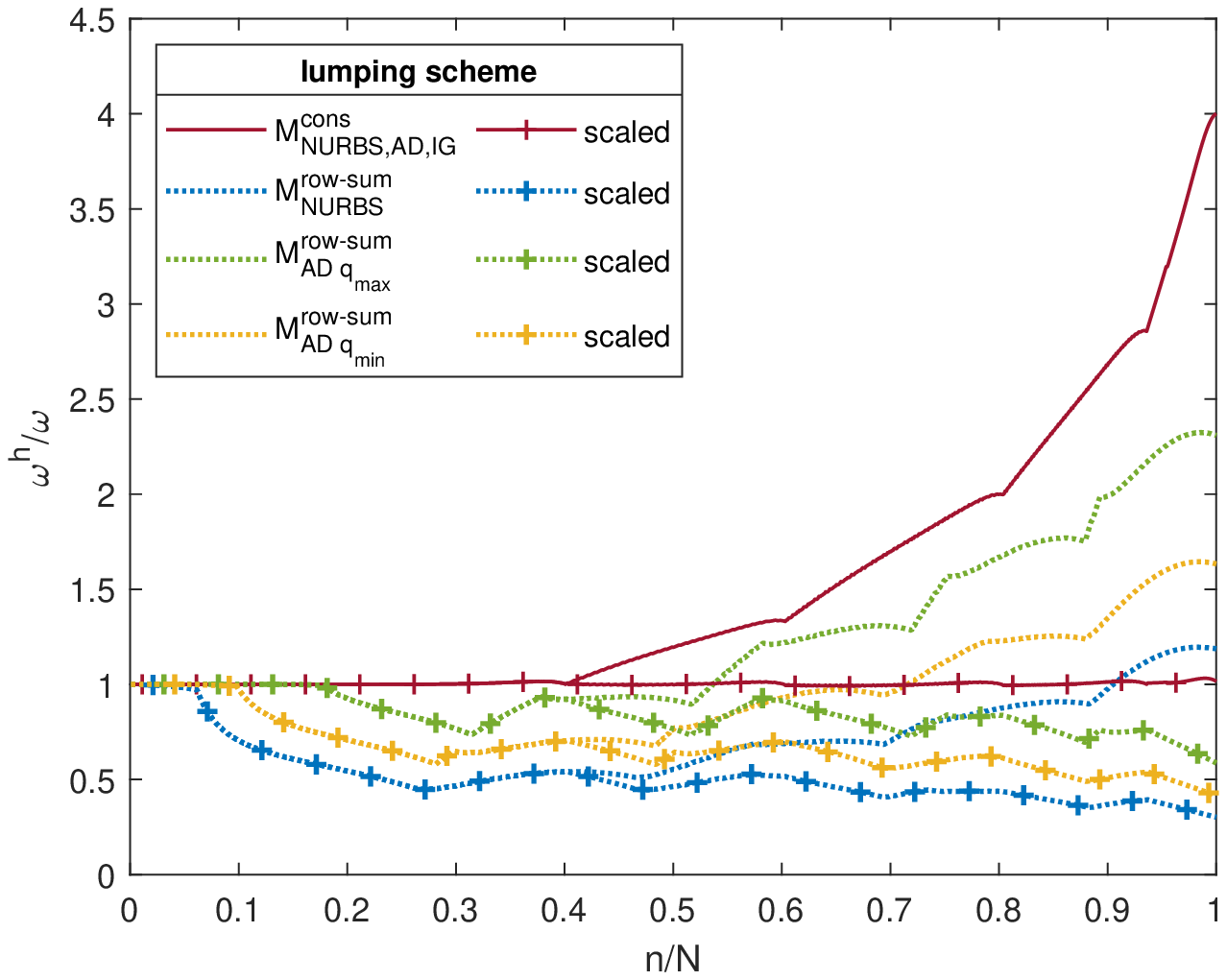}
	\caption{$p=4$}
\end{subfigure}
\quad
\begin{subfigure}[ht]{0.40\textwidth}
	\centering
	\includegraphics[width=\textwidth]{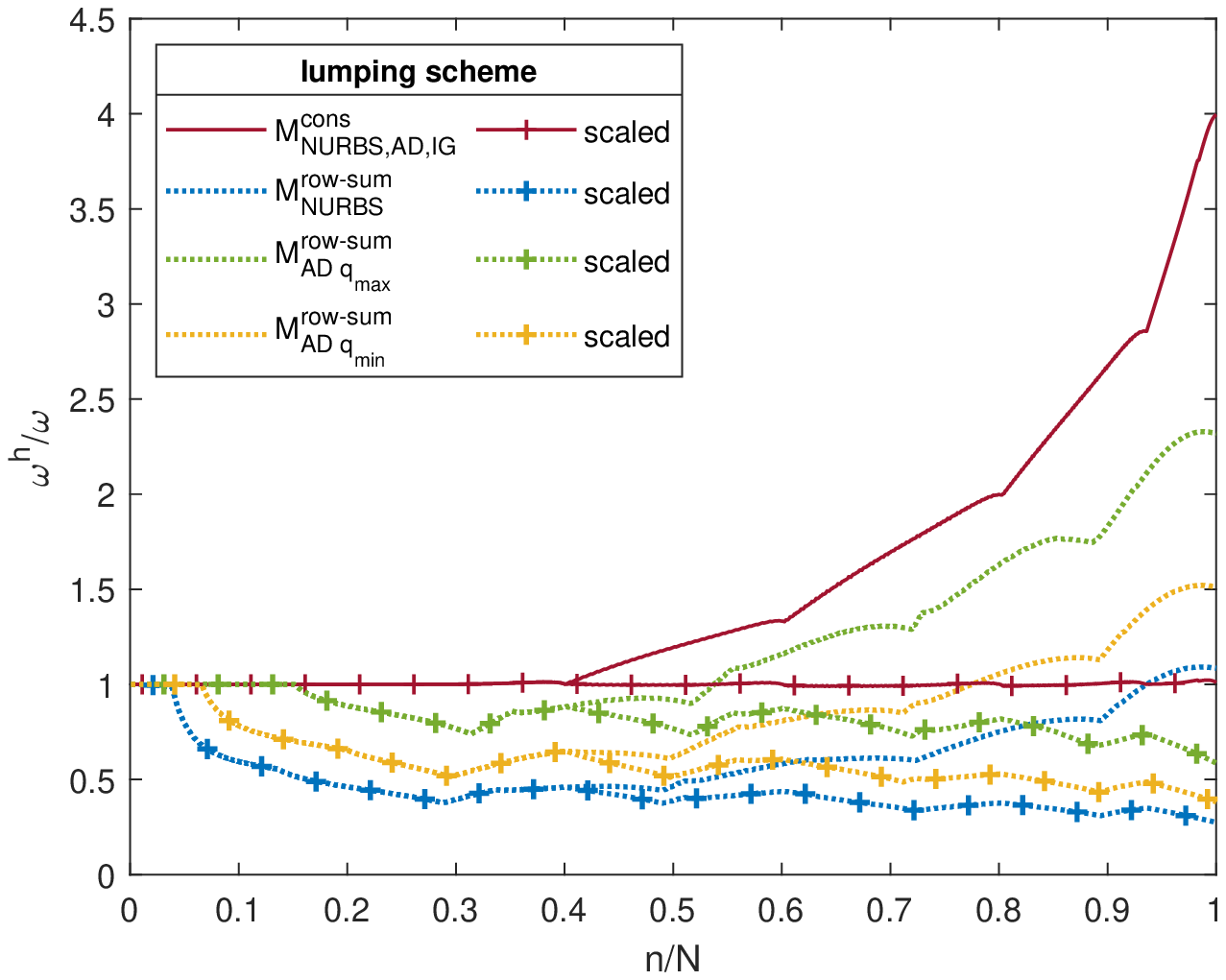}
	\caption{$p=5$}
\end{subfigure}
\caption{Numerical spectra of the consistent in comparison with the lumped Bubnov-Galerkin formulation and the dual lumping approach AD with minimal and maximum reproduction degree for polynomial orders $p=$ 2 to 5. Computation was done on non-uniform mesh~C (Fig.~\ref{fig:mesh3}). Each spectrum is opposed to a scaled spectrum considering the actual approximated eigenfrequency.}
\label{fig:e2_spec_netC_scaled}
\end{figure}

In addition to the spectral plots from Fig.~\ref{fig:e2_spec_dual}, the proposed formulation is also tested on non-uniform meshes B and C. The results, depicted in Figs.~\ref{fig:e2_spec_netB} and \ref{fig:e2_spec_netC_scaled}, differ from the findings reported so far. If mesh~B is used, the spectrum can be divided into 4 branches, where each branch shows similar characteristics as previously reported, i.e., moderately rising for about 80\% of the branch and then slightly going down at the latest frequencies. The slope of each branch is increased with every ensuing one. As also the formulations with additional row-sum lumping evolve this way, their former downwards trend is now balanced to capturing the exact eigenfrequencies very well. The plot might suggest that in case of non-uniform meshes the standard lumped approach performs better than the consistent and lumped dual formulations. But it is just a lucky coincidence for this particular numerical value. The corresponding mode shapes and the system response are still far from exact. The outcome is comparable to an overall applied mass scaling, where the higher eigenmodes evaluated as irrelevant for the application and thus, also the outliers are mitigated. That allows to choose larger time steps within explicit time integration procedures \cite{Hartmann.2014, Tkachuk.2013, Stoter.2022}, but accepts loosing the physical meaning of the new enforced lower modes. Here, the non-uniform spacing of knots results in a mismatch between the not subsequent list of approximated eigenfrequencies and the continuous set of the first $n$ analytical eigenfrequencies, which are the reference solution of the spectral plots. In Fig.~\ref{fig:e2_spec_netC_scaled}, the same behavior occures for 5 branches as mesh~C was defined as 5 segments of unique size, while mesh~B consists of 5 sections, but two of them of same size. Being aware of some intermediate eigenfrequencies, which are not considered in the system response at all, for mesh~C in Fig.~\ref{fig:e2_spec_netC_scaled} the spectrum was already piecewise scaled to the approximated set of eigenfrequencies of each branch. That scaling still leads to visible subdivisions as the characteristics of the non-uniform mesh cannot be denied, but it refutes the assumption of improving accuracy through irregular spacing of knots. 

The quality of the captured eigenfrequencies is actually determined by the placement of the control points as the degrees of freedom correspond to specific control points and not to the chosen allocation of knots. To achieve accuracy for a gapless set of captured eigenfrequencies, the knots can be placed arbitrarily, but the control points have to be spaced uniformly. This procedure was already proposed by \citet{Cottrell.2006} and within \cite{Hughes.2008} to avoid outliers, which are a result of non-uniformly distributed control points at the edges of a structure caused by open knot vectors.

\subsection{Approximation of mode shapes applying NURBS}
\label{sec:examples_NURBS}
\begin{figure}[t]
\centering
	\includegraphics[width=0.5\textwidth]{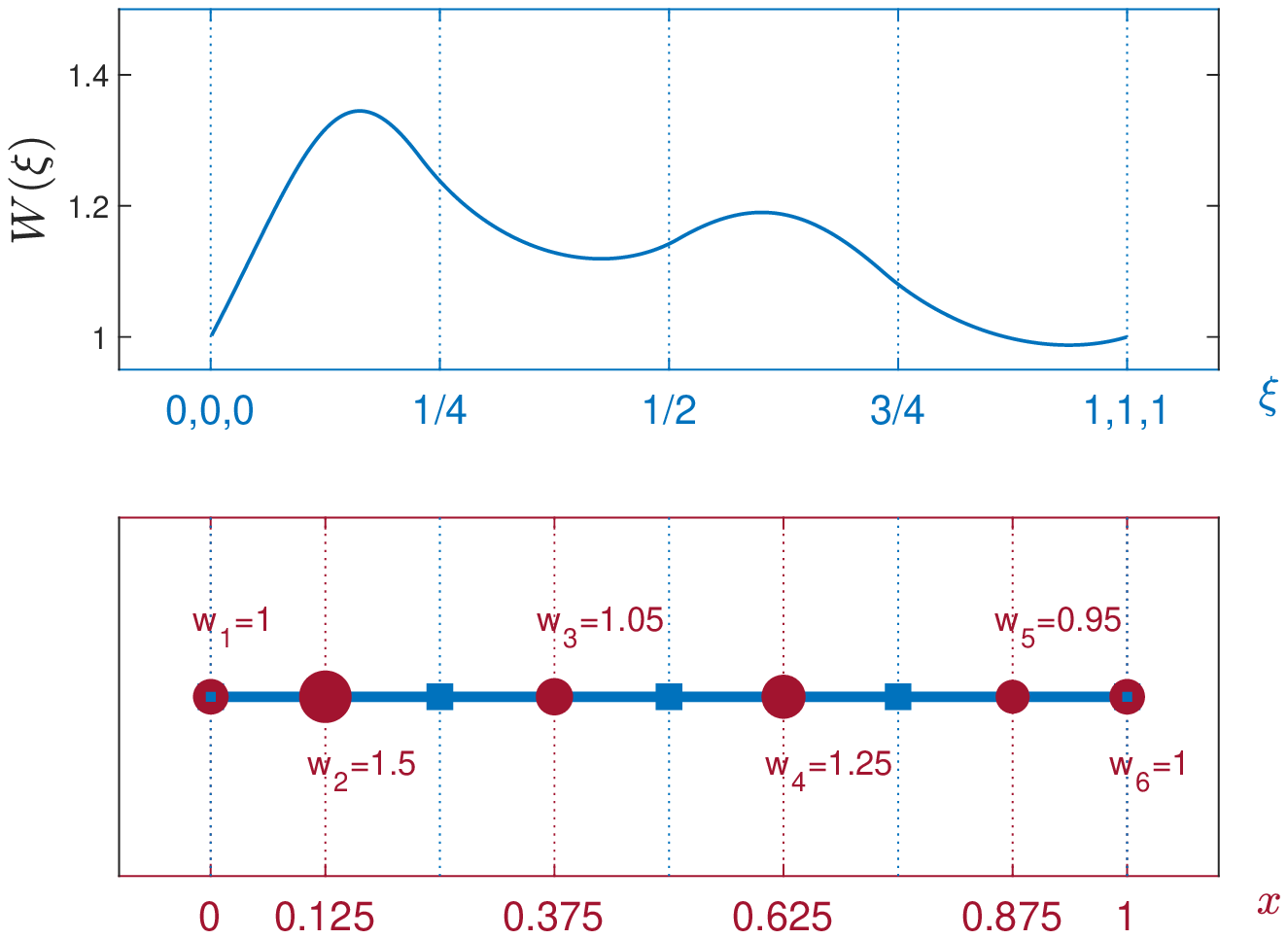}
\caption{NURBS curve of order $p=2$ and corresponding weighting function $W$. Squares~(\textcolor{blue}{$\blacksquare$}) denote knots of knot vector $\mathbf{\Xi}=\left[0,0,0,1/4,1/2,3/4,1,1,1\right]$. Size of the dots~(\textcolor{red}{$\medbullet$}) relates to the weight $w_i$ of each control point. Meshes for polynomial orders $p>2$ are constructed using $k$-refinement. Further $h$-refinement is applied uniformly within each initial knot span.}
\label{fig:sketch_NURBS}
\end{figure}

\begin{figure}[t]
\centering
\begin{subfigure}[ht]{0.40\textwidth}
	\centering
	\includegraphics[width=\textwidth]{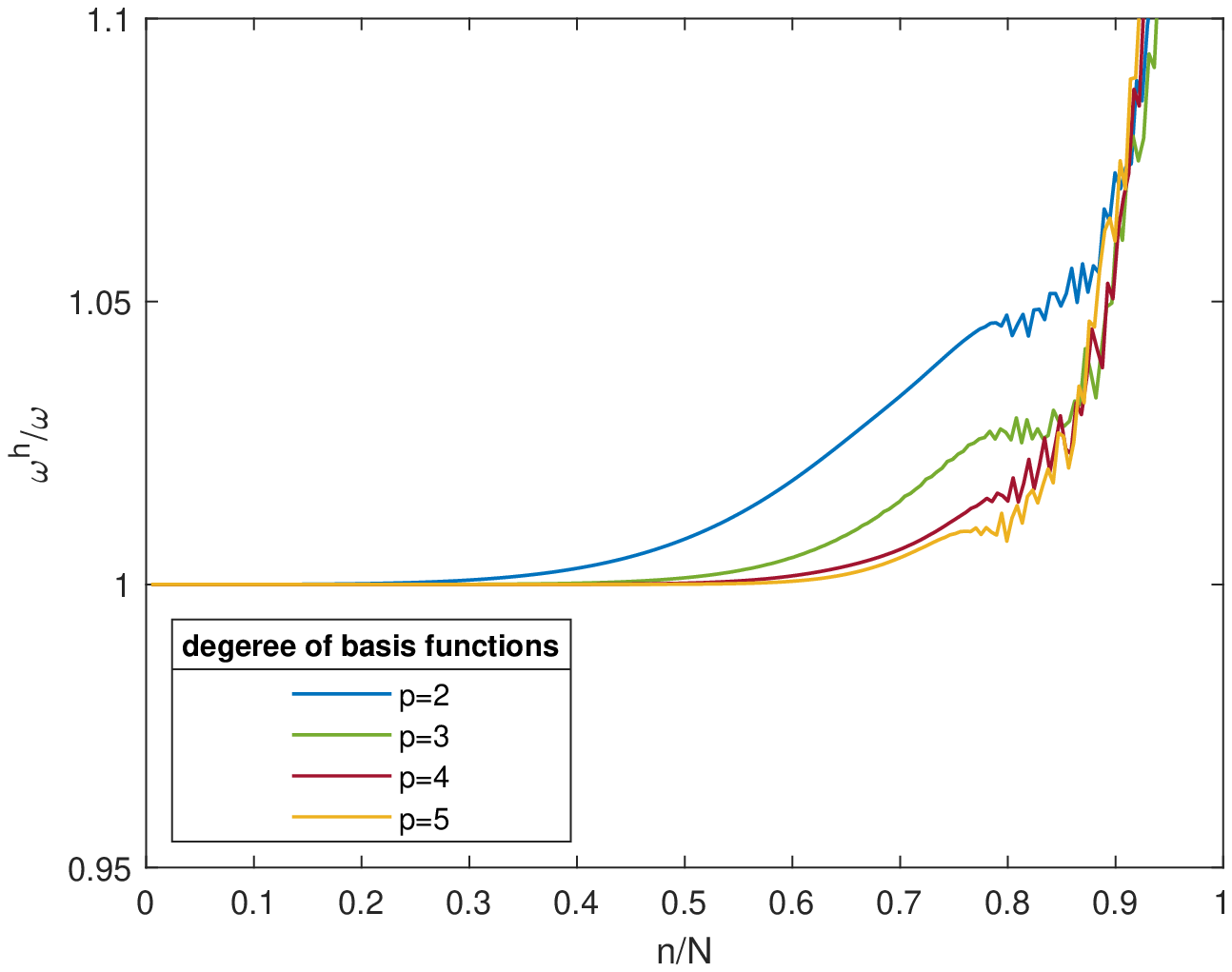}
	\caption{$M_{\mathrm{NURBS}}^{\mathrm{cons}}$}
\end{subfigure}
\quad
\begin{subfigure}[ht]{0.40\textwidth}
	\centering
	\includegraphics[width=\textwidth]{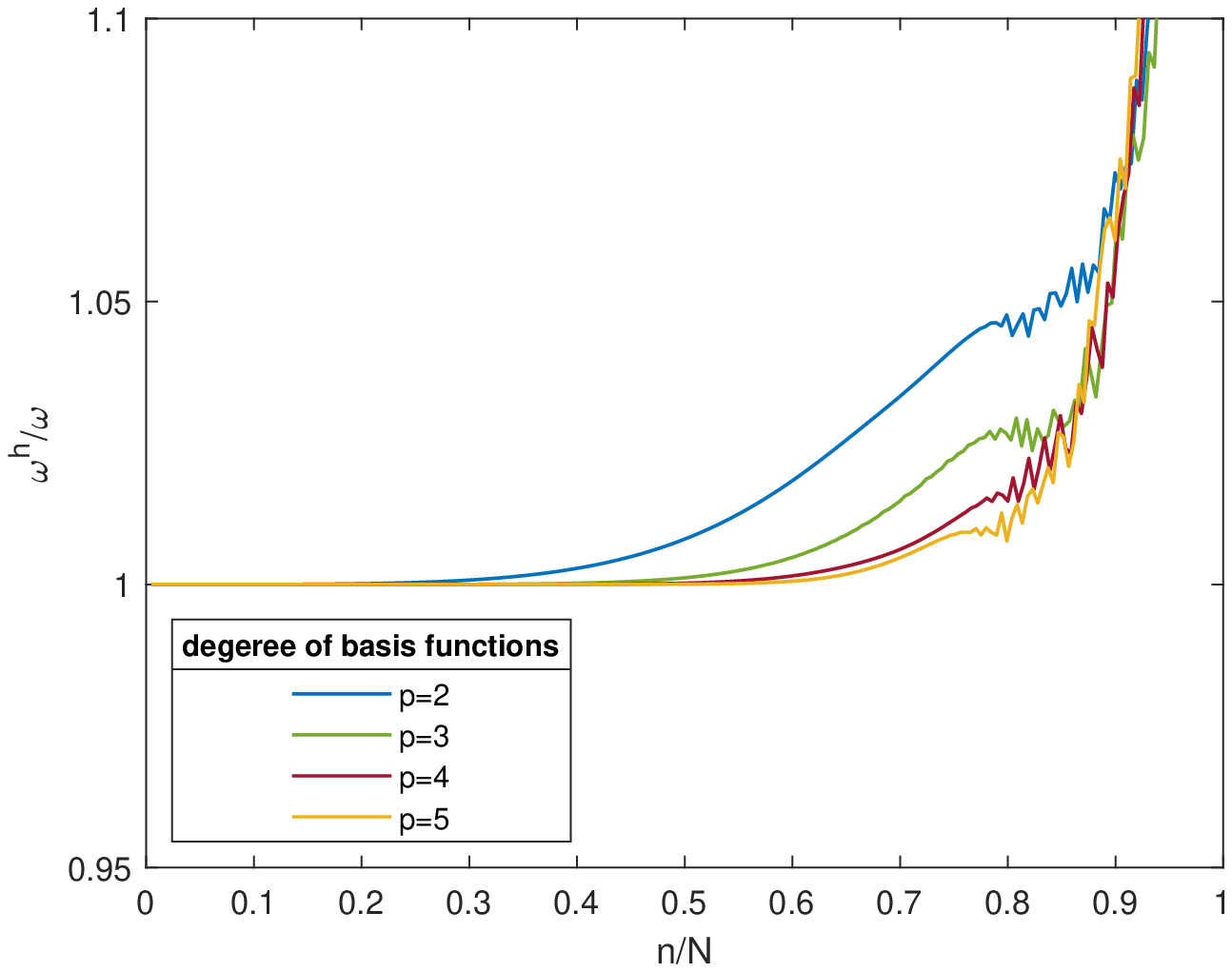}
	\caption{$M_{\mathrm{IG}}^{\mathrm{cons}}$}
\end{subfigure}\\
\begin{subfigure}[ht]{0.40\textwidth}
	\centering
	\includegraphics[width=\textwidth]{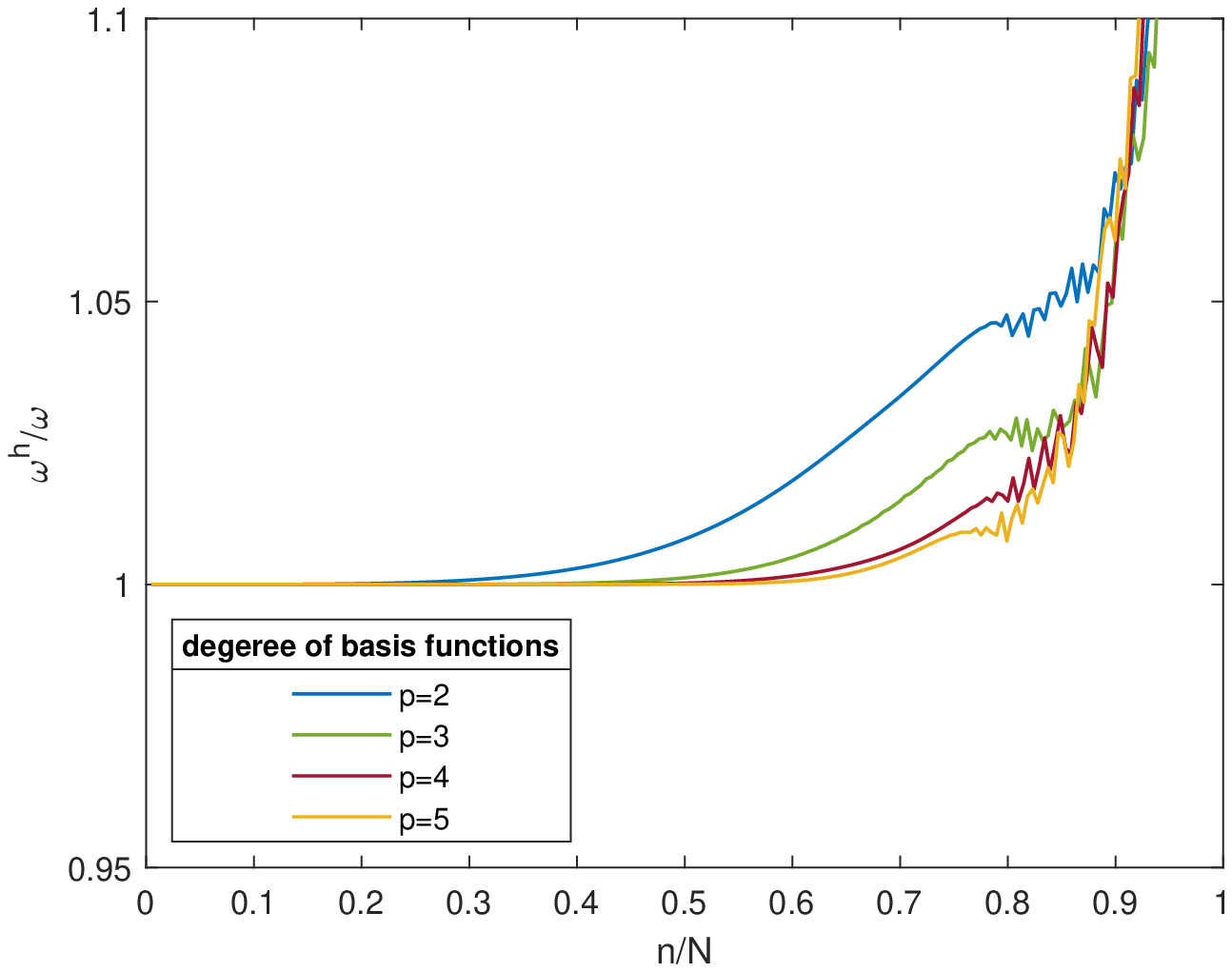}
	\caption{$M_{\mathrm{AD}~q_{\mathrm{min}}}^{\mathrm{cons}}$}
\end{subfigure}
\quad
\begin{subfigure}[ht]{0.40\textwidth}
	\centering
	\includegraphics[width=\textwidth]{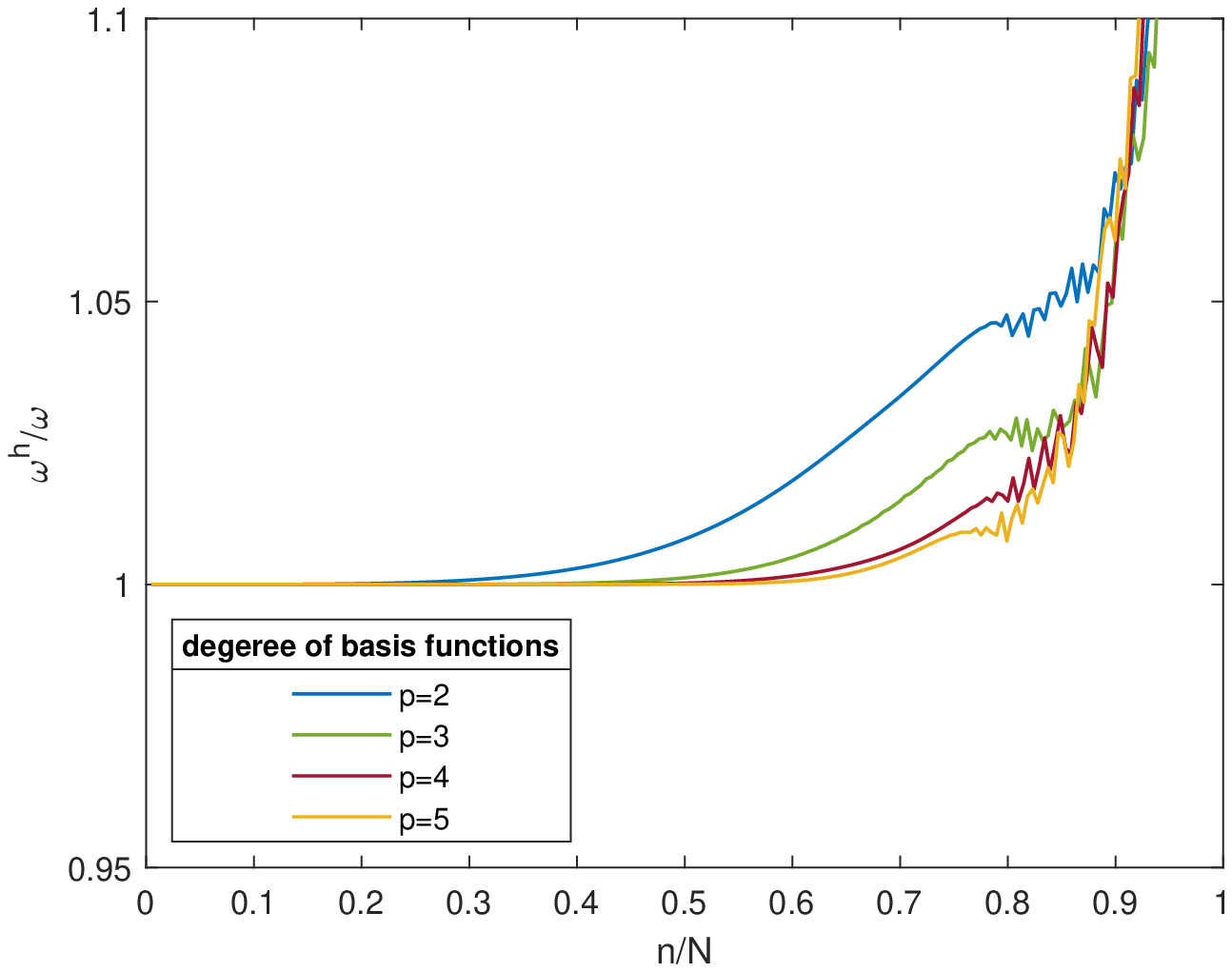}
	\caption{$M_{\mathrm{AD}~q_{\mathrm{max}}}^{\mathrm{cons}}$}
\end{subfigure}
\caption{Numerical spectra of consistent formulations. NURBS of polynomial orders $p$ are used as shape functions, type of test functions varies. Computation was done on mesh with varying weights of control points as shown in Fig.~\ref{fig:sketch_NURBS}.}
\label{fig:e2_spec_NURBS_cons}
\end{figure}

\begin{figure}[t]
\centering
\begin{subfigure}[ht]{0.40\textwidth}
	\centering
	\includegraphics[width=\textwidth]{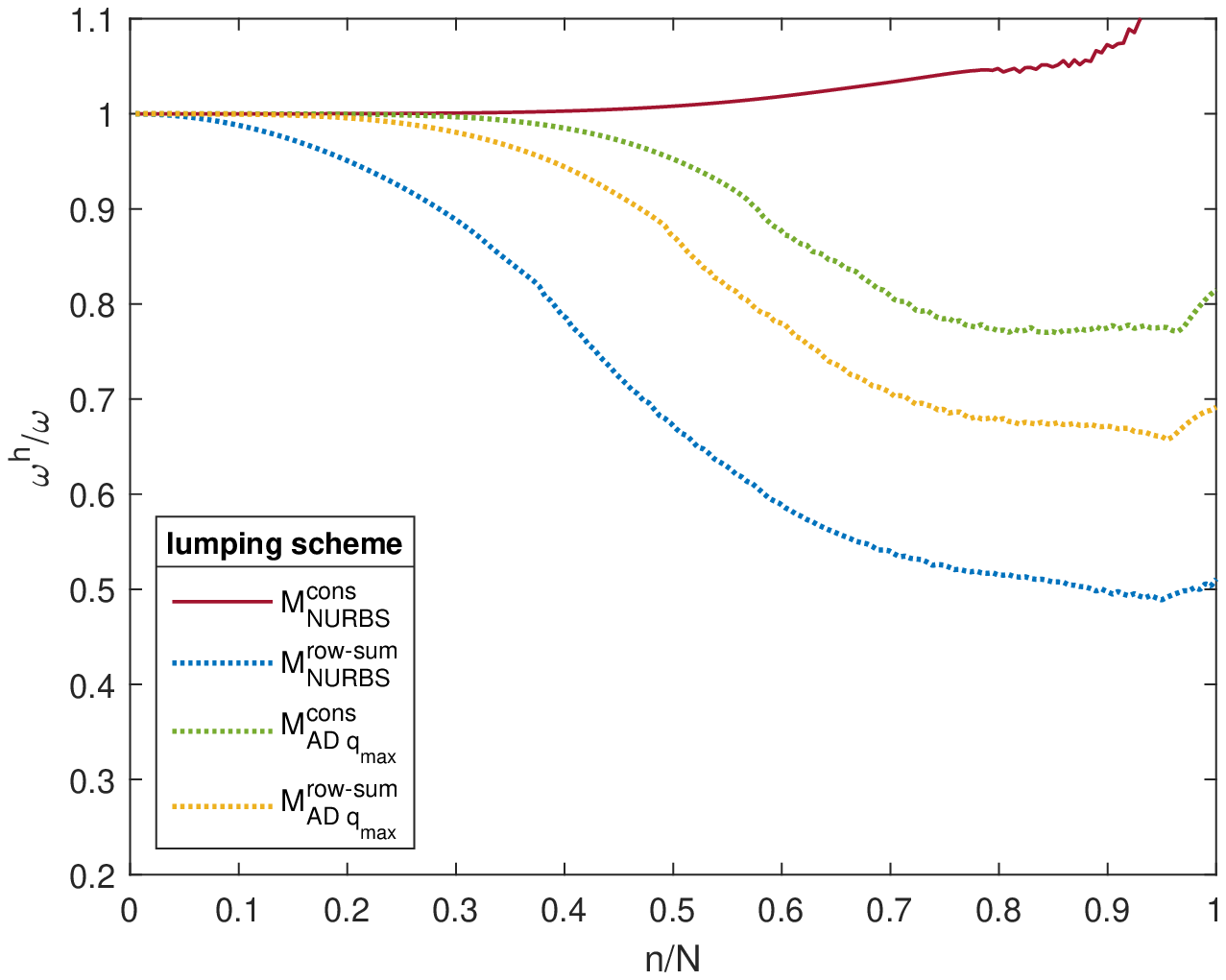}
	\caption{$p=2$}
\end{subfigure}
\quad
\begin{subfigure}[ht]{0.40\textwidth}
	\centering
	\includegraphics[width=\textwidth]{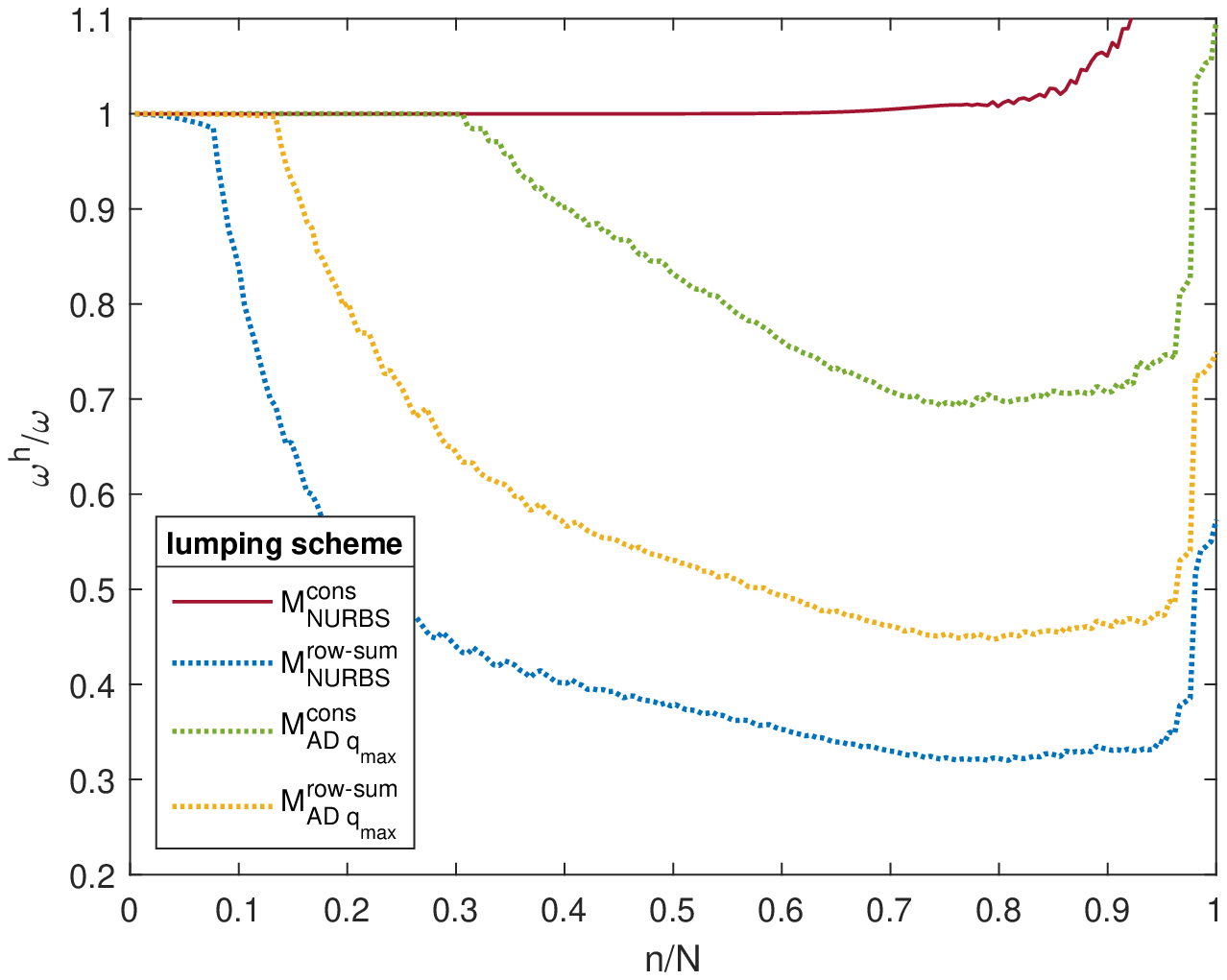}
	\caption{$p=5$}
\end{subfigure}
\caption{Numerical spectra of the consistent formulation in comparison with the lumped Bubnov-Galerkin formulation and the dual lumping approach AD with minimal and maximum reproduction degree for polynomial orders $p=2$ and 5. Computation was done on mesh with varying weights of control points as shown in Fig.~\ref{fig:sketch_NURBS}.}
\label{fig:e2_spec_NURBS_diag}
\end{figure}

We introduce another mesh in Fig.~\ref{fig:sketch_NURBS}, still related to a one-dimensional NURBS curve, but with varying weights of the control points. To ensure a smooth $C^1$ weighting function, we choose an initial polynomial order of $p=2$. Dividing the curve uniformly into 4 segments, we than apply the weights $w_i = \left\lbrace1,1.5,1.05,1.25,0.95,1\right\rbrace$ to the 6 control points shown in Fig.~\ref{fig:sketch_NURBS}. The spectra received through further $k$- and $h$-refinement until a total amount of 200 knot spans are depicted in Fig.~\ref{fig:e2_spec_NURBS_cons}. The non-equal weights of the initial control points enforce a series of non-uniformly spaced control points after refinement. This results into the appearance of even more outlier frequencies, which can be exemplified by approximately the last 20\% of the data. But this cannot be claimed as a drawback of our proposed method. As mentioned before, some techniques have already been developed to remove these spurious modes and furthermore, these difficulties also appear using common IGA formulations. Figure~\ref{fig:e2_spec_NURBS_cons} demonstrates that even for the general case of NURBS with non-constant weighting, dual test functions can be applied successfully without loosing general properties. Comparing the row-sum lumped dual approach in Fig.~\ref{fig:e2_spec_NURBS_diag}, again with the consistent and the simple row-sum lumped formulation, none of the former findings is changed. Applying row-sum lumping to the formulation with AD test functions provides better results than pure row-sum lumping but still suffers from less accuracy of higher modes.

\subsection{Explicit dynamic analysis of a fixed-fixed 1D truss}
\label{sec:examples_dyn2}
As we do not consider a removal technique for outlier frequencies until now, for further examples we will stick with the simplest case of the uniform mesh~A using B-Splines. In this section, we want to assess the accuracy and the convergence rates of the different approaches in an explicit dynamic analysis. The impact of the lumping schemes on results of dynamic analyses is especially noticeable for the undamped or mass-proportional damped case within explicit time integration schemes, where an inversion of the mass matrix, see Eqs.~(\ref{eq:SOE}) and (\ref{eq:K_eff}), is required. Thus, the following numerical examples are calculated by using the Central Differences Method (CDM) \cite{Hughes.1987}. Starting from the equation of motion for the undamped case evaluated at time $t$
\begin{equation}
\mathbf{M}\ddot{\mathbf{u}}_t+\mathbf{K}\mathbf{u}_t=\mathbf{F}_t~,
\label{eq:motion}
\end{equation} 
the underlying system of equations to solve for the next time step `$t+\Delta t$' is
\begin{equation}
\mathbf{K}^*_t\mathbf{U}_{t+\Delta t}=\mathbf{F}^*_t
\label{eq:SOE}
\end{equation}
with the effective stiffness matrix
\begin{equation}
\mathbf{K}^*_t = \frac{1}{\Delta t^2}\mathbf{M}_t
\label{eq:K_eff}
\end{equation}
and the effective force vector
\begin{equation}
\mathbf{F}^*_t = \mathbf{F}_t-\left(\mathbf{K}_t-\frac{2}{\Delta t^2}\mathbf{M}_t\right)\mathbf{U}_t-\frac{1}{\Delta t^2}\mathbf{M}_t\mathbf{U}_{t-\Delta t}~.
\label{eq:F_eff}
\end{equation}
To start the computation, $\mathbf{U}_{-\Delta t}$ has to be calculated, based on the values of the displacements $\mathbf{U}_0$, velocities $\mathbf{\dot{U}}_0$ and accelerations $\mathbf{\ddot{U}}_0$. While $\mathbf{U}_0$ and $\mathbf{\dot{U}}_0$ are prescribed initial conditions within our numerical examples, $\mathbf{\ddot{U}}_0$ is computed by solving Eq.~(\ref{eq:motion}) for $t=0$. Therefore, the displacements at $t=-\Delta t$ are determined by
\begin{equation}
\mathbf{U}_{-\Delta t} = \mathbf{U}_0-\Delta t\mathbf{\dot{U}}_0+\frac{\Delta t^2}{2}\mathbf{\ddot{U}}_0~.
\end{equation}
The time step is chosen as $\Delta t<\Delta t_{\mathrm{crit}}$ with the critical time step computed by $\Delta t_{\mathrm{crit}}=\frac{2}{\omega_{\mathrm{max}}}$, where $\omega_{\mathrm{max}}$ corresponds to the largest eigenfrequency approximated by the system.

To study the convergence, an example with a known analytic solution is chosen. An initial sinusoidal velocity field 
\begin{equation}
\frac{\mathrm{d}}{\mathrm{d}t}u(x,t=0)=2\pi\sin\left(\frac{2\pi}{L}x\right)
\end{equation}
is applied to a fixed-fixed truss in Fig.~\ref{fig:sysplot_dyn}, while the initial displacement $u_0=u(x,t=0)=0$ equals the undeformed geometry. A unit 1D truss ($\{EA,\mu,L\}=1$) yields the analytical solution
\begin{equation}
u(x,t)=\sin\left(\frac{2\pi}{L}x\right)\sin\left(2\pi t\right)~,
\end{equation}
which corresponds to a harmonic oscillation of the structure with $f=1\,$Hz. The computation is done for two periods until $t=2\,$s. 

\begin{figure}[t]
\centering
\begin{subfigure}[ht]{0.47\textwidth}
	\centering
	\includegraphics[width=\textwidth]{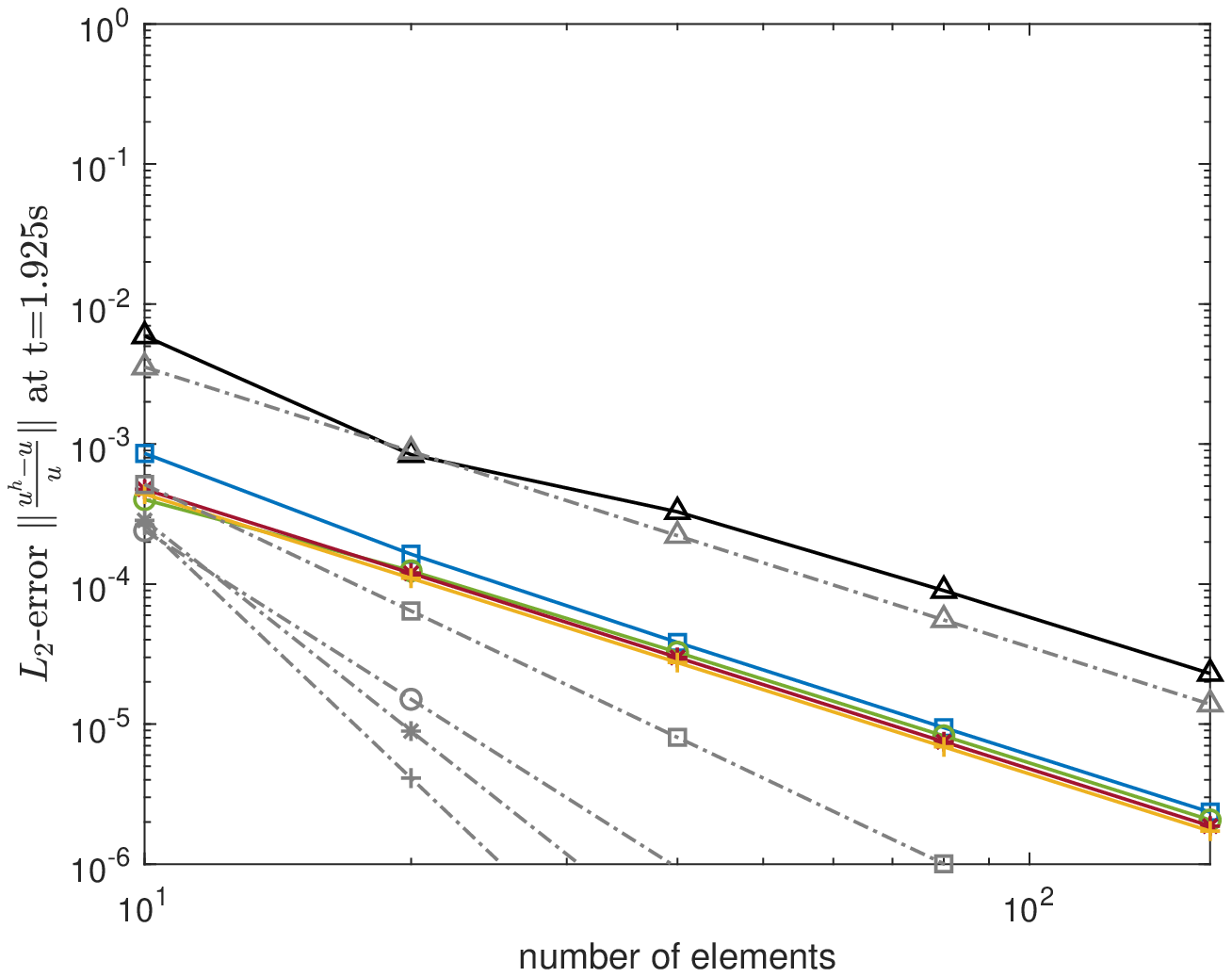}
	\caption{$M_{\mathrm{NURBS,AD,IG}}^{\mathrm{cons}}$}
	\label{fig:e4_L2_NURBSc}
\end{subfigure}
\hfill
\begin{subfigure}[ht]{0.47\textwidth}
	\centering
	\includegraphics[width=\textwidth]{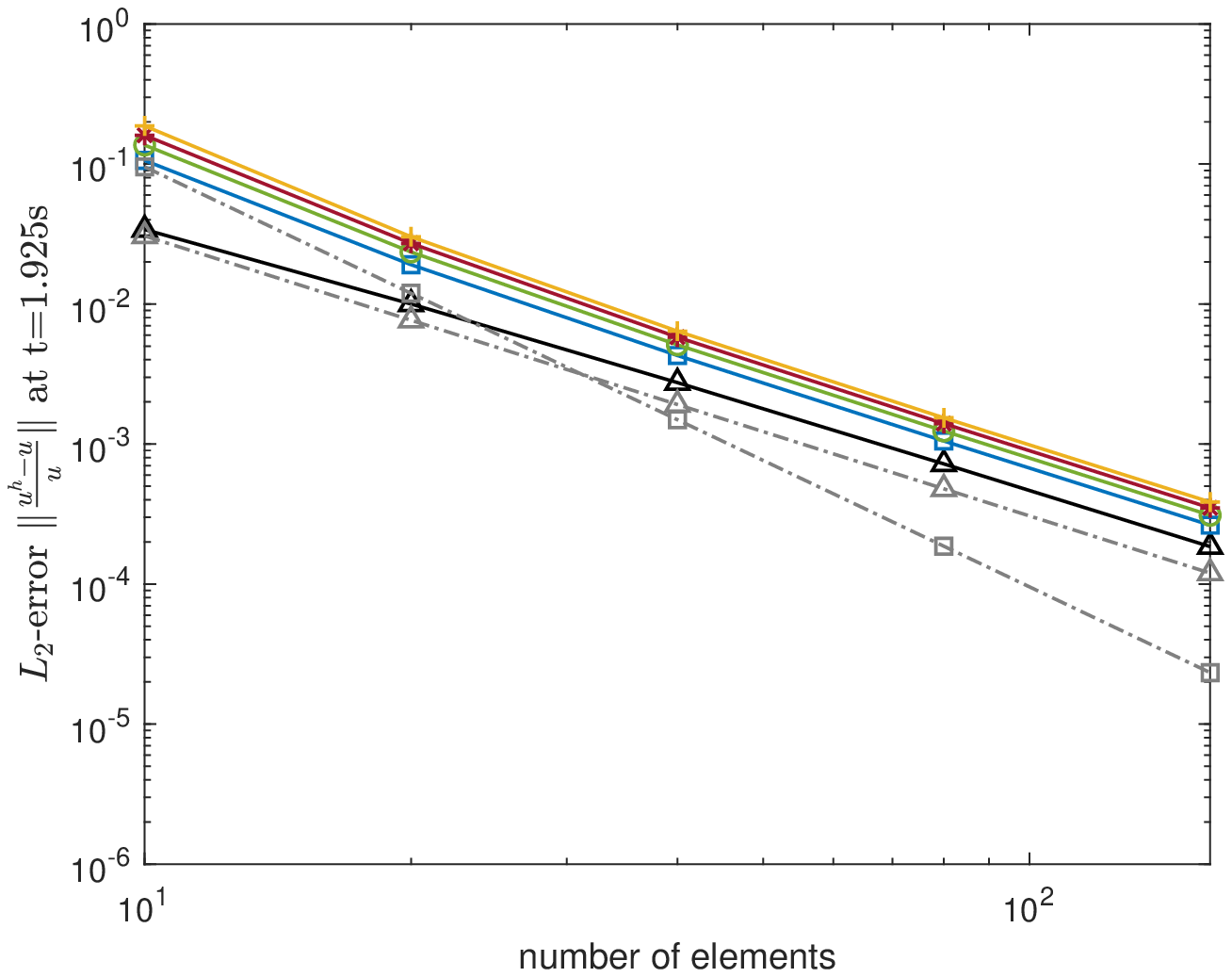}
	\caption{$M_{\mathrm{NURBS}}^{\mathrm{row-sum}}$}
	\label{fig:e4_L2_NURBSd}
\end{subfigure}\\
\begin{subfigure}[ht]{0.47\textwidth}
	\centering
	\includegraphics[width=\textwidth]{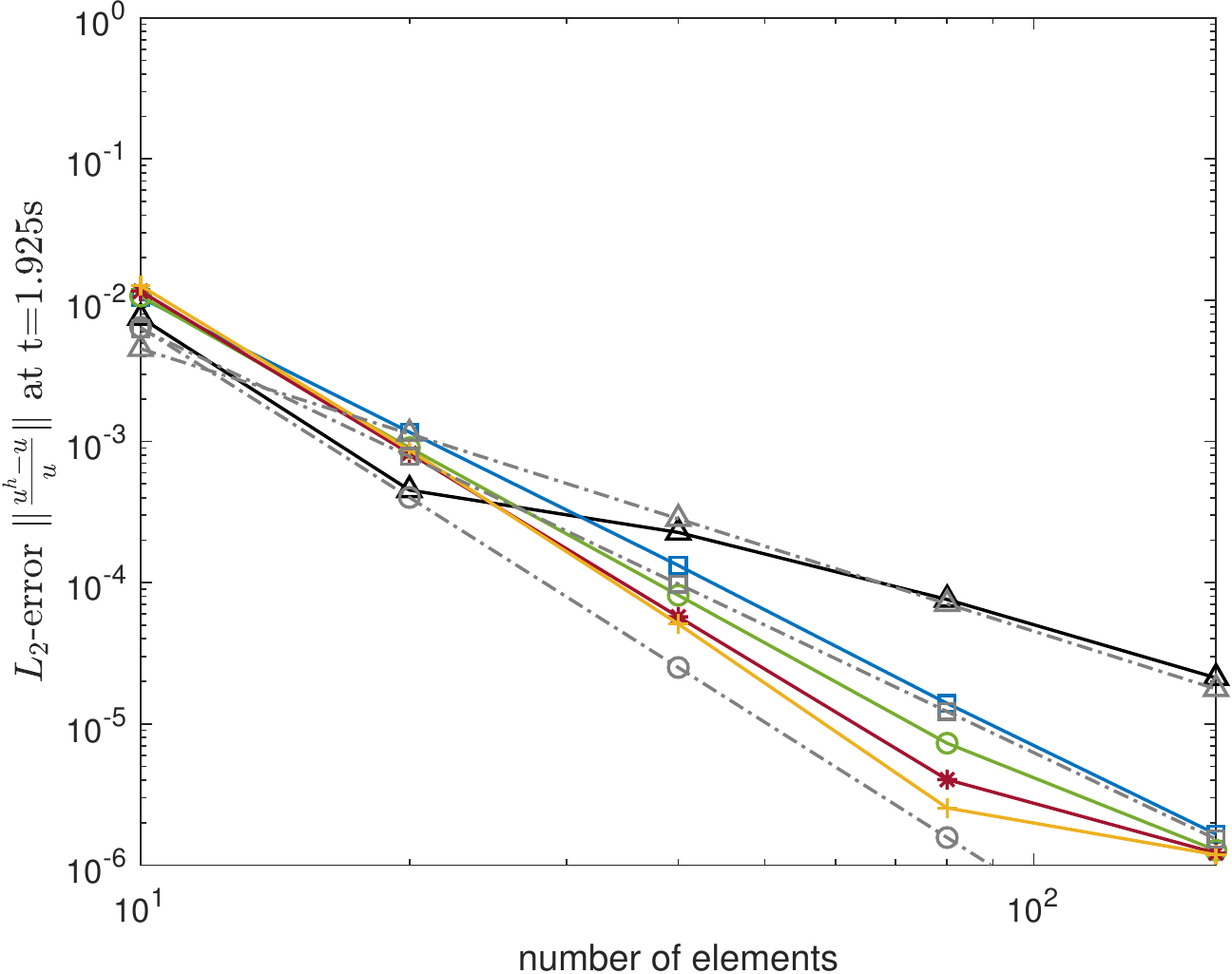}
	\caption{$M_{\mathrm{AD}~q_{\mathrm{min}}}^{\mathrm{row-sum}}$}
	\label{fig:e4_L2_ADmin}
\end{subfigure}
\hfill
\begin{subfigure}[ht]{0.47\textwidth}
	\centering
	\includegraphics[width=\textwidth]{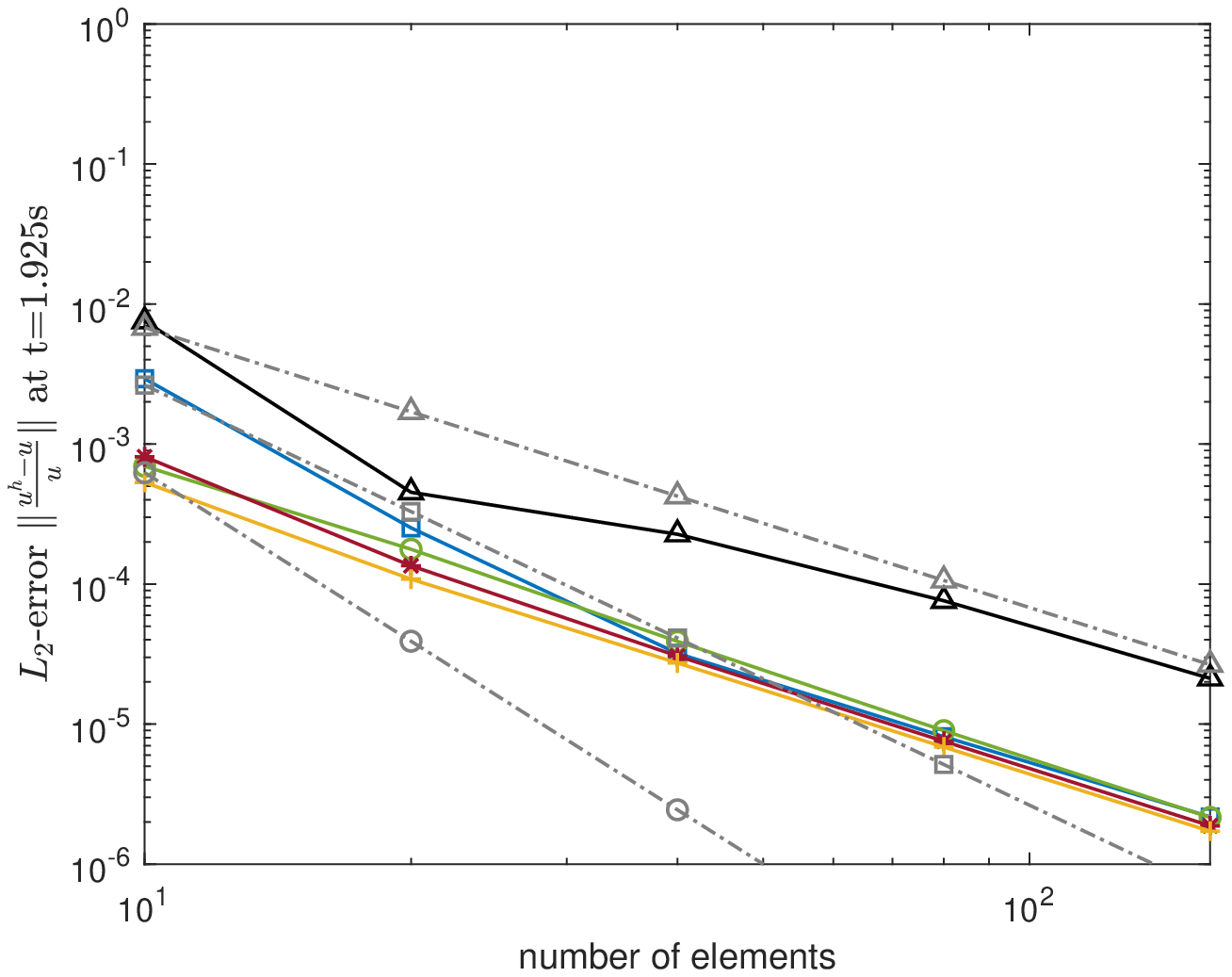}
	\caption{$M_{\mathrm{AD}~q_{\mathrm{max}}}^{\mathrm{row-sum}}$}
	\label{fig:e4_L2_ADmax}
\end{subfigure}\\
\begin{subfigure}[ht]{1.00\textwidth}
	\centering
	\includegraphics[width=\textwidth]{legend_example2_1}
\end{subfigure}
\caption{Fixed-fixed 1D truss: $L_2$-error norm of displacement $u(t=1.925\,\text{s})$, NURBS of order $p$ are used as shape functions, type of test functions varies. Computation was done on uniform mesh~A (Fig.~\ref{fig:mesh1}). Time integration was applied using CDM with $\Delta t=\frac{h}{10}$.}
\label{fig:e4_error}
\end{figure}
\begin{figure}[t]
\centering
\begin{subfigure}[ht]{0.47\textwidth}
	\centering
	\includegraphics[width=\textwidth]{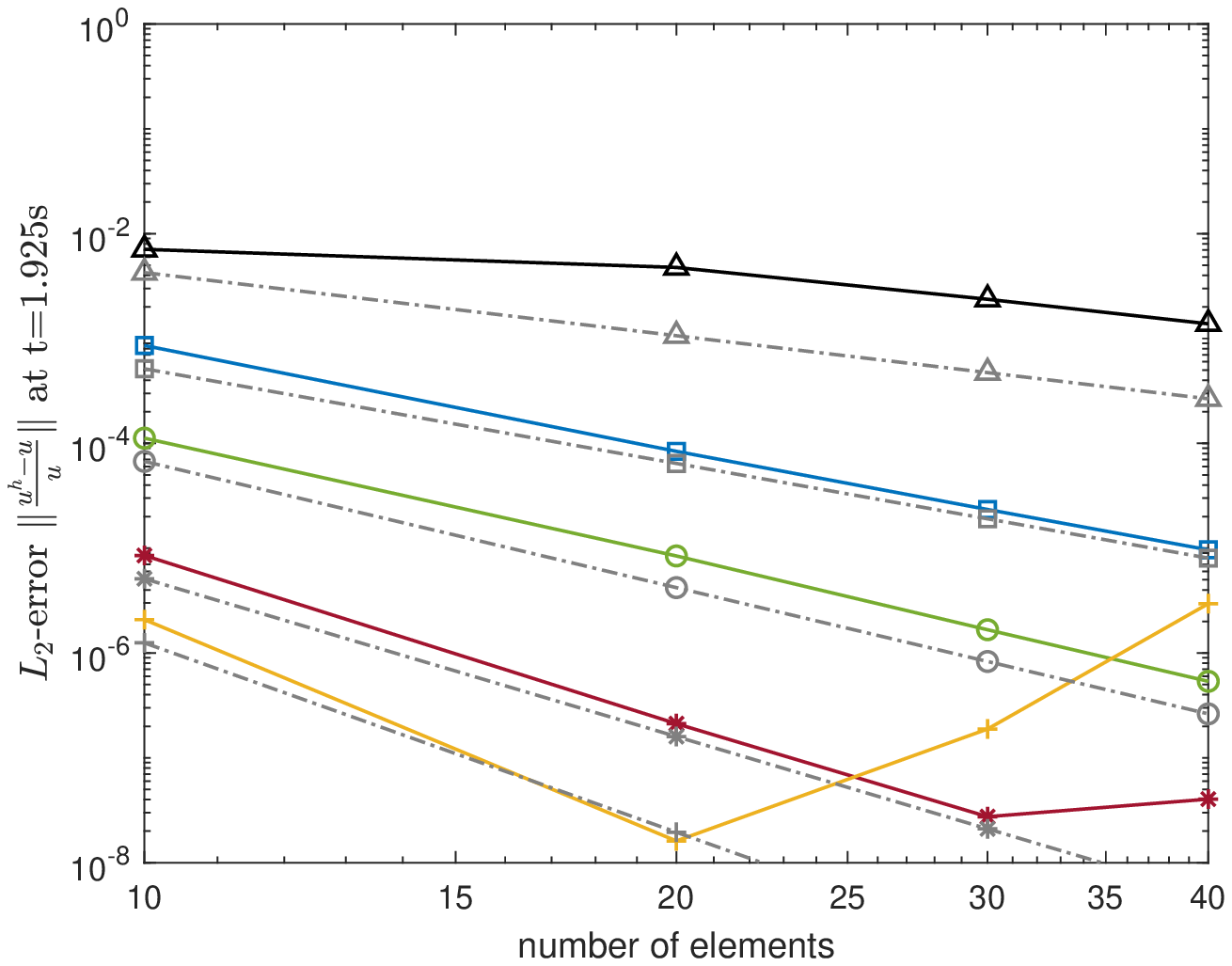}
	\caption{$M_{\mathrm{NURBS,AD,IG}}^{\mathrm{cons}}$}
	\label{fig:e4_L2_NURBSc2}
\end{subfigure}
\hfill
\begin{subfigure}[ht]{0.47\textwidth}
	\centering
	\includegraphics[width=\textwidth]{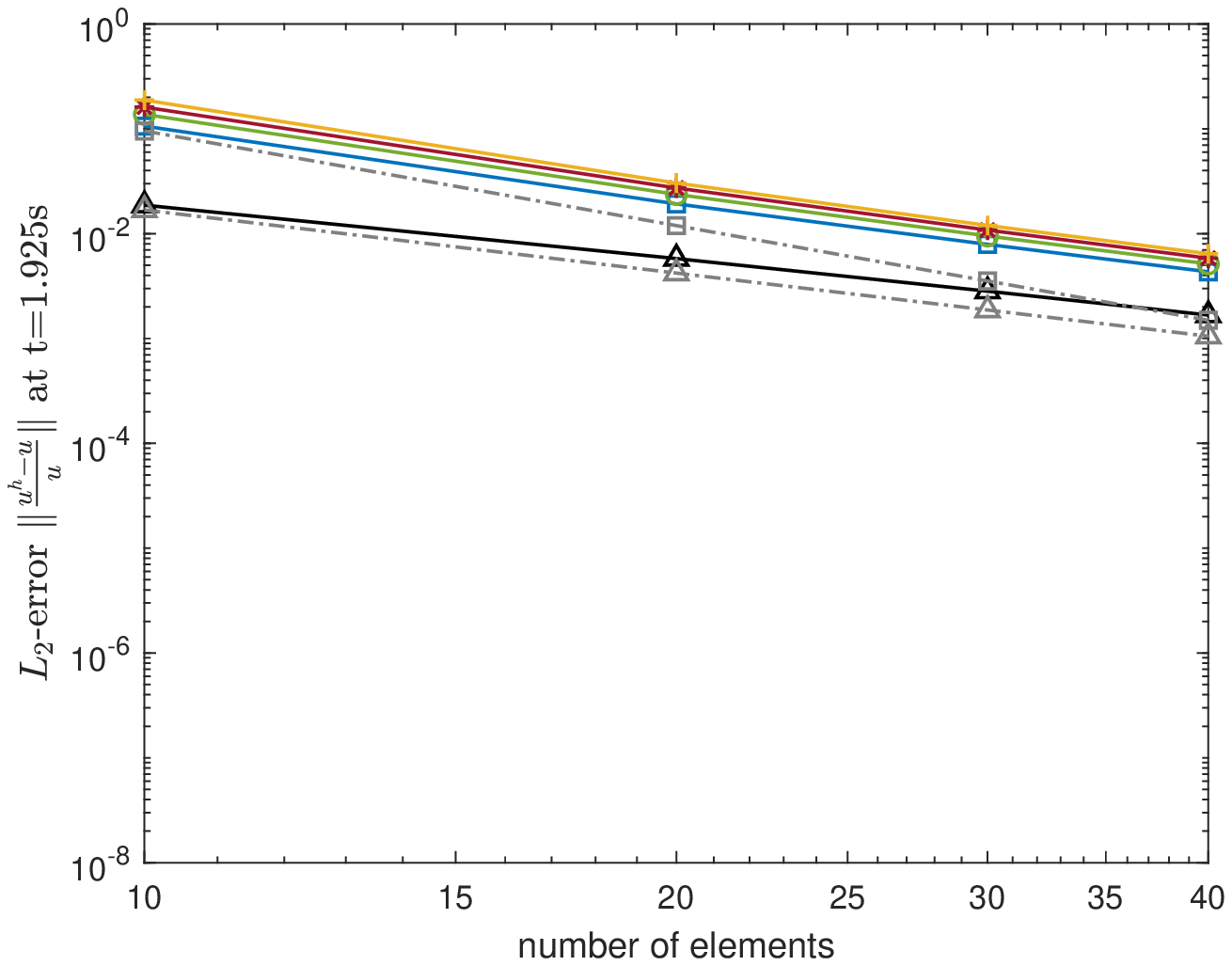}
	\caption{$M_{\mathrm{NURBS}}^{\mathrm{row-sum}}$}
\end{subfigure}\\
\begin{subfigure}[ht]{0.47\textwidth}
	\centering
	\includegraphics[width=\textwidth]{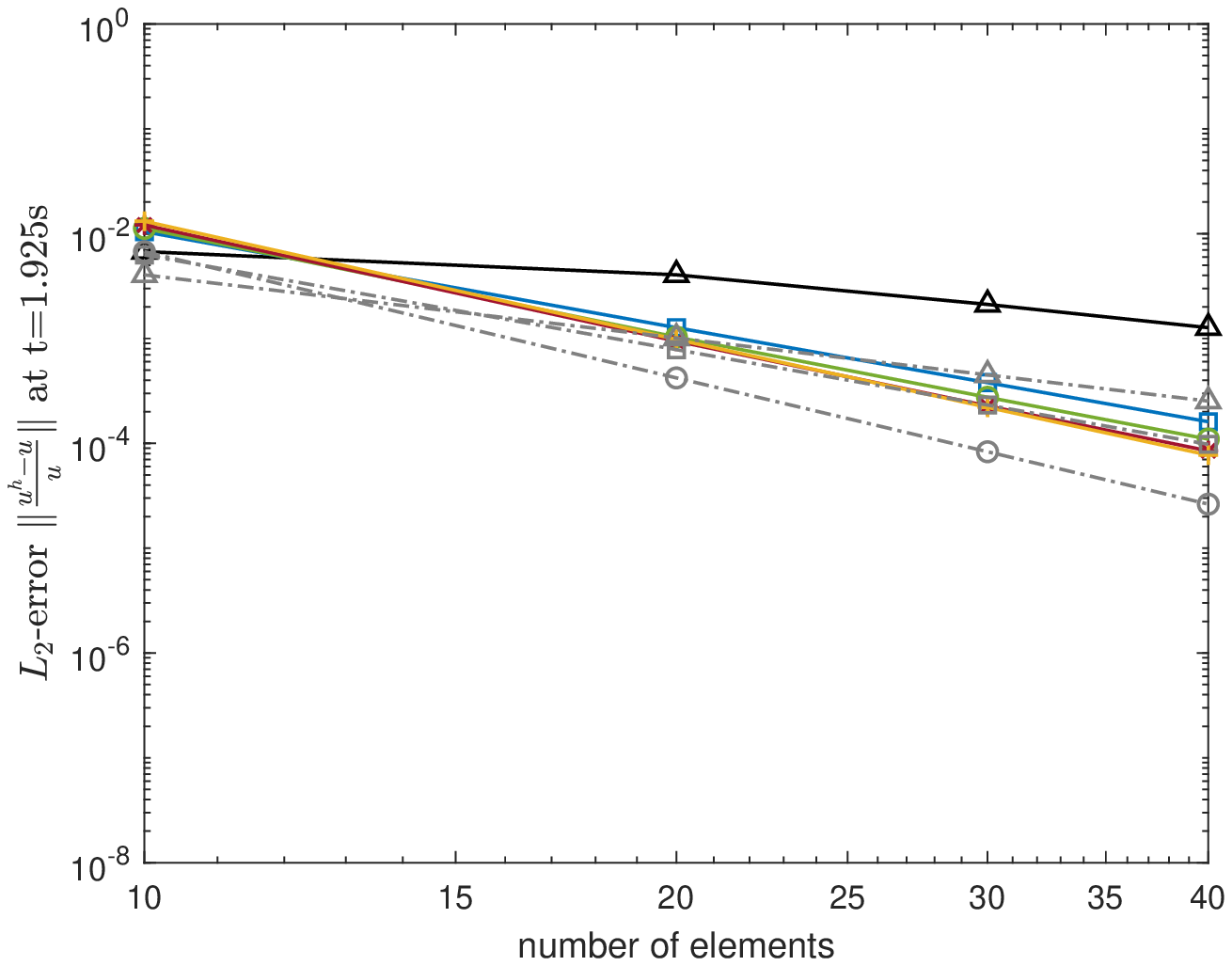}
	\caption{$M_{\mathrm{AD}~q_{\mathrm{min}}}^{\mathrm{row-sum}}$}
\end{subfigure}
\hfill
\begin{subfigure}[ht]{0.47\textwidth}
	\centering
	\includegraphics[width=\textwidth]{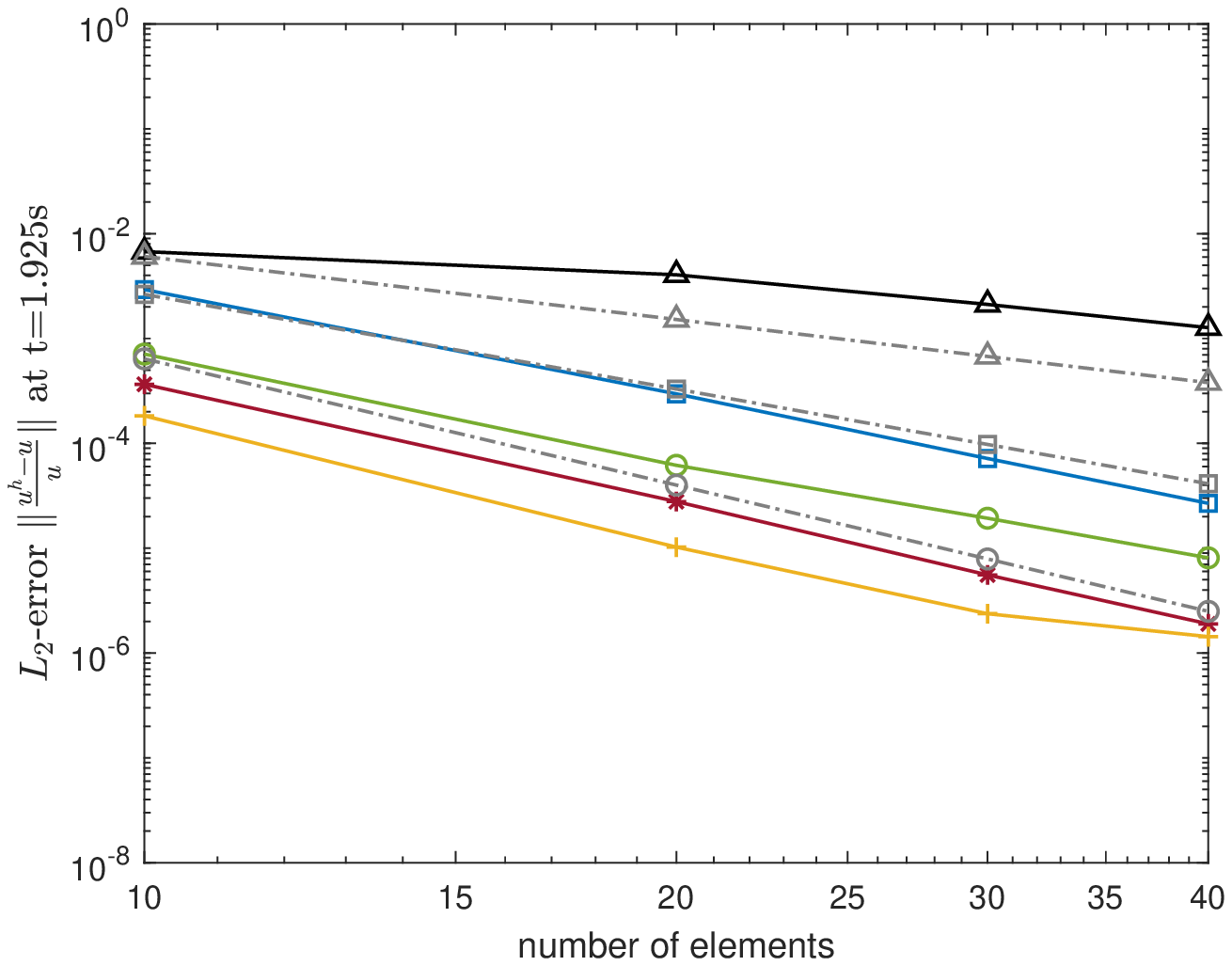}
	\caption{$M_{\mathrm{AD}~q_{\mathrm{max}}}^{\mathrm{row-sum}}$}
	\label{fig:e4_L2_ADmax2}
\end{subfigure}\\
\begin{subfigure}[ht]{1.00\textwidth}
	\centering
	\includegraphics[width=\textwidth]{legend_example2_1}
\end{subfigure}
\caption{Fixed-fixed 1D truss: $L_2$-error norm of displacement $u(t=1.925\,\text{s})$, NURBS of order $p$ are used as shape functions, type of test functions varies. Computation was done on uniform mesh~A (Fig.~\ref{fig:mesh1}). Time integration was applied using CDM with $\Delta t=\left(\frac{p}{2n_{el}}\right)^p$.}
\label{fig:e4_error_pN}
\end{figure}
\begin{figure}[t]
\centering
\begin{subfigure}[ht]{0.47\textwidth}
	\centering
	\includegraphics[width=\textwidth]{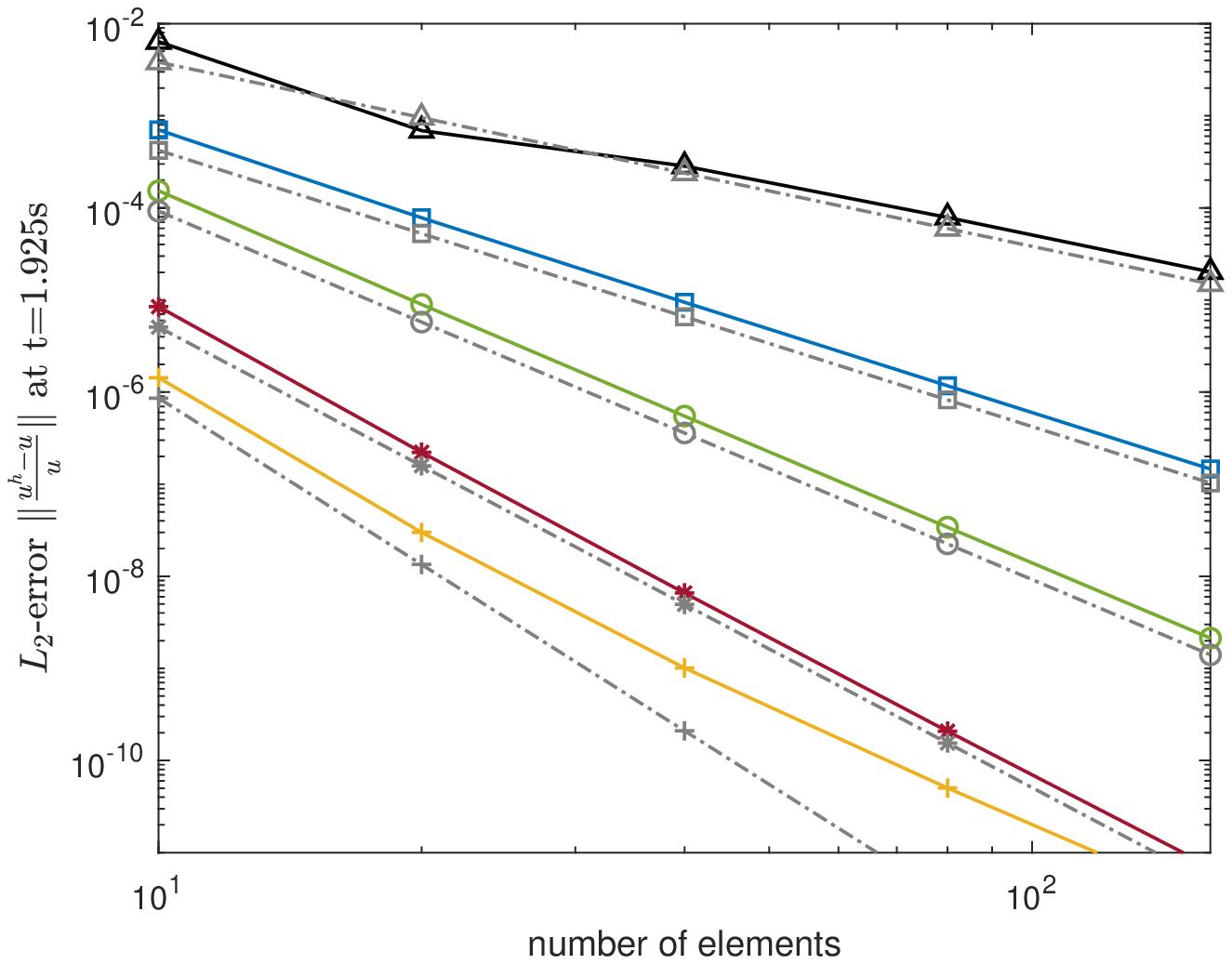}
	\caption{$M_{\mathrm{NURBS,AD,IG}}^{\mathrm{cons}}$}
\end{subfigure}
\hfill
\begin{subfigure}[ht]{0.47\textwidth}
	\centering
	\includegraphics[width=\textwidth]{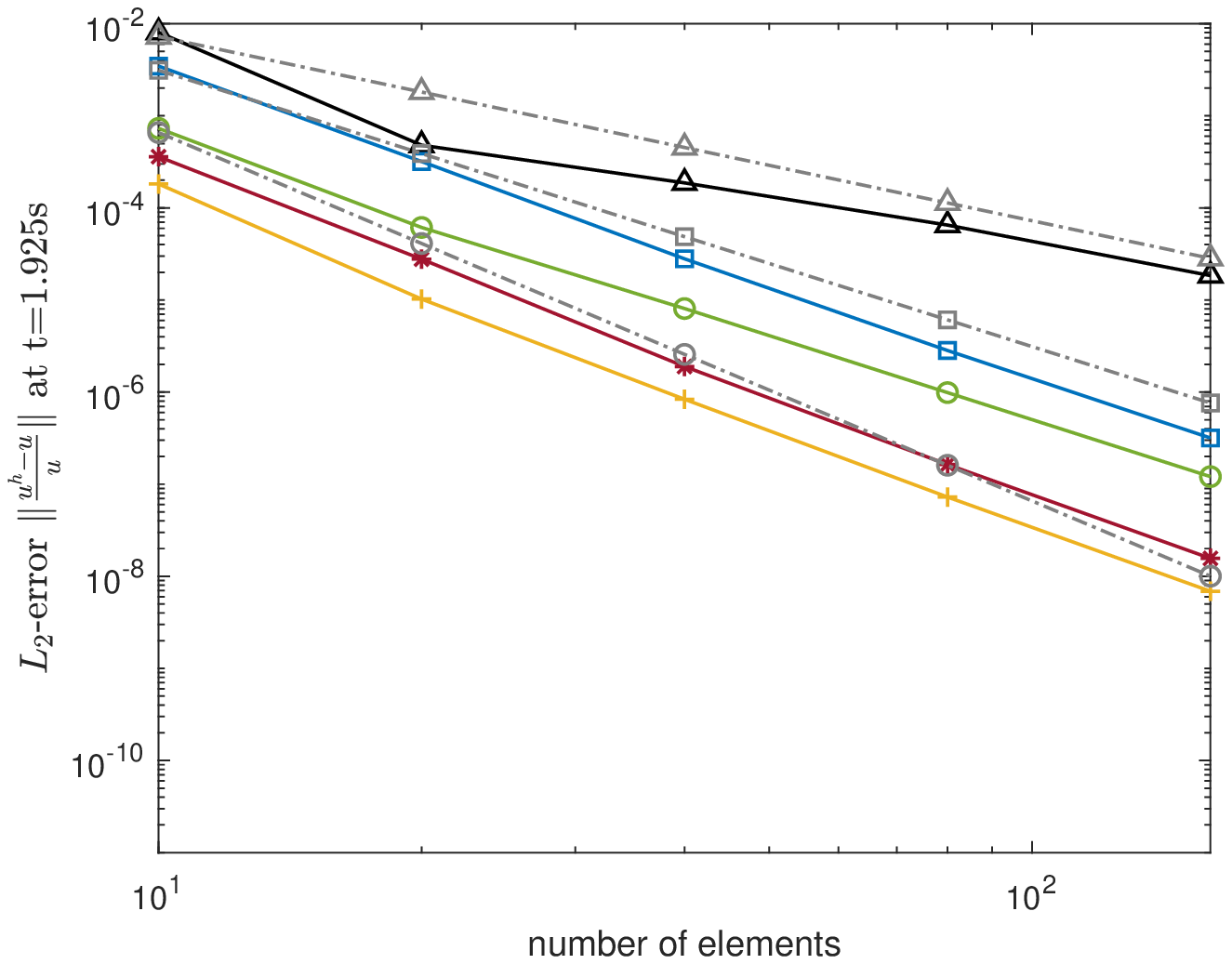}
	\caption{$M_{\mathrm{AD}~q_{\mathrm{max}}}^{\mathrm{row-sum}}$}
\end{subfigure}\\
\begin{subfigure}[ht]{1.00\textwidth}
	\centering
	\includegraphics[width=\textwidth]{legend_example2_1}
\end{subfigure}
\caption{Fixed-fixed 1D truss: $L_2$-error norm of displacement $u(t=1.925\,\text{s})$, NURBS of order $p$ are used as shape functions, type of test functions varies. Computation was done on uniform mesh~A (Fig.~\ref{fig:mesh1}). Time integration was applied using RK4 with $\Delta t=\frac{h}{10}$.}
\label{fig:e4_error_RK4}
\end{figure}

The $L_2$-error of the deformation according to
\begin{equation}
e^{\mathrm{u}}_{L_2}(t)=\frac{\lVert u^{\mathrm{ref}}(x,t)-u^{\mathrm{h}}(x,t)\rVert}{\lVert u^{\mathrm{ref}}(x,t)\rVert}
\end{equation}
is calculated at $t=1.925\,$s. The results are presented in Fig.~\ref{fig:e4_error} and confirm the findings of the prior examples. Applying additional row-sum lumping to mass matrices based on AD test functions leads to more accurate results compared to the standard lumped NURBS approach (Fig.~\ref{fig:e4_L2_NURBSd}). For a stable computation, the time step for the results in Fig.~\ref{fig:e4_error} was estimated with $\Delta t = \frac{h}{10}$, where $h$ refers to the element length. While, one more time it can be noticed that simple row-sum lumping (Fig.~\ref{fig:e4_L2_NURBSd}) deteriorates the results with increasing the polynomial order $p$, the lumped AD scheme, see Figs.~\ref{fig:e4_L2_ADmin} and \ref{fig:e4_L2_ADmax}, improves noticeably until $p=2$ and for higher orders the accuracy is only slightly improved, but convergence rates stagnate. As the CDM time integration itself is of low approximation order, the convergence of the $L_2$-error is limited to second order accuracy for the selected time step size in this example. Even for the common consistent Bubnov-Galerkin formulation, cf. Fig.~\ref{fig:e4_L2_NURBSc}, the convergence rate is not further elevated, raising the polynomial order $p$. Surprisingly, a minor improvement of convergence can be noticed in Fig.~\ref{fig:e4_L2_ADmin} for the lumped AD approach with minimum reproduction. This behavior is due to the chosen time step and examination at $t=1.925\,$s and cannot be observed for all time evaluation points in general. 

The temporal discretization error dominates the spatial discretization error. To improve the results, an adapted time step $\Delta t=\left(\frac{p}{2n_\mathrm{el}}\right)^p$ depending on the polynomial order $p$ and the number of elements $n_\mathrm{el}$ is considered in Fig.~\ref{fig:e4_error_pN}. The results of pure row-sum technique and lumped AD approach with minimum reproduction do not improve; their accuracy is directly limited by applying row-sum lumping. A minimal change in the slope of the convergence graphs can be spotted for AD with $q_{\mathrm{min}}$ if the polynomial order is raised, but in contrast to that, the ones belonging to standard lumped NURBS approach run clearly parallel. Thus, the common lumping procedure will only reach the accuracy of the dual schemes, if a considerably larger amount of elements is used, as the initial accuracy is already lower. In Fig.~\ref{fig:e4_L2_NURBSc2}, the assumed `$p+1$' convergence rate of the common Bubnov-Galerkin formulation can be achieved. The adapted time step size overcomes the limitations of CDM with coarser time stepping. From Fig.~\ref{fig:e4_L2_ADmax2}, we notice that the convergence of lumped AD formulation with maximum degree of reproduction is also improved, raising the polynomial order $p$. The achieved maximum fourth order accuracy confirms that the proposed formulation with AD $q_{\mathrm{max}}$ test functions and additional row-sum lumping is still high order accurate.

Instead of increasing the computational costs by choosing the small adapted time step, it is computational more efficient to use a higher order time integration scheme. Making use of high order time integration schemes as the explicit $4^{th}$-order Runge-Kutta method (RK4) or the implicit high order Padé-based time integration scheme \cite{Song.2022,Song.2023}, results of better accuracy can be achieved with coarser time stepping.

For Fig.~\ref{fig:e4_error_RK4}, the larger time step $\Delta t=\frac{h}{10}$ was used within RK4. The results are similar to the adapted time step within the CDM. The lumped AD scheme with maximum reproduction is again limited to $4^{th}$ order accuracy, which is not optimal but a significant improvement over standard row-sum lumping.

Regarding accuracy, the results of applying NURBS as shape and test function and enforcing diagonal mass matrices by row-sum technique strongly detach from all other investigated approaches and confirms the assessment of waiving it. Although the simple implementation and the gains in time may be convincing, they do not excuse the quite huge losses in accuracy. 

\subsection{Explicit dynamic analysis of a fixed-free 1D truss}
\label{sec:examples_dyn1}
\begin{figure}[t]
\centering
\includegraphics[width = 0.8\textwidth]{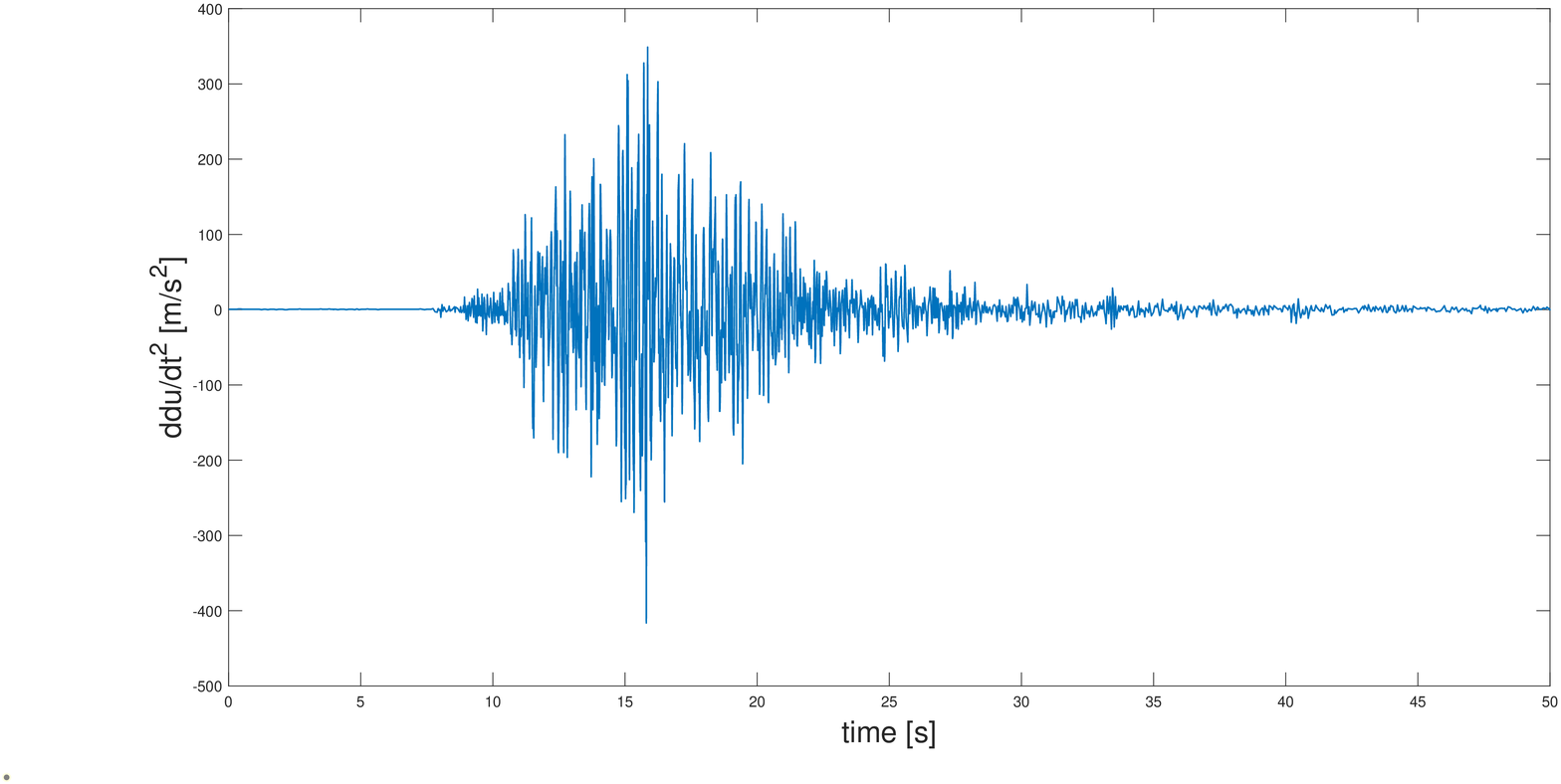}
\caption{Horizontal acceleration during Loma Prieta earthquake (1989).}
\label{fig:e3_signal}
\end{figure}

\begin{figure}[p]
\centering
\begin{subfigure}[ht]{\textwidth}
	\centering
	\includegraphics[width=0.6\textwidth]{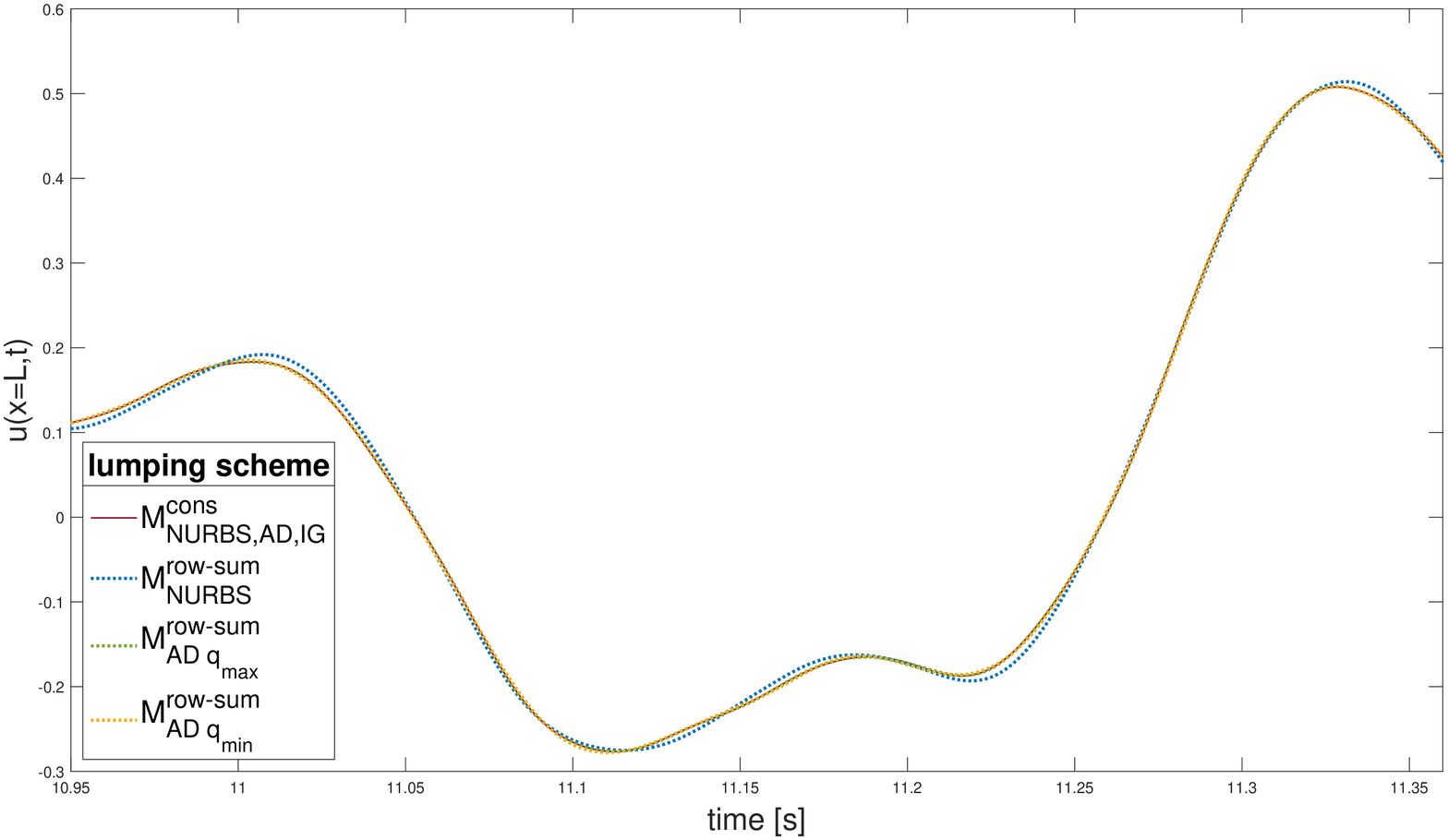}
	\caption{$u(x=L,t)$}
	\label{fig:e3_u1}
\end{subfigure}\\
\begin{subfigure}[ht]{\textwidth}
	\centering
	\includegraphics[width=0.6\textwidth]{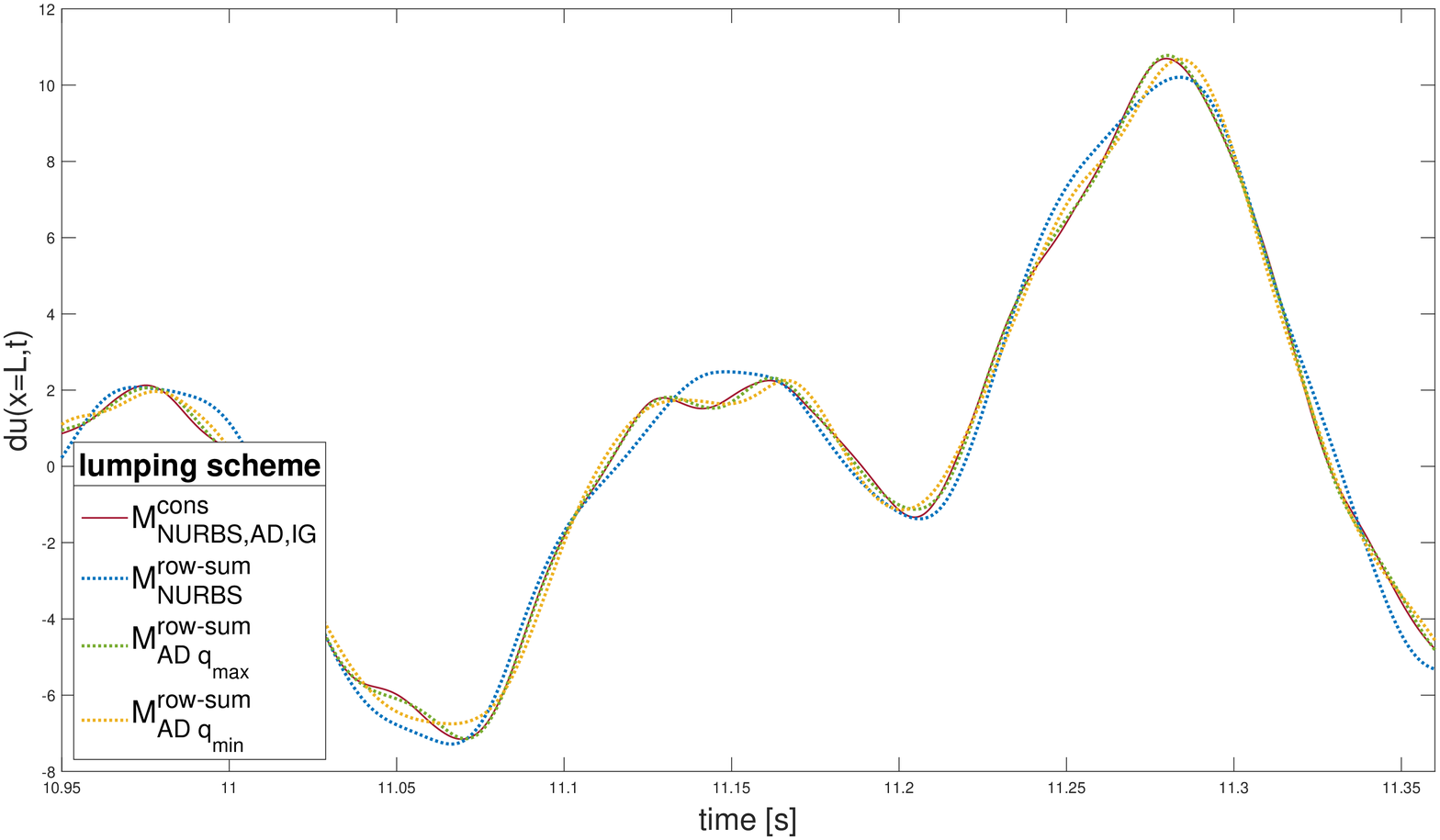}
	\caption{$\dot{u}(x=L,t)$}
	\label{fig:e3_du1}
\end{subfigure}\\
\begin{subfigure}[ht]{\textwidth}
	\centering
	\includegraphics[width=0.6\textwidth]{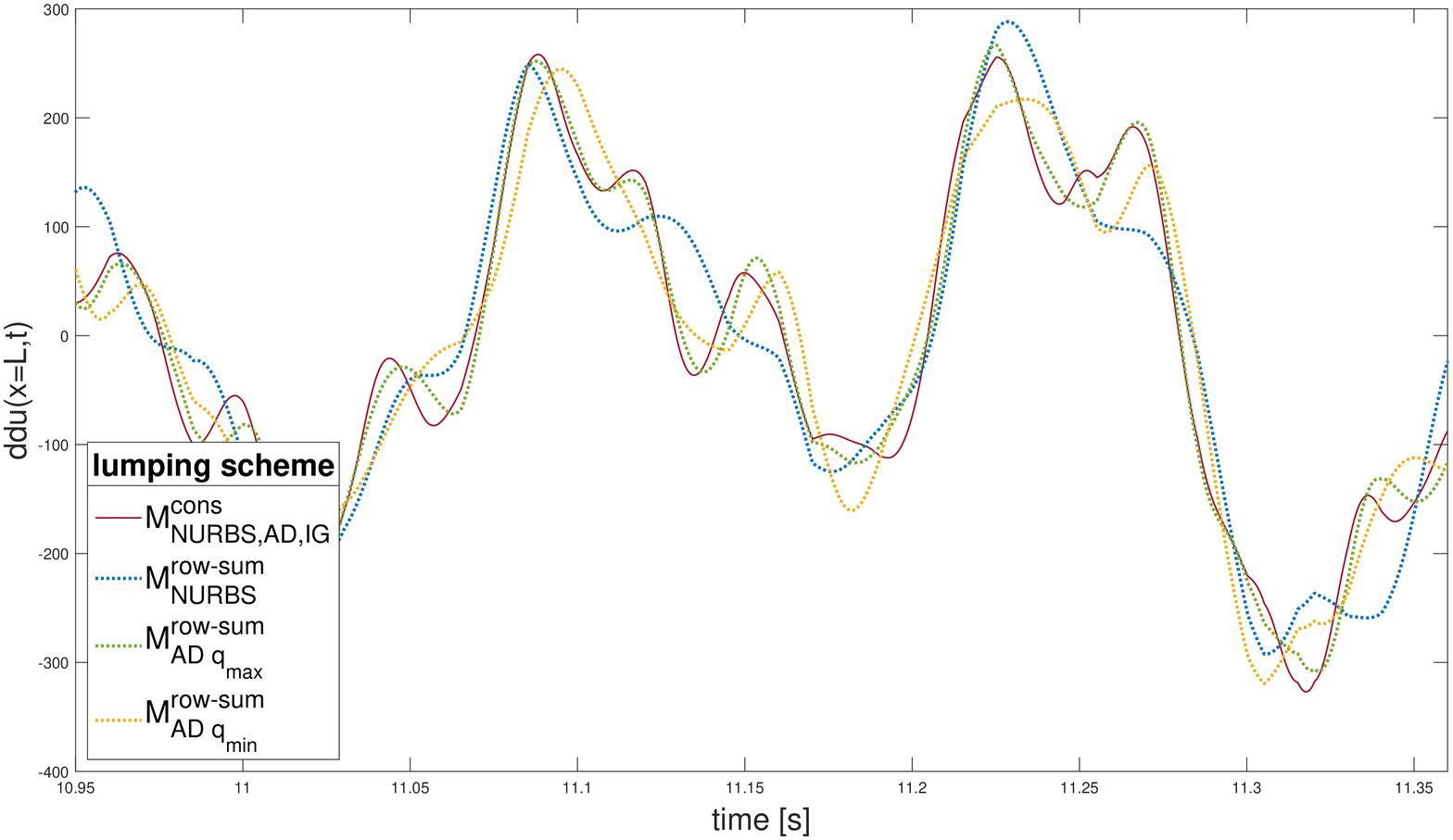}
	\caption{$\ddot{u}(x=L,t)$}
	\label{fig:e3_ddu1}
\end{subfigure}
\caption{Fixed-free 1D truss: Response of system with $p=2$ and 20 elements to signal shown in Fig.~\ref{fig:e3_signal}. Displacement $\mathrm{u}$, velocity $\mathrm{du}$ and acceleration $\mathrm{ddu}$ at position $x=L$ are compared for different formulations with consistent and lumped mass matrices.}
\label{fig:e3_response2}
\end{figure}

\begin{figure}[p]
\centering
\begin{subfigure}[ht]{\textwidth}
	\centering
	\includegraphics[width=0.6\textwidth]{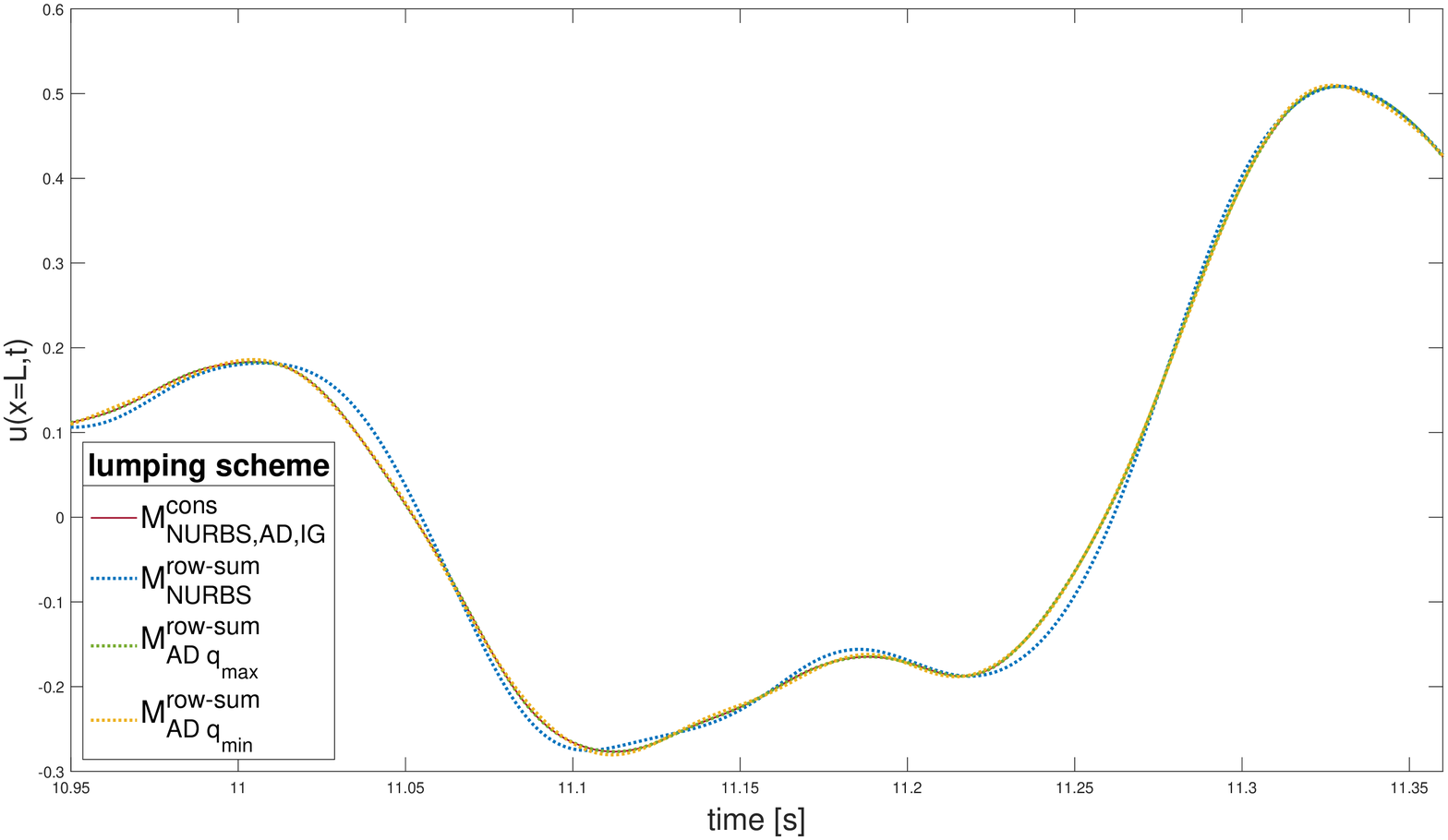}
	\caption{$u(x=L,t)$}
	\label{fig:e3_u5}
\end{subfigure}\\
\begin{subfigure}[ht]{\textwidth}
	\centering
	\includegraphics[width=0.6\textwidth]{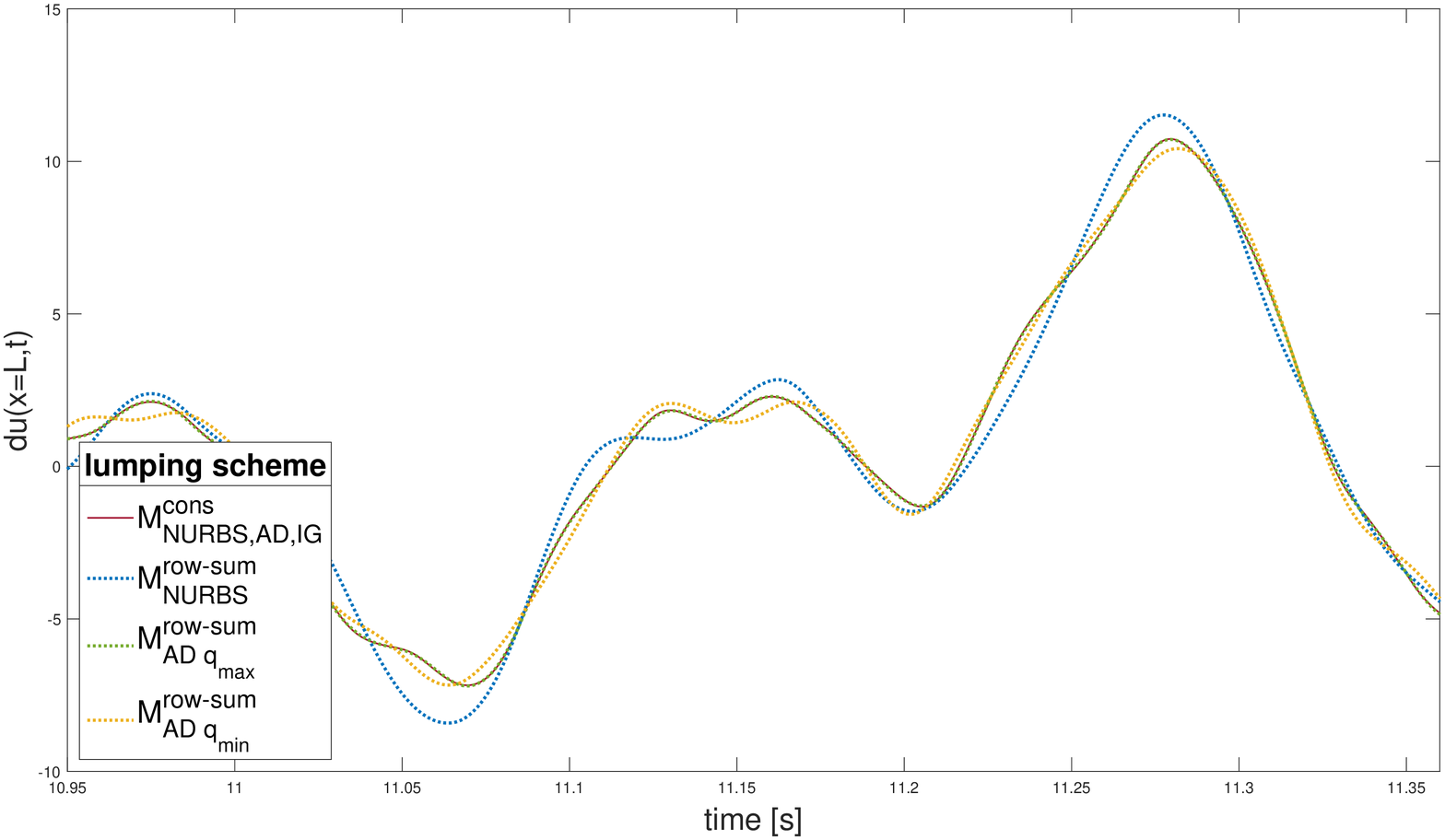}
	\caption{$\dot{u}(x=L,t)$}
	\label{fig:e3_du5}
\end{subfigure}\\
\begin{subfigure}[ht]{\textwidth}
	\centering
	\includegraphics[width=0.6\textwidth]{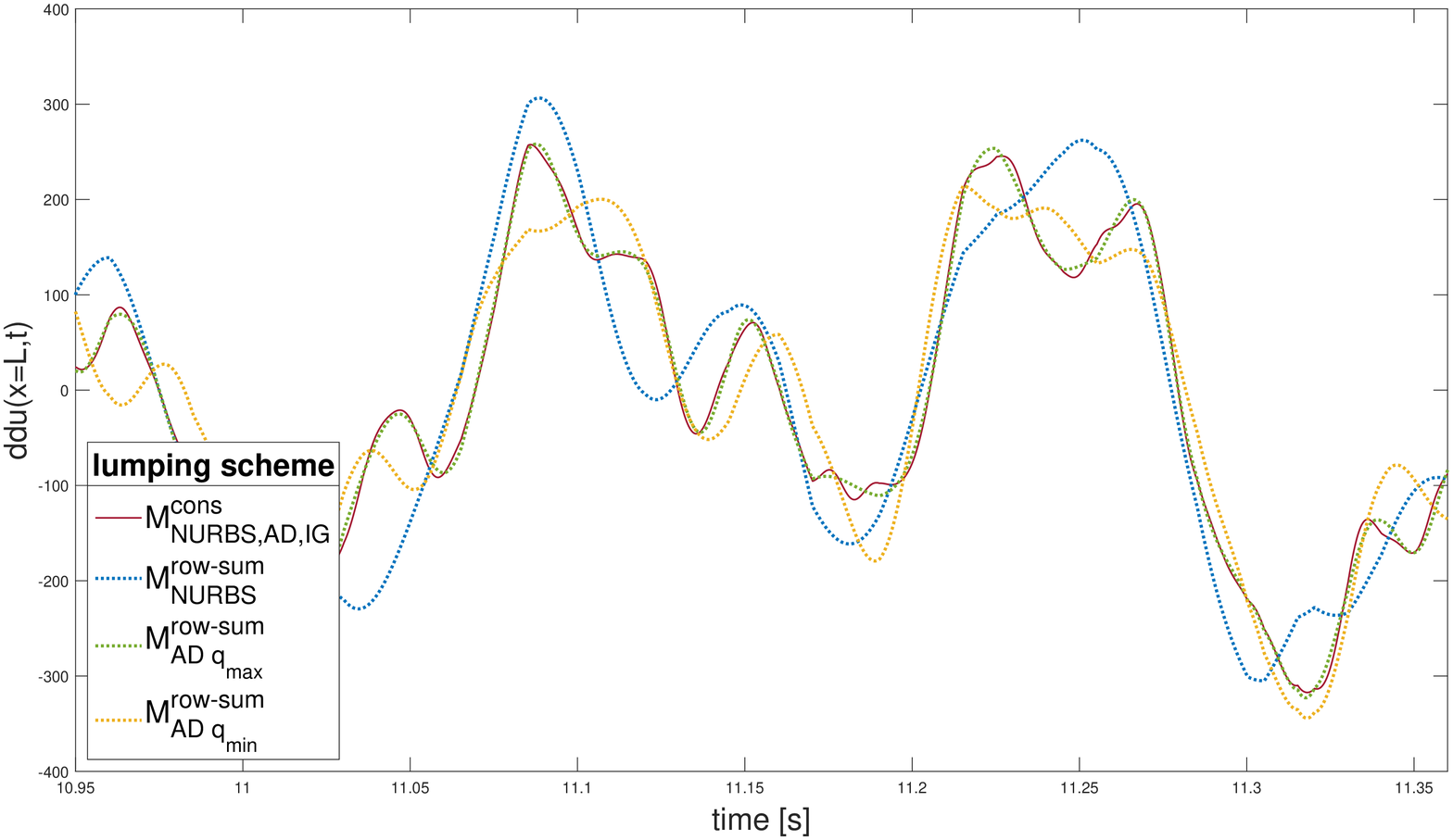}
	\caption{$\ddot{u}(x=L,t)$}
	\label{fig:e3_ddu5}
\end{subfigure}
\caption{Fixed-free 1D truss: Response of system with $p=5$ and 20 elements to signal shown in Fig.~\ref{fig:e3_signal}. Displacement $\mathrm{u}$, velocity $\mathrm{du}$ and acceleration $\mathrm{ddu}$ at position $x=L$ are compared for different formulations with consistent and lumped mass matrices.}
\label{fig:e3_response5}
\end{figure}

Finally, we study the systems response to a non-deterministic transient excitation signal. Therefore, the horizontal east-west ground acceleration $\ddot{\mathbf{u}}_g$ (see Fig.~\ref{fig:e3_signal}) of the Loma Prieta earthquake in 1989 \cite{Matlab.2019} is applied as an external force 
\begin{equation}
\mathbf{F}=\mathbf{M}\ddot{\mathbf{u}}_g
\end{equation}
to the structure, a simple supported truss as shown in Fig.~\ref{fig:sysplot}, but without any static loading. Note that the recordings of the accelerometers located at the Natural Sciences building at the University of California in Santa Cruz are readily available in \texttt{MATLAB}, using the command \texttt{load quake}. An identical time step size $\Delta t$ was chosen for all investigated formulations and polynomial orders $p$, to ensure the same input for each analysis. Note that we linearly interpolate between the recorded sampling points and therefore, different time step sizes would also alter the external force input. The structure is discretized into 20 elements using several polynomial orders $p$. This discretization ensures very accurate results for $p>1$ and is in line with the guidelines proposed in \cite{Willberg.2012}. The displacement, velocitiy and acceleration results are depicted in Figs.~\ref{fig:e3_response2} and ~\ref{fig:e3_response5} for $p=2$ and $p=5$, respectively. The values have been obtained for the right end of the truss at $x=L$. The calculated displacements in Figs.~\ref{fig:e3_u1} and \ref{fig:e3_u5} do not vary a lot. The AD lumping scheme with additional row-sum lumping almost captures the results of the consistent formulations within the whole depicted time interval. For standard NURBS test functions and row-sum lumping increasing deviations are observed when the polynomial order is raised. The same behavior is also observed for the velocities in Figs.~\ref{fig:e3_du1} and \ref{fig:e3_du5}. Here, the AD lumping scheme with minimum reproduction degree deviates as well. As the acceleration constitutes the highest investigated derivative within these plots, in Figs.~\ref{fig:e3_ddu1} and \ref{fig:e3_ddu5}, all results achieved by means of lumped approaches differ from the consistent formulation notably. As for the approximation of the eigenmodes in Sec.~\ref{sec:examples_eig}, the different aproaches can be categorized in terms of the attainable accuracy as follows: the Bubnov-Galerkin formulation with NURBS as shape and test functions and applied row-sum lumping results in the worst approximation of the actual results, followed by AD formulations with additional lumping from lowest to highest degree of reproduction. An additional remarkable fact is the different behavior of the AD approach with maximum reproduction degree. In contrast to the other row-sum lumped results, which get worse while increasing the polynomial degree of the basis functions, it improves slightly by raising the order.

\begin{figure}[t]
\centering
\includegraphics[width=0.7\textwidth]{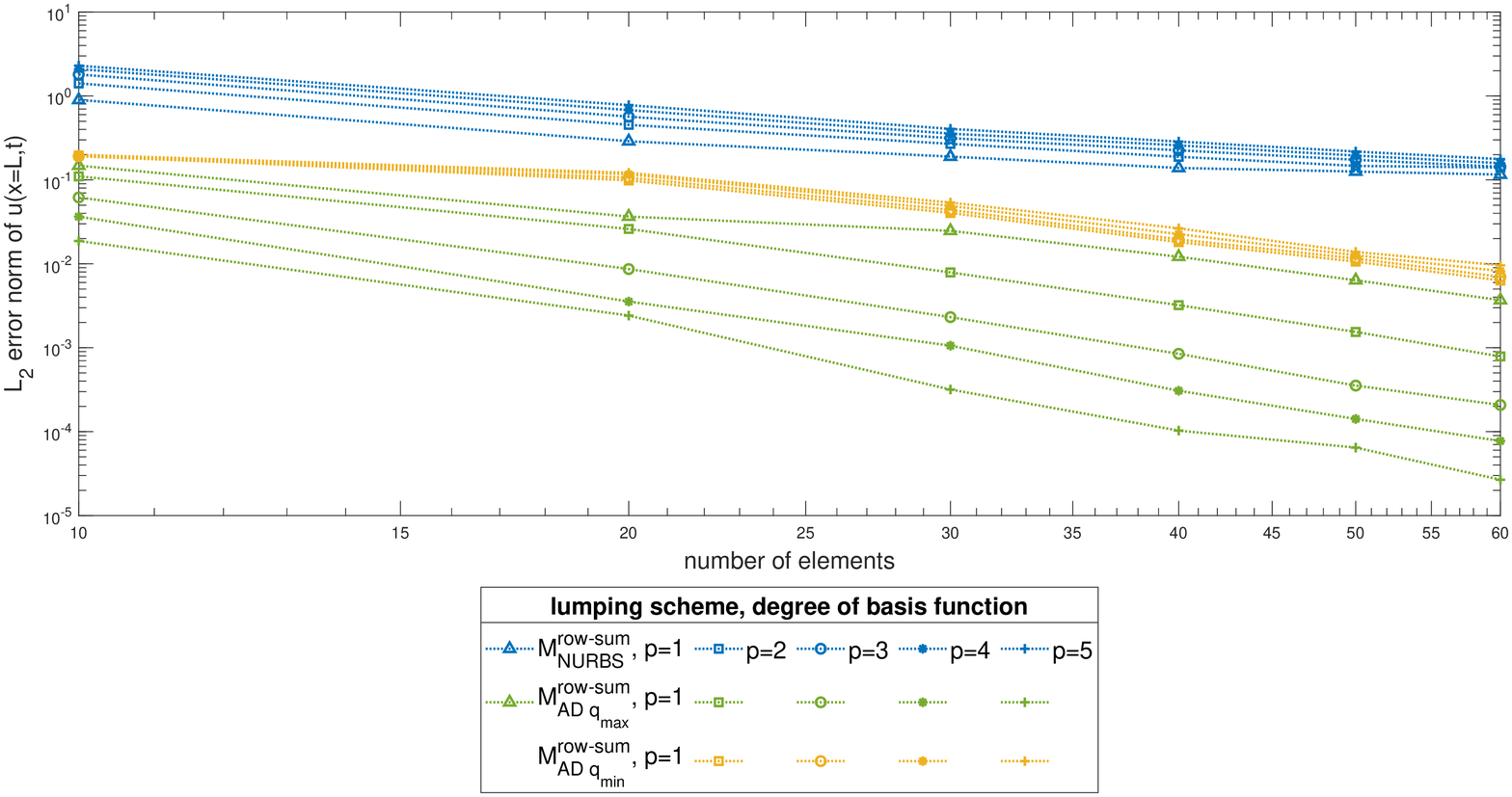}
\caption{Fixed-free 1D truss: $L_2$ error norm of the three lumping approaches generating its diagonal masses by row-sum method measuring the deviations of displacement $u(x=L,t)$ to the reference solution $u^{\mathrm{ref}}(x=L,t)$ received from a computation with consistent mass matrices. The error is evaluated over time for the whole period of the signal shown in Fig.~\ref{fig:e3_signal}. Accuracy is compared for polynomial orders $p=$ 1 to 5.}
\label{fig:e3_L2}
\end{figure}

That behavior can be detected not only for the small selected time sample, but also for the whole observation period. It is even more obvious in Fig.~\ref{fig:e3_L2}, where the $L_2$-error 
\begin{equation}
e^{\mathrm{u}}_{L_2}(x)=\sqrt{\frac{\sum\limits_{t=t_0}^{T}{\left[u^{\mathrm{ref}}(x,t)-u^{\mathrm{h}}(x,t)\right]^2}}{\sum\limits_{t=t_0}^{T}{\left[u^{\mathrm{ref}}(x,t)\right]^2}}}
\label{eq:L2_time}
\end{equation}
for displacements at $x=L$ over time $t\in [0,50]\,$s was calculated with the system response of the consistent formulations as reference $u^{\mathrm{ref}}$, as no analytical solution is available. The different lumping approaches can be easily distinguished by the different colors and the fact that they are clustered in certain error levels, appearing in the plot of the $L_2$-error. The green dashed lines represent the AD scheme with $q_{\mathrm{max}}$ and approximate the system response of the consistent formulation best. If a higher polynomial order $p$ is applied, the corresponding graph is shifted towards smaller error levels, i.e., downwards in the plot. The standard row-sum lumping approach is displayed by blue dashed lines and marks the upper bound of accuracy levels. Raising $p$ is equivalent to increasing the error, which is in stark contrast to the AD scheme. The yellow dashed lines of the AD scheme with $q_{\mathrm{min}}$ show a behavior between the previously reported cases. Although they are clustered close together and seem to be almost unaffected by the underlying degree of basis functions, a slight but almost imperceptible upwards shift through increasing $p$ can be detected. \\

Especially in the field of structural dynamics, the efficiency of solving procedures cannot only be measured by accuracy. The main interest is on reducing the computational costs, accepting a minor increase in error. The largest impact on computational time of the proposed dual lumping schemes can be expected for solving the SOE (\ref{eq:SOE}), where an inversion of the mass matrix is required. The effect on the computational costs in particular depends on the structure of the system matrices and thus, every studied lumping scheme has to be evaluated separately, combining the results for all consistent approaches is no longer reasonable. 

To estimate the overall computational time for a specific formulation during an explicit dynamic analysis, several impacts have to be considered. A diagonal mass matrix is always beneficial regarding computational aspects. The time for solving the occurring SOE in each time step is much lower than for non-diagonal mass matrices. In our geometrically linear example, this effort is reduced to the first time step, where the inverse of the effective stiffness matrix has to be calculated only once. All succeeding steps can reuse that pre-inverted matrix and thus (\ref{eq:SOE}) is reduced to a matrix vector multiplication, which is indeed an inexpensive operation. Thus, all considered lumped mass approaches and the IG formulation are computational less costly for solving the SOE than the consistent formulation based on NURBS test functions. When elevating the polynomial degree $p$, the gap between formulations with and without diagonal mass matrices will increase due to the growing density of $\mathbf{M}$. Since the inversion of the diagonalized $\mathbf{M}$ turns into a simple element by element inversion, also the multiplication with $\mathbf{F}^*_t$ in Eq.~(\ref{eq:SOE}) requires only `$n_\mathrm{np}-k$' multiplications. The highest computational effort will be required for the multiplication $\mathbf{K}_t \mathbf{U}_t$ in Eq.~(\ref{eq:F_eff}), since $\mathbf{K}_t$ is the most dense matrix. The number of multiplications is equal to the number of non-zero entries of $\mathbf{K}^*$. Here, the lumped NURBS formulation will be fastest, followed by AD with $q_\mathrm{min}$ and then AD with $q_\mathrm{max}$. The IG dual formulation would require by far the highest resources.

The computational time in general also depends on the number of time steps, which is required for a stable computation and is directly linked to the highest eigenfrequency as we already described in the beginning of this section. Hence, the IG approach with consistent but diagonal mass matrices should be more costly in time than the formulations with diagonal mass matrices obtained through additional row-sum lumping. Lumped formulations based on NURBS and AD test functions require comparatively less time steps for a stable solution, because, as already mentioned in Sec.~\ref{sec:examples_eig}, they underestimate the eigenvalues and thus, enable a larger time step.

\begin{figure}[t]
\centering
\begin{subfigure}[ht]{0.47\textwidth}
	\centering
	\includegraphics[width=\textwidth]{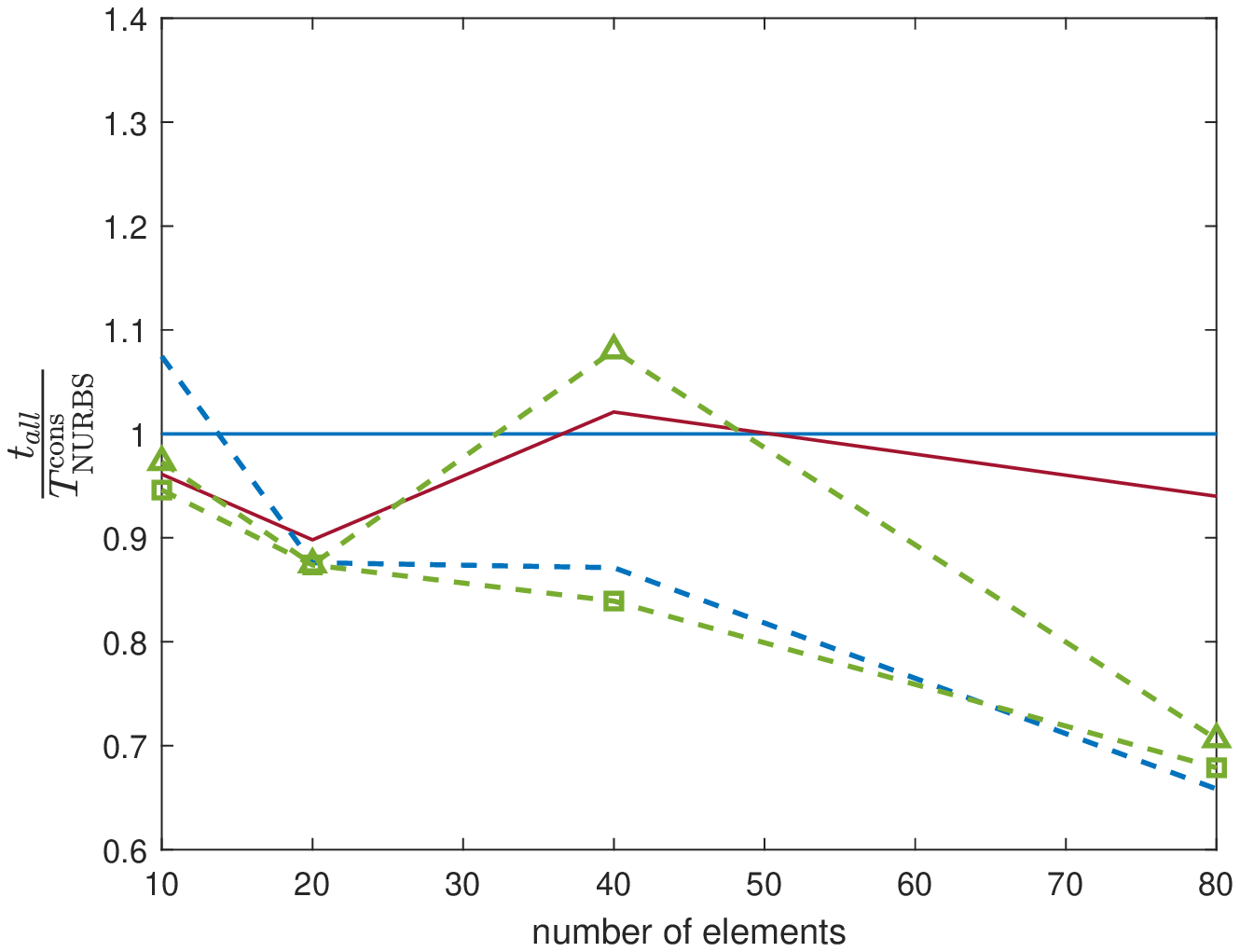}
	\caption{$p=2$}
	\label{fig:e3_time_all_2}
\end{subfigure}
\hfill
\begin{subfigure}[ht]{0.47\textwidth}
	\centering
	\includegraphics[width=\textwidth]{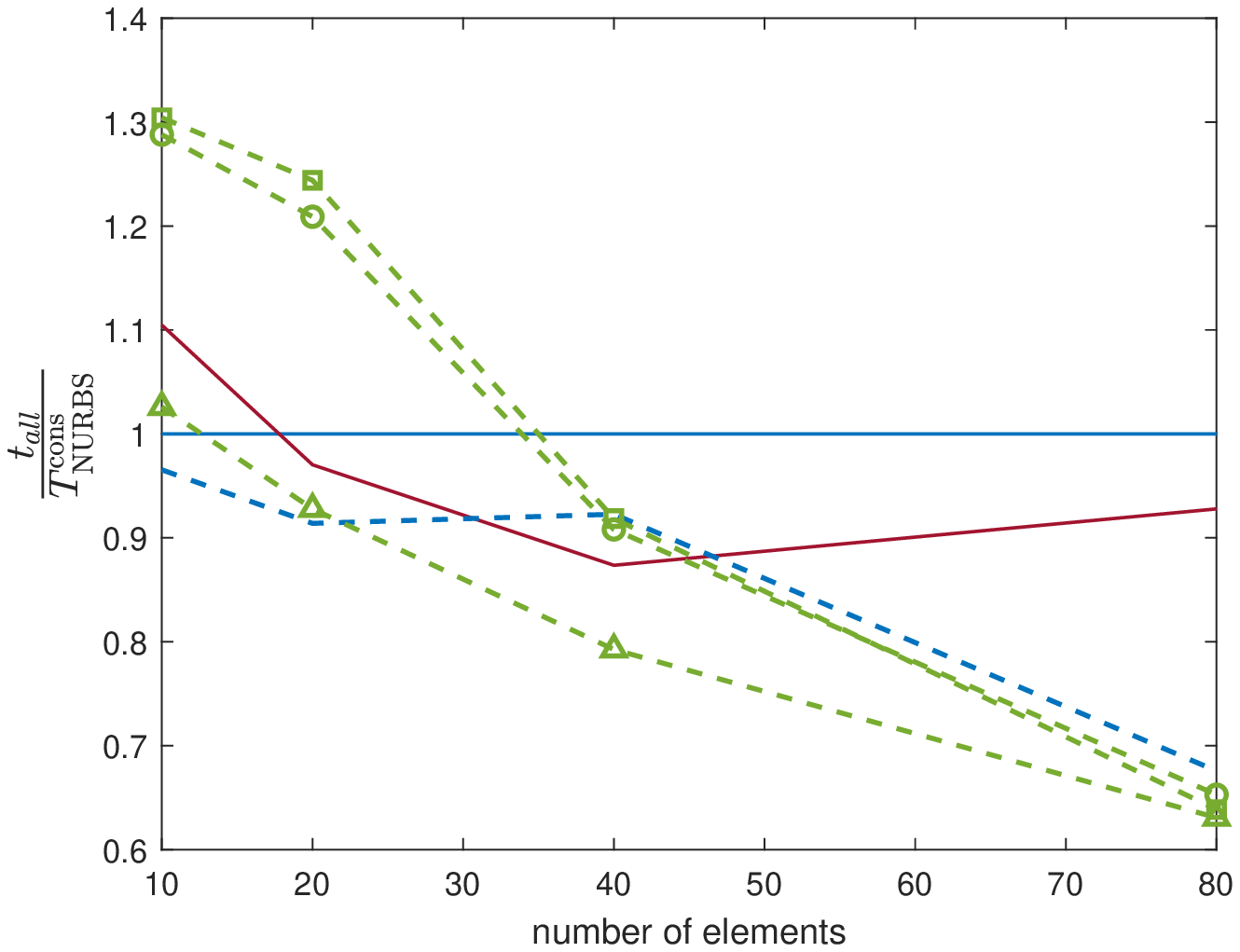}
	\caption{$p=3$}
	\label{fig:e3_time_all_3}
\end{subfigure}\\
\begin{subfigure}[ht]{0.47\textwidth}
	\centering
	\includegraphics[width=\textwidth]{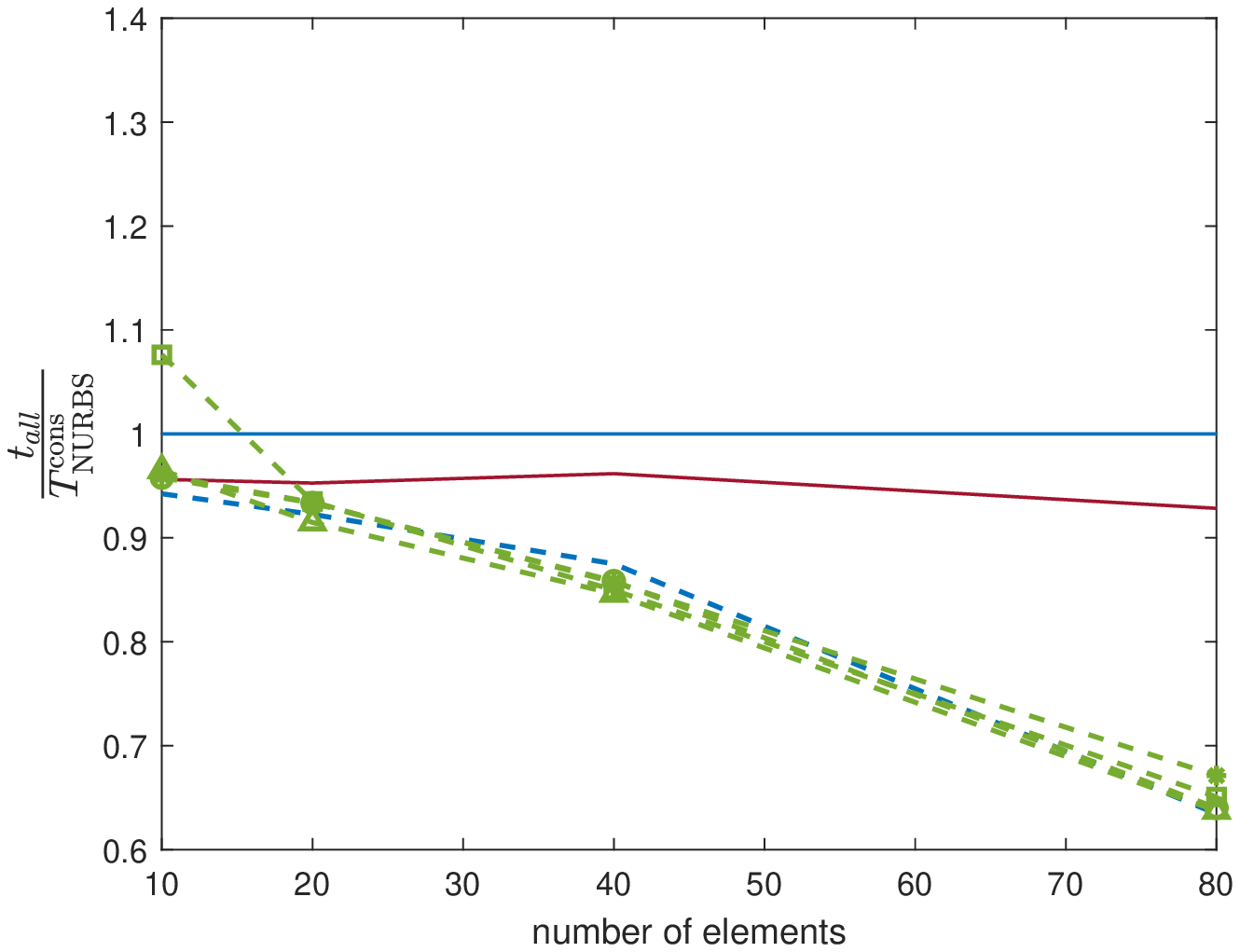}
	\caption{$p=4$}
	\label{fig:e3_time_all_4}
\end{subfigure}
\hfill
\begin{subfigure}[ht]{0.47\textwidth}
	\centering
	\includegraphics[width=\textwidth]{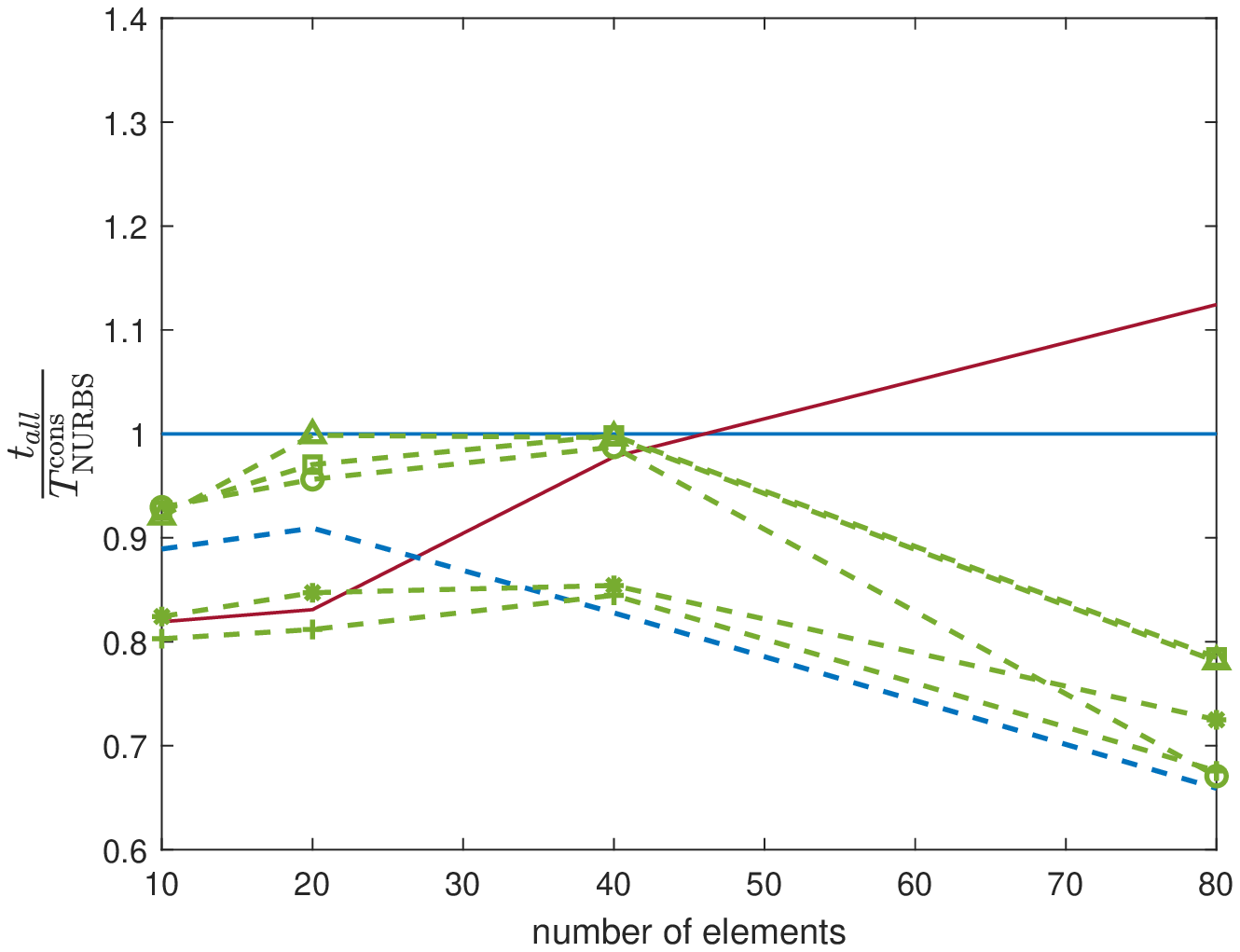}
	\caption{$p=5$}
	\label{fig:e3_time_all_5}
\end{subfigure}\\
\begin{subfigure}[ht]{0.85\textwidth}
	\centering
	\includegraphics[width=\textwidth]{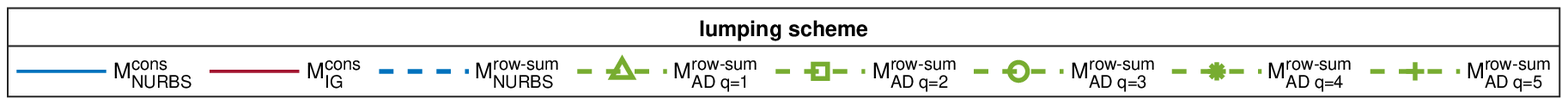}
\end{subfigure}
\caption{Ratio of computational time $t_{\mathrm{all}}$ for performing whole explicit dynamic analysis and computational time of common Bubnov-Galerkin formulation with consistent mass matrix. Computational time was measured for system response to signal of Loma Prieta earthquake, shown in Fig.~\ref{fig:e3_signal}. NURBS of order $p$ are used as shape functions, type of test functions varies. Computation was done on uniform mesh~A (Fig.~\ref{fig:mesh1}).}
\label{fig:e3_time_all}
\end{figure}

Figure~\ref{fig:e3_time_all} shows a rough first estimate for the overall computational time $t_{\mathrm{all}}$. The results depict the ratio of $t_{\mathrm{all}}$ of each investigated formulation and the computational time for the dynamic analysis with the same amount of elements using the common formulation based on NURBS shape and test functions. As assumed, it can be clearly noticed that the formulations with additional row-sum lumping are least costly in time.

The IG formulation is a consistent formulation with diagonal mass matrix and fully populated stiffness matrix, and thus very similar to the standard consistent formulation based on NURBS test functions with a pre-inverted mass matrix and less populated stiffness matrix. If we compare the computational time $t_{\mathrm{all}}$ for the overall time integration procedure (Fig.~\ref{fig:e3_time_all}), between the standard approach and the IG approach, we do not get big differences. The gap between the corresponding continuous lines is comparatively small and for the largest investigated order $p=5$ in Fig.~\ref{fig:e3_time_all_5}, the IG approach is strongly related to the number of elements and gets even costlier in time than the common consistent formulation. Although, Fig.~\ref{fig:e3_time_all} is merely a rough estimate of computational costs and not as reliable as an error norm, the raise in computational costs can be explained by the fully populated stiffness matrix when using IG test functions. The computational time for the whole time integration procedure is not only determined by the density of the mass matrix and the total number of time steps, but also by the bandwidth of the stiffness matrix as well. To be able to solve for the next time step $\mathbf{U}_{t+\Delta t}$, the effective matrices $\mathbf{K}^*$ and $\mathbf{F}^*$ have to be computed first. As we can notice from Eqs.~(\ref{eq:K_eff}) and (\ref{eq:F_eff}), simply matrices and vectors have to be multiplied, involving all system matrices. Thus, the computational effort is not as high as suggested for an inversion, but these calculations have to be performed for every single time step. It mainly depends on the allocated memory and therefore, on the bandwidths of the matrices. 

The computations based on the lumped mass matrices, represented by the dashed lines, differ also only in the population of the accompanying stiffness matrices. The bandwidth of $\mathbf{K}$ based on AD rises equally for $p$ and $q$, whereas the bandwidth for the standard lumped approach rises only with the polynomial order $p$. Additionally, the critical time step of the lumped NURBS approach is the largest of all investigated ones, as the highest eigenfrequency is harshly underestimated. Hence, $M_{\mathrm{NURBS}}^{\mathrm{row-sum}}$ should be least expensive with respect to the overall computational time. Figure~\ref{fig:e3_time_all_5} provides the best inside on this expected behavior.

If we only consult Fig.~\ref{fig:e3_time_all}, we may end up with the assumption that the already established row-sum lumping technique applied to standard Bubnov-Galerkin formulation is also most efficient within IGA methods. However, we also noticed that the common row-sum lumped approach is far away from the accuracy level of the consistent formulations and requires just less time steps as it underestimates the decisive eigenfrequency. Thus, we conclude this numerical example with a discussion on efficiency, measuring accuracy and computational costs at once. Simply setting the focus on the need in time to solve the SOE~(\ref{eq:SOE}), the assumption arises that the IG approach is most beneficial. As it is a consistent formulation, the loss of accuracy compared to the common consistent formulation is actually zero and the diagonal mass matrix enables the simplest possible inversion. However, due to the fully populated stiffness matrix, this approach is as unsuitable for dynamic computations as the common consistent one. Regarding the overall computational time $t_{\mathrm{all}}$ for the whole time integration procedure, it turns out to be even more expensive for high polynomial orders $p$. Despite the favorable diagonal mass matrix, the standard NURBS formulation with additional row-sum lumping performs far from reasonable accuracy, as it was already stated during motivation of the new lumping procedures, and hence is not preferable to the other lumping schemes. That leads to the choice of AD test functions and additional row-sum lumping. As we have seen so far, the computational time is almost the same and only slightly higher in comparison to the common row-sum lumping, but the accuracy is much better. The accuracy level can be even adjusted by the degree of reproduction $q$. A small $q$ is accompanied by a slightly higher error level, but less computational costs and raising $q$ behaves the other way round. Considering the domination of computational costs, the minimal degree is preferable to the maximum one. Balancing accuracy level and computational costs equally, maximum reproduction should be considered.

The influence of the number of time steps cannot be captured within Fig.~\ref{fig:e3_time_all}, because of the applied identical time step size $\Delta t$, as explained earlier for this example.\\

These assumptions hold for the studied geometrically linear case with constant mass and stiffness matrices. If we consider geometrically nonlinear applications, the stiffness matrix has to be updated and computed again at the beginning of each time step. Within an explicit dynamic formulation, the stiffness matrix is only used in a matrix-vector product, but never has to be inverted. Thus, the higher density of the AD stiffness matrix will yield slightly higher computational costs of the required matrix-vector product for each time step in comparison to the standard approach. But here we should remember, that matrix products can be parallelized effectively and scale perfectly, while matrix inversions, as required for the consistent mass formulation, do not scale well for a large number of parallel processes. Thus, if the mass matrix has to be updated within every single time step as well, the gap between diagonal and non-diagonal formulations, which is already observed in our linear application, will be significantly larger. As effect of the ongoing inversions, the common consistent formulation based on NURBS test functions will be by far more expensive than any formulation with diagonal mass matrices. For total Lagrangian formulations the increase of computational costs will not be as tremendous as for updated ones, as they provide constant mass matrices. However, for practical applications requiring a huge amount of DOFs, a lumped $\mathbf{M}$ should be considered even for a total Lagrangian formulation, as the required memory for the matrix inversion in the beginning and storing the inverse will easily exceed the available memory, as a full occupied matrix has to be stored. In the case of the AD lumping approach, the sparse transformation matrix is multiplied with the sparse mass matrix, resulting again in a sparse matrix. This sparse matrix can be easily stored and reused in all time steps.

Processing geometrically nonlinear analyses, it is possible to rebuild and store just the symmetric stiffness and mass matrix. The symmetric transformation operator remains constant and hast to be computed and stored just once in the beginning of the dynamic calculation. Thus, the occupied memory is still of similar magnitude as the standard consistent or row-sum lumped formulation. It is to  be noted, that the proposed AD lumping does not interfere with a matrix-free implementation of explicit dynamics, where the global stiffness and mass matrices are never assembled. The multiplications with the transformation operator can also be performed on element level.

\section{Conclusion}
\label{sec:conclusion}
Within this work a lumping approach for the explicit dynamic analysis of one-dimensional linear elasticity problems using dual test functions has been obtained. This method was proposed to overcome the weak approximation behavior obtained through standard row-sum lumping, and concurrently lower the computational costs in comparison to dynamic analyses based on consistent mass matrices. Since the focus is set on the fundamental aspects, the consideration of the 1D case is sufficient. The applicability of the method in general to the higher dimensional case and higher order PDEs has been shown concurrently in \citet{Nguyen.2023}. In upcoming work, we will also extend our method to the 2D case of shells, which will allow a more robust study on efficiency. So far, the detailed findings of this study are: 

\begin{itemize}
\item The investigated IG duals provide a consistent diagonal mass, but result in a fully populated stiffness matrix. Through the use of AD as test functions neither diagonal nor full, but banded system matrices are obtained. The bandwidth depends on the initial polynomial order $p$ of the corresponding NURBS basis functions and on the chosen degree of reproduction $q$. Although the mass matrix is not a diagonal one, its entries decrease rapidly from the main to the secondary diagonals, because they are diagonally-dominant.

\item It has been shown, that both formulations can be applied to existing B-Spline based IGA formulations by simply applying a transformation to the system matrices after assembly based on Bubnov-Galerkin routine. A novel contribution to the topic of dual basis functions within IGA is the demonstration that the method can be successfully applied for the general case of NURBS basis functions. When using NURBS with non-constant weighting of the control points, small adjustments on element level have to be considered. The transformation operator, a constant matrix, can be extracted from the resulting static and dynamic equations of equilibrium and ensures a similar accuracy for computations in comparison to standard formulations based on NURBS as shape and test functions. 

\item The proposed dual lumping scheme overcomes the drawbacks of the dual basis proposed by \citet{Anitescu.2019}, which are complicated numerical integration due to jumps in the test functions. The implementation of our proposed method is straightforward and the additional computational effort can be shifted from operations on element level to a global treatment, which is probably a main advantage of this method over the work of \citet{Nguyen.2023}. Furthermore, the AD approach yields a continuous basis, which does not provide restrictions for the employed quadrature schemes, and the ability of reproducing the initial basis function degree $p$ is provided. 

\item As the intent of our study was to enable computationally efficient explicit dynamic computations, the row-sum method was also used for the investigated AD approach to obtain diagonal mass matrices. Applying additional row-sum lumping, a loss of accuracy has to be accepted. But as effect of the diagonally-dominant mass matrices, this additional lumping procedure is accompanied by a smaller error, which is still of reasonable magnitude. Thus, computations with the lumped AD approach are more accurate than the standard row-sum lumped approach and more efficient in time than the common consistent formulation. The influence of the degree of reproduction $q$ is noticeable on accuracy and computational costs. Maximum reproduction $q_\mathrm{max}$ guaranties maximum accuracy and minimal reproduction $q_\mathrm{min}$ yields lowest computational times and occupies a little less memory due to the smaller bandwidth of the stiffness matrix.

\item The investigated IG scheme can easily exceed the accuracy of \cite{Anitescu.2019}, providing consistent diagonal mass matrices with full reproduction. Applying IG test functions is almost identical to a common IGA formulation, using a pre-inverted mass matrix within the dynamic computation. It guaranties the same accuracy and does not require a costly inversion of the mass matrix. However, the computational efficiency is restricted by the fully populated stiffness matrix and thus prohibits a use of IG duals in explicit dynamics  due to the high amount of required memory.

\item Time measurements within this work are not as accurate as calculating the $L_2$-error and therefore, a more precise measurement and an optimized code might lead to a slightly different picture. So far, the applied dual lumping scheme and time integration method are not parallelized. After optimization of the new implemented algorithms, we conjecture that the results will allow to directly oppose computational time and accuracy to obtain a reliable value for declaring the efficiency of a specific lumping scheme. However, the main findings will not change. 

\item As stated before, the measurement of time is less reliable than the measurement of accuracy. A further look on other significant computational issues, e.g., the allocated memory for the system matrices, is necessary to decide between these approaches. That may be part of future work. So far, we can assume that the additional computational costs for implementing the AD lumping scheme will not neutralize the gains through the faster time integration based on a diagonal mass matrix. In comparison to the common Bubnov-Galerkin approach, where next to the system matrices for an efficient computation the inverse of the mass matrix is stored, only a few changes appear. $\mathbf{K}$ is turned into a non-symmetric matrix with slightly increased bandwidth, but remains banded. In return, only a diagonal mass matrix has to be stored and waives the computational effort for its inversion.

\item Measuring the efficiency by accuracy and computational costs is also related to the time integration method and the corresponding critical time step. In Sec.~\ref{sec:examples_dyn2}, it was shown that the lumped AD approach with maximum reproduction is not limited in accuracy by the applied row-sum lumping. Considering other techniques for time integration, like RK4, improves the results for higher polynomial orders $p$, while simple CDM restricts them to second order accuracy. Thus, for future work, the proposed formulation should be tested on even higher order time integration procedures, e.g. the mentioned Padé-based scheme \cite{Song.2022,Song.2023}, to examine the maximum accuracy level that can be obtained.

\item As the critical time step for explicit dynamics in IGA is mainly determined by the occurring outliers, it can also be worth considering a similar study combined with mass scaling approaches or the ideas published in \cite{Hiemstra.2021, Nguyen.2022}, which will definitely lower the computational costs of the consistent formulations as the amount of time steps is reduced. An outlier-free or in the case of mass scaling at least less outlier-affected formulation will obviously also raise the efficiency of all consistent formulations. The accuracy of the AD lumped approach may remain unchanged, but right now, we cannot predict the effects on that interaction with the computational costs in comparison to the improved consistent formulations.

\item The choice of an appropriate degree of reproduction within AD test functions probably depends on the studied kind of application. Thus, more complex numerical examples such as beam structures with multiple degrees of freedom per control point or two-dimensional plate or shell formulations have to be investigated subsequently to this work. However, this work already establishes important insights into the use of AD in structural dynamics. 

\end{itemize}



%%       \citet{<label>} ==> Jones et al. [21]
%%       \citep{<label>} ==> [21]
%%

  %\bibliographystyle{elsarticle-num-names} 
  %\bibliography{references}


%\begin{thebibliography}{00}
%
%
%%% \bibitem[Author(year)]{label}
%%% Text of bibliographic item
%\end{thebibliography}
\end{document}